\documentclass[12pt]{article}
\pdfoutput=1
\usepackage{hyperref}
\usepackage{graphicx}
\usepackage{subcaption}
\graphicspath{{./picture/}}
\usepackage{empheq}
\usepackage{amssymb}
\usepackage{mathtools}
\usepackage{commath}
\usepackage{verbatim}
\usepackage{xcolor}
\usepackage{soul}
\usepackage{booktabs}
\usepackage{amsmath}
\usepackage{hhline}
\usepackage[scale={0.75,0.8}]{geometry}
\usepackage{url}
\usepackage{caption}
\usepackage[numbers, sort&compress]{natbib}
\usepackage{epsfig}
\usepackage{lscape}
\usepackage{mathrsfs}
\usepackage{nicefrac} 
\usepackage{upgreek}
\usepackage{bbm}
\usepackage{psfrag}
\usepackage{hyperref}
\newcommand{\p}{\textbf{p}}
\renewcommand{\k}{\textbf{k}}
\renewcommand{\l}{\textbf{l}}

\usepackage{hyperref}

\renewcommand{\k}{\textbf{k}}

\newcommand{\V}{\mathcal{V}}
\newcommand{\tildeg}{\tilde{g}}

\newcommand{\Rrad}{{\rm R}_{\rm rad}}

\newcommand{\aend}{a_{\rm end}}
\newcommand{\rhoend}{\rho_{\rm end}}
\newcommand{\rhoreh}{\rho_{\rm re}}

\newcommand{\wrehbar}{\bar{w}_{\rm re}}
\newcommand{\wreh}{w_{\rm re}}
\newcommand{\Nreh}{N_{\rm re}}

\begin{document}

\title{ 
\vspace{-2cm}
\begin{flushright}
{\scriptsize TUM-HEP-1093/17}
\end{flushright}
\vspace{0.5cm}
{\bf \boldmath CMB constraints on the inflaton couplings and reheating temperature in $\alpha$-attractor inflation} \\[8mm]}
\date{}

\author{Marco Drewes$^{a,b}$, Jin U Kang$^{c, d}$, Ui Ri Mun$^{d}$ \\ \\
{\normalsize \it$^a$ Physik Department T70, Technische Universit\"at M\"unchen,} \\ {\normalsize \it James Franck Stra\ss e 1, D-85748 Garching, Germany}\\
{\normalsize \it$^b$ Centre for Cosmology, Particle Physics and Phenomenology,}\\ {\normalsize \it Universit\'{e} catholique de Louvain, Louvain-la-Neuve B-1348, Belgium}\\
{\normalsize \it$^c$ Abdus Salam International Centre for Theoretical Physics,} \\ {\normalsize \it Strada Costiera 11, Trieste 34014, Italy}\\ 
{\normalsize \it$^d$ Department of Physics, Kim Il Sung University,} \\ {\normalsize \it RyongNam Dong, TaeSong District, Pyongyang, DPR.\ Korea} 
} 
\maketitle
\thispagestyle{empty}
\begin{abstract}
\noindent We study reheating in $\alpha$-attractor models of inflation in which the inflaton couples to other scalars or fermions. We show that the parameter space contains viable regions in which the inflaton couplings to radiation can be determined from the properties of CMB temperature fluctuations, in particular the spectral index. This may be the only way to measure these fundamental microphysical parameters, which shaped the universe by setting the initial temperature of the hot big bang and contain important information about the embedding of a given model of inflation into a more fundamental theory of physics. The method can be applied to other models of single field inflation.
\end{abstract}
\newpage
\tableofcontents

\section{Introduction}

The question about the origin of the cosmos has puzzled humans for millennia.
Modern cosmology allows us to understand most properties of the observable universe as the result of processes that occurred during the early stages of its evolution, when it was filled with a hot and dense plasma of elementary particles. 
This picture is supported by numerous observations that cover many orders of magnitude in length scales and time. It is, however, not known which mechanism set the initial conditions for this ``hot big bang'' or, more precisely, the \emph{radiation dominated epoch} in the cosmic history. In this paper, we discuss the possibility to obtain information about this mechanism from observations of the cosmic microwave background (CMB).

Observations of the CMB show that the primordial plasma was homogeneous and isotropic up to small temperature fluctuations \cite{Ade:2015xua} at temperatures of a few thousand Kelvin.
The most popular explanation is \emph{cosmic inflation} \cite{Starobinsky:1980te,Guth:1980zm,Linde:1981mu}, i.e., the idea that the universe underwent a period of exponential growth of the scale factor.
Indeed, the power spectrum of these temperature fluctuations has confirmed several predictions of cosmic inflation \cite{Ade:2015lrj}.
This makes the idea that the observable universe underwent a phase of accelerated cosmic expansion during its very early history very appealing. 
However, it is not known what was the driving force behind this rapid acceleration. Moreover, inflation dilutes matter and radiation and leaves a cold and empty universe. In contrast to that, the good agreement of the observed light element abundances with the predictions from \emph{big bang nucleosynthesis} (BBN) indicates that the universe was filled with a dense medium of relativistic particles in thermal equilibrium, which we in the following refer to as ``radiation'', and puts a lower bound of roughly 10 MeV on the temperature in the early universe \cite{Olive:2016xmw}.
Thus any viable theory for cosmic inflation should address at least following two questions: 
\begin{itemize}
\item[I)] What mechanism drove the inflationary growth of the scale factor?  
\item[II)] How did the transition to the hot radiation dominated epoch occur? 
\end{itemize}
These are fundamentally important questions not only for cosmologists, but also for particle physicists, who would like to understand how the idea of inflation can be embedded into a more general theory of nature. 
In the present work, we focus on the second question above. Regarding the first question, we adopt the viewpoint that inflation was caused by 
a scalar \emph{inflaton} field $\phi$ with a flat potential, 
which dominated the energy density of the universe and led to a negative equation of state. There exist many models of this single field inflation, see e.g. Ref. \cite{Martin:2013tda} for a partial overview. 

The rapid expansion during inflation diluted all pre-inflationary matter and radiation,
leaving a cold and empty universe.
The transition to the radiation dominated epoch occurred when the inflaton's energy density 
was transferred into relativistic particles via dissipative effects, see Ref. \cite{Amin:2014eta} for a recent review. 
This process is called \emph{cosmic reheating}.\footnote{
We use the term reheating in the general sense described below, which includes a possible ``preheating'' phase.  }
The reheating process lasts for a finite amount of time, which should be regarded as a separate era in the cosmic history, i.e. the \emph{reheating era}.
\begin{figure}[h!]
\begin{center}
\includegraphics[scale=0.6]{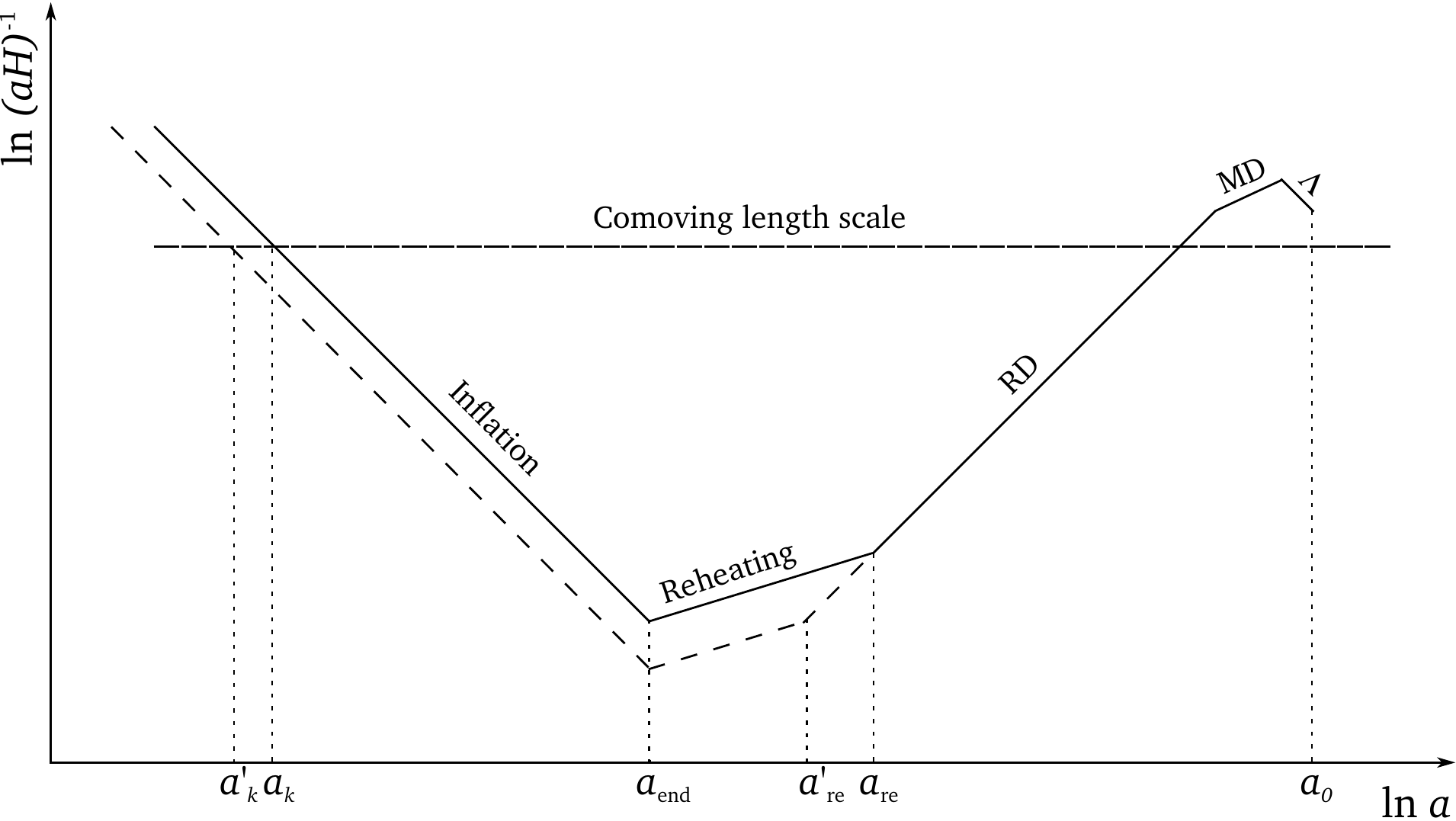}
\caption{Evolution of the comoving Hubble horizon in different epochs of cosmic history. 
RD, MD and $\Lambda$ indicate the radiation dominated era, matter dominated era and the current era of accelerated cosmic expansion , respectively. 
$H$ is Hubble parameter and $a$ the scale factor. $a_k, a_{\rm end}, a_{\rm re}$ and $a_0$ are the values of the scale factor at the horizon crossing of a reference mode with a comoving wave number $k$, the end of inflation, the end of reheating and the present time, respectively. 
The different slopes are a result of the different equation of state in different epochs. In order to visualise the effect 
of the reheating era
on the horizon crossing point, we used two line styles: solid line corresponds to the small dissipation rate $\Gamma$, i.e. the large $e$-fold number of reheating $N_{\rm re}$ and dashed line corresponds to the relatively large value of $\Gamma$, i.e. small $N_{\rm re}$. 
In both cases we have assumed that the equation of state parameter $\wreh$ remains approximately constant during reheating; if this were not the case, the slope of the lines would change during reheating. Our analysis, however, does not rely on this assumption because the effect on the CMB only depends on the average value $\wrehbar$, cf. Eq. (\ref{Rrad}).
For the larger $\Gamma$, the end of reheating lies further back in time.  
As a result, the inferred values for the scale factor at that moment and at the moment of horizon crossing decrease, as can be seen by comparing $a_{\rm re}$ and $a_{k}$ to $a'_{\rm re}$ and $a'_k$. 
This implies that the horizon crossing happens at the larger field value $\phi_k$ if $\Gamma$ is larger, where the slow-roll parameters are more suppressed, and hence the spectral index $n_s$ gets closer to 1. Conversely, a larger $n_s$ implies a larger $\Gamma$. Assuming that $\Gamma$ is a monotonically increasing function of the inflaton's coupling constant to radiation, the coupling constant is an increasing function of $n_s$.}
\label{Fig1}
\end{center}
\end{figure} 
The most important effect of the reheating era on the CMB lies in the modified expansion rate, which is illustrated in Fig.~\ref{Fig1}.
This affects the red-shifting of cosmological perturbations, 
and therefore the relation between physical scales of the CMB mode at the present time and at Hubble crossing of the modes during inflation.\footnote{Other possible signatures of reheating in the CMB include non-Gaussianities \cite{Bond:2009xx,Cicoli:2012cy} and curvature-perturbations \cite{Ringeval:2013hfa,Mazumdar:2014haa,Mazumdar:2015xka}.} 
This effect can be parametrised by a single number $\Rrad$ \cite{Martin:2006rs,Martin:2010kz}, which can be expressed in terms of the averaged equation of state during reheating $\wrehbar$ and
the ratio of the cosmic energy densities at the end of inflation $\rhoend$ and reheating $\rhoreh$ as
\begin{eqnarray}
\label{Rrad}
\ln \Rrad=
\frac{1-3\wrehbar}{12\left(1+\wrehbar\right)}
\ln\left(\frac{\rhoreh}{\rhoend}\right)\,,
\end{eqnarray}
or in terms of $N_{\rm re}$, the number of $e$-folds from the end of inflation until the end of reheating, as
\begin{eqnarray}\label{someRformula}
\ln \Rrad = \frac{N_{\rm re}}{4}\left(3\wrehbar-1\right)\,.
\end{eqnarray}
If $\wrehbar$ can be somehow fixed, which is often the case in a given model, then $N_{\rm re}$ or $\rhoreh$ can be used instead of $\Rrad$ to parametrise the effect of reheating on the CMB. 
In Ref.~\cite{Martin:2014nya} it has been shown that the CMB indeed contains enough information to treat $\Rrad$ as a meaningful independent fit parameter when constraining models of inflation. 
In view of various upcoming CMB observations,\footnote{ An overview of realistic sensitivities of various proposed experiments can be found in Ref.~\cite{Errard:2015cxa}. The CORE collaboration has already studied their potential to gather information about the reheating period \cite{Finelli:2016cyd}.} this provides strong motivation to study the potential of these measurements to say something about reheating. 

The derivation of constraints on inflationary models from CMB observations in principle requires knowledge of $\Rrad$, which depends not only on the inflaton potential, but also on the interactions between $\phi$ and the radiation. The lack of knowledge about these interactions imposes a systematic uncertainty on the derived constraints \cite{Kinney:2005in,Peiris:2006sj}, which can be quantified by the deviation of $\Rrad$ from unity.
One may, however, turn the tables and use this dependency to impose constraints on the reheating epoch \cite{Martin:2010kz,Adshead:2010mc,Easther:2011yq,Dai:2014jja,Munoz:2014eqa,Martin:2014nya,Cook:2015vqa,Cai:2015soa,Ueno:2016dim,Hardwick:2016whe,Lozanov:2016hid,Dalianis:2016wpu,Kabir:2016kdh,Choi:2016eif,Cabella:2017zsa,Nozari:2017rta,DiMarco:2017zek,Bhattacharya:2017ysa}.
Most of the previous works on CMB constraints on the reheating epoch has focused on constraints on the macroscopic parameters such as $N_{\rm re}$,  $\wrehbar$ and the \emph{reheating temperature} $T_{\rm re}$ at the onset of the radiation dominated era.  However, from the viewpoint of particle physics it is interesting 
to derive constraints on microphysical parameters, such as inflaton coupling to radiation. This is the main goal of the present paper. 

In Ref.~\cite{Drewes:2015coa} it has been pointed out that constraints on $\Rrad$ can be converted into constraints on the inflaton couplings to radiation, 
and that this relation can have a simple analytic form if reheating is primarily driven by perturbative processes. 
In the present work, we apply this idea to $\alpha$-attractor models \cite{Kallosh:2013xya,Kallosh:2013hoa,Kallosh:2013yoa,Kallosh:2013wya,Kallosh:2013maa,Galante:2014ifa} that cover a wide class of inflationary scenarios. 
A lower bound on the inflaton coupling in this class of models has previously been studied in Ref.~\cite{Ueno:2016dim}, where it was assumed that the inflaton couples to another scalar field $\chi$ via the interaction of the type $g\phi\chi^2$. We extend this analysis to different kinds of interactions and include the feedback of the produced radiation on the reheating process, which can severely limit the range of validity of the results.
This is indeed a non-trivial requirement. If the coupling constant is too large, the particle production is efficient enough to trigger a parametric resonance, and perturbative techniques cannot be applied. For very small values, reheating is not efficient enough to heat the universe to temperatures above 10 MeV, which is required for consistency with BBN. 
We also show that thermal corrections to the perturbative decay do not affect the CMB constraints in the perturbative regime. 
This allows us to establish analytic relations between the inflaton couplings and observable quantities, and to identify their range of applicability.
 
This article is organised as follows. 
In Section \ref{Section2} we briefly review the CMB-constraints on reheating and set up the theoretical framework and describe the methodology adopted in this work. 
In Section \ref{Section3} we apply this to derive constraints on interactions via which the inflaton $\phi$ may couple to other scalars $\chi$ or fermions $\psi$. 
We consider interactions of the form $g\phi\chi^2$,  $h\phi\chi^3$ and $y\phi\bar{\psi}\psi$ and establish relations between the coupling constants $g$, $h$ and $y$ and CMB observables. 
These relations hold in the parameter regime in which reheating is entirely driven by perturbative processes. 
We show that there exists a range of values of the coupling constants for which this is the case. 
We conclude in Section \ref{Conclusions}. In the appendix, we present the expressions for dissipation rates 
used in the main text.

\section{Reheating in \texorpdfstring{$\alpha$}{Lg}-attractor E-model}\label{Section2}

\subsection{General considerations}\label{GeneralConsiderations}

The expectation value of the inflaton field $\varphi\equiv\langle\phi\rangle$ is often assumed to follow an equation of motion of the form
\begin{equation}\label{EOM}
\ddot{\varphi} + (3H + \Gamma)\dot{\varphi} + \partial_\varphi \V(\varphi)=0,
\end{equation}
see e.g. Ref. \cite{Cheung:2015iqa} and references therein for a recent derivation.
Here $a$ is the scale factor, $H=\dot{a}/a$ the Hubble rate and  
$\V(\varphi)$ the effective potential for $\varphi$, which includes all quantum and thermodynamic corrections  
to the bare potential $V(\phi)$ appearing in the Lagrangian. $\Gamma$ is an effective dissipation rate that leads to the transfer of energy from $\varphi$ to radiation. 
The flatness for the potential is often expressed in terms of the small ``slow roll parameters''
\begin{eqnarray}
\epsilon = \frac{1}{2} M_{pl}^2\left(\frac{\partial_\varphi\V}{\V}\right)^2 \  , \ 
\eta = M_{pl}^2\frac{\partial_\varphi^2\V}{\V} \,.\label{SlowRollDef}
\end{eqnarray}
Here $M_{pl}=2.435\times 10^{18}$ GeV is the reduced Planck mass.
The inflationary stage or \emph{slow roll stage}, during which $\V(\varphi)$ dominates the energy budget, is characterised by $\epsilon,\eta\ll 1$ and an exponential growth of the scale factor.
Inflation roughly ends (and reheating begins) when the $\partial_\varphi\V(\varphi)$ term in Eq. (\ref{EOM}) exceeds $3H\dot{\varphi}$. 
Reheating may commence with a \emph{fast roll} phase during which $\varphi$ quickly moves towards the minimum of its potential.
The  beginning of this phase can be defined as the moment when the universe stops accelerating (the equation of state exceeds $w>-1/3$),
which approximately happens when the slow roll parameter $\epsilon$ exceeds unity. 
It is followed by an \emph{oscillatory phase} during which one typically observes the hierarchy $\Gamma\dot{\varphi}\ll 3H\dot{\varphi} \lesssim \partial_\varphi\V(\varphi)$.
Due to the relative smallness of $\Gamma\dot{\varphi}$, $\varphi$ loses only a small fraction of its energy per oscillation. However, the total amount of energy transferred from $\varphi$ to radiation is the largest at early times because $\rho_\varphi$ is huge in the beginning and red-shifted at later times. 
The oscillations end when $\Gamma=H$, and shortly afterwards the energy density $\rho_\gamma$ of the radiation exceeds $\rho_\varphi$. 
This moment is often referred to as the onset of the radiation dominated era, which is obviously true from an energy budget viewpoint.\footnote{From a particle physics viewpoint the phase space distribution functions of the produced particles can be very important, as they affect the rate at which microscopic processes occur. Because of this, particle physicists often define the onset of the radiation dominated era as the moment when the radiation has reached thermal equilibrium and can be characterised by a well-defined \emph{reheating temperature}, 
see e.g. Refs. \cite{Mazumdar:2013gya,Harigaya:2013vwa,Harigaya:2014waa,Mukaida:2015ria} for some recent discussions.
The \emph{thermalisation period} between 
$\rho_\gamma=\rho_\varphi$ 
and this moment should then be counted as a part of the reheating era. 
However, from a viewpoint of the expansion history (which is what primarily affects CMB modes), ``radiation domination" is usually defined as a period with an equation of state $w=1/3$. For relativistic particles, this is in good approximation fulfilled (almost) independently of their phase space distribution.} 
For the present purpose, we, therefore, define the reheating era as the period between the moments when $\epsilon=1$ and $H=\Gamma$. 
Moreover, we assume that the conversion of $\rho_\varphi$ into
 $\rho_\gamma$ occurs instantaneously at the end of the reheating era and define the reheating temperature via the relation 
\begin{equation}\label{RHOofT}
\rho_{\rm re}=\frac{\pi^2 g_*}{30}T_{\rm re}^4\equiv \rho_\gamma\big|_{\Gamma=H} \,.
\end{equation}
Here $g_*$ is the effective number of relativistic degrees of freedom. $T_{\rm re}$ should therefore not be understood as a temperature in the microphysical sense of phase space distribution functions, but as a convenient effective parameter to associate the onset of the radiation dominated era with an energy scale. It corresponds to a physical temperature if the equilibration is very rapid.

Some remarks on the effective potential are in order. 
The functional form of the effective potential $\V(\varphi)$ 
in general differs from the potential  $V(\phi)$ 
appearing in the Lagrangian. $\V(\varphi)$ includes quantum and thermodynamic corrections, through which it is sensitive to the way how $\phi$ couples to other fields, see Ref. \cite{Cheung:2015iqa} for a recent discussion. 
It is well-known that there exist models of inflation in which these corrections are crucial in the regime of field values $\varphi$ where inflation takes place, see e.g. Refs.~\cite{Marzola:2016xgb,Artymowski:2016dlz,Choi:2016eif} for some explicit examples.
This means that the parameters in $\V(\varphi)$ that can be constrained from CMB data are in principle not independent of the inflaton's couplings to other fields.
In addition, there are corrections from gravity that may not be negligible for $\varphi/M_{pl}>1$. 
We assume that these effects are negligible during reheating, which usually occurs at small (sub-Planckian) field values $\varphi$ in the scenarios that we consider. 
We may therefore treat the inflaton couplings to radiation near $\varphi\simeq 0$ (which drive reheating)
and the parameters in $\V(\varphi)$ at the field values $\varphi_k$ where the observable $\phi$-mode $k$ crosses the horizon during inflation as independent fit parameters. \footnote{There is a small caveat in this argument because we express the inflaton mass near the potential minimum $\varphi\simeq 0$ in terms of the parameters in the potential obtained at a different scale $\varphi\simeq \varphi_k$, cf. Eq.~(\ref{m_phi}). In principle we should take proper account of the ``running'' of the parameters. However, given the current observational error bars and the wide range of inflaton couplings consistent with them, such a refined treatment does not seem to be necessary at this stage.} 
To simplify the notation, we in the following do not distinguish between $\V$ and $V$ and between $\varphi$ and $\phi$.

\subsection{Relation to CMB parameters}\label{RelaToCMB}

The most important parameters in this context that can be extracted from the CMB are the amplitude of the scalar perturbations $A_s$, the tensor-to-scalar ratio $r$ and the spectral index $n_s$, which are to be evaluated at some reference scale, i.e., for a specific mode of 
the inflaton fluctuations with a comoving wave number $k$.
In our analysis, we use CMB data at a pivot scale $k/a_0=0.05\, {\rm Mpc}^{-1}$, where $a_0$ is the scale factor at the present time.
The observables can be related to the slow roll parameters and value of $H$ at the moment when the mode $k$  crosses the horizon,
\begin{eqnarray}
n_s=1-6\epsilon_k+2\eta_k \ , \quad  r=16\epsilon_k \ , \quad H_{k}=\frac{\pi M_{pl}\sqrt{r A_s}}{\sqrt{2}} \,,\label{CMBvsSlowRoll}\end{eqnarray}
where we have used the slow roll approximation and $A_s=H^4/(4\pi^2\dot\phi^2)$. From the slow-roll approximation, we also obtain
\begin{eqnarray} 
H_k^2&\simeq&\frac{V(\phi_k)}{3M_{pl}^2}\label{Feq} \,.
\end{eqnarray} 
Here $\phi_k$ is the value of the scalar field $\phi$ at the Hubble crossing of mode $k$.
Eqns. \eqref{CMBvsSlowRoll}-\eqref{Feq} and \eqref{SlowRollDef} establish the relation between the observable quantities $n_s$, $A_s$ and $r$ and the parameters in the potential at leading order in $\epsilon_k$ and $\eta_k$.\footnote{More precisely, quantities $n_s$, $r$ and $A_s$ can determine $\phi_k$ and two parameters in $\V(\varphi)$ in a given model.}
We work at this order in the following, which is sufficient in view of the present data. The interpretation of future CMB observations may require the inclusion of higher order terms \cite{Martin:2016iqo}.
The sensitivity of the CMB to the reheating era primarily comes from the fact that the equation of state parameter $w$ during reheating is different from inflation or radiation domination. 
The energy density redshifts as 
\begin{equation} \label{rho-w}
\rho(N) = \rhoend \exp\left(-3 \int_0^N[1+w(N')]\, dN'\right) \,,
\end{equation}
where $N$ is the $e$-folding number from the end of inflation, i.e. $N=\ln(a/\aend)$.
Then we can write the Friedmann equation during reheating as 
\begin{eqnarray} \label{H2}
H^2 = \frac{\rhoend}{3 M_{pl}^2}\exp\left(-3 \int_0^{N}[1+\wreh(N')] dN'\right) 
\,.
\end{eqnarray}
Using the fact that reheating ends when $\Gamma=H$ at $N=N_{\rm re}$ and using Eq. \eqref{H2}, 
we find
\begin{eqnarray} \label{Nre-GammaConstraint}
\Nreh=\frac{1}{3(1+\wrehbar)}\ln\left(\frac{\rho_{\rm end}}{3\Gamma^2 M_{pl}^2}\right) \qquad 
\end{eqnarray}
or 
\begin{eqnarray}
\Gamma=\frac{1}{ M_{pl}}\left(\frac{\rhoend}{3}\right)^{1/2} e^{-3(1+\wrehbar) \Nreh/2}\, \label{GammaConstraint}
\end{eqnarray}
where $\wrehbar$ is the averaged equation of state parameter during reheating, which is defined as
\begin{equation} \label{wrehbar}
\wrehbar= \frac{1}{N_{\rm re}}\int_0^{N_{\rm re}} w(N) dN \,.
\end{equation}
Using Eq. \eqref{Rrad} for the definition of the reheating parameter $\Rrad$, we can rewrite Eq. \eqref{GammaConstraint} as   
\begin{equation}
\Gamma= \frac{1}{ M_{pl}}\left(\frac{\rhoend}{3}\right)^{1/2} \Rrad^{6(1+\wrehbar)/(1-3 \wrehbar)} \,. \label{GammaConstraint-Rrad}
\end{equation}
The energy density at the end of inflation $\rho_{\rm end}$ is
\begin{eqnarray}
  \rho_{\rm end} \simeq \left(1+\frac{\epsilon_{\rm end}}{3}\right)V(\phi_{\rm end}) = {4\over 3}V(\phi_{\rm end})\equiv {4\over 3}V_{\rm end}\,,
\label{rhoend}
\end{eqnarray}
where $\epsilon_{\rm end}=1$ and $\phi_{\rm end}$ are values of the slow-roll parameter $\epsilon$ and the scalar field $\phi$, respectively, at the end of inflation.
Then Eq. \eqref{GammaConstraint} (or \eqref{GammaConstraint-Rrad}), once $\wrehbar$ is specified, allows us to convert a constraint on $\Nreh$ (or $\Rrad$) into a constraint on the damping rate $\Gamma$ in the moment when it equals $H$, and hence on microphysical parameters. 
Due to the feedback of the produced radiation on the inflaton dynamics,
the effective damping rate $\Gamma$ in general is a function of time.\footnote{
In addition to the feedback effect from the produced radiation, $\Gamma$ may also exhibit a time dependence due to the coupling of $\phi$-modes to the rapidly oscillating condensate $\langle\phi\rangle$, which has e.g. been studied in Ref.~\cite{Nurmi:2015ema}. For the interactions that we study here (which depend on $\phi$ linearly) this effect is of higher order in the inflaton couplings. 
In cases where it is significant, it is generally not justified to use an effective kinetic equation of the form (\ref{EOM}), the validity of which is based on a strong hierarchy between the microscopic timescale $\sim 1/m_\phi$ and the macroscopic time scales $1/\Gamma$ and $1/H$, cf. e.g.~\cite{Drewes:2012qw,Cheung:2015iqa}.
} 
This is true even in the perturbative regime, i.e., when reheating is driven by decays and scatterings of individual inflaton quanta \cite{Boyanovsky:1995ema} (cf. \cite{Mukaida:2012qn,Mukaida:2012bz,Drewes:2013iaa,Mukaida:2013xxa} for explicit recent results).
While the feedback can significantly modify the \emph{thermal history} during perturbative reheating \cite{Drewes:2014pfa}, it was argued in Ref. \cite{Drewes:2015coa} that it has no big effect on the \emph{expansion history} (which is what the CMB is sensitive to) for the interactions we consider here,
and that in the absence of a parametric resonance the dissipation rate $\Gamma$ 
in Eq. \eqref{GammaConstraint} can be approximated by the vacuum decay rate. 
The reason is that even large relative changes in the radiation density do not modify the expansion rate $H$ significantly as long as the radiation density is subdominant in comparison to the inflaton energy.
We confirm this statement in Section \ref{Section3}.     



We now establish relations expressing the parameter $N_{\rm re}$ (or $\Rrad$) in Eq. (\ref{GammaConstraint}) (or (\ref{GammaConstraint-Rrad})) 
and hence $\Gamma$ in terms of observable quantities and potential parameters. 
Using $3H\dot{\phi} + \partial_\phi V\simeq0$ and $H^2\simeq V/ (3 M_{pl}^2)$ during the slow roll, 
the $e$-folding number $N_k$ between the horizon crossing of a perturbation with wave number
$k$ and the end of inflation can be estimated as
\begin{eqnarray}
N_k=\ln\biggl({a_{\rm end}\over a_{k}}\biggr)
= \int_{\phi_k}^{\phi_{\rm end}} \frac{H d\phi}{\dot{\phi}}
\simeq -{1\over M_{\rm pl}^2}\int_{\phi_k}^{\phi_{\rm end}} d\phi \, {V(\phi)\over  \partial_\phi V(\phi)} \,.
\label{defNk}
\end{eqnarray}
where $a_k$ and $a_{\rm end}$ are the scale factors at horizon crossing of mode $k$ and at the end of inflation, respectively.
From Eqns. \eqref{wrehbar} and \eqref{rho-w} we can write the $e$-folding number of the reheating epoch as
\begin{eqnarray}
N_{\rm re}
&=&\ln\biggl({a_{\rm re}\over a_{\rm end}}\biggr)
=-{1\over 3(1+\wrehbar)}
\ln\biggl(\frac{\rho_{\rm re}}{\rho_{\rm end}}\biggr),
\label{NreFormula}
\end{eqnarray}
where $a_{\rm re}$ and $\rho_{\rm re}$ are the scale factor and energy density, respectively, at the end of reheating.
Using the fact that $k a_k=H_k$ at horizon crossing we can write
\begin{eqnarray}
0=\ln\biggl({k\over a_kH_k}\biggr)=\ln\biggl({a_{\rm end}\over a_k}{a_{\rm re}\over a_{\rm end}}
{a_0\over a_{\rm re}}{k \over a_0 H_k}\biggr).
\label{NNN0}
\end{eqnarray}
Here $a_0$ is the scale factor at the present time. 
From Eqns. (\ref{defNk})-(\ref{NNN0}), it follows that
\begin{eqnarray}
N_k+N_{\rm re}+\ln\biggl({a_{0}\over a_{\rm re}}\biggr)+\ln\biggl({k\over a_0H_k}\biggr)=0. 
\label{NNN2}
\end{eqnarray}
To proceed further, we need to assume that the universe was dominated by radiation after the end of reheating until the time of radiation-matter equality of the standard cosmology,\footnote{ This e.g. excludes, from the scope of this paper, scenarios in which the energy density of the universe is dominated by some heavy particle or field \cite{Kane:2015jia} and ``reheated''  again by its decay \cite{Scherrer:1984fd}.} 
and that there was no significant release of entropy into the primordial plasma.
The latter assumption about entropy is needed to actually ``measure'' $\Gamma$ from the CMB (rather than just obtaining an upper bound).\footnote{If entropy was released into the plasma after reheating, then back-extrapolation of the present CMB temperature leads to an overestimate of the temperature before the moment of the release.} 
Under this assumption, we can write
\begin{eqnarray}
{a_{\rm re}\over a_0}=\left(\frac{43}{11g_{s*}}\right)^{{1}/{3}}{T_0\over T_{\rm re}}=\left(\frac{43}{11g_{s*}}\right)^{{1}/{3}}
\left({\pi^2 g_{*} T_0^4\over 30 \rho_{\rm re}}\right)^{1/4},
\label{2.5.4}
\end{eqnarray}
where $g_{s*}$ is the number of relativistic degrees of freedom for the entropy density. 
$T_0=2.725$~K is  the temperature of the CMB at the present time. 
Together with Eqns. (\ref{RHOofT}), (\ref{rhoend}) and (\ref{NreFormula}), this allows us to express the reheating temperature in terms of $\wrehbar$ and $N_{\rm re}$, 
\begin{eqnarray}
T_{\rm re}=\mathrm{exp}\Bigg[-\frac{3}{4}(1+\wrehbar)N_{\rm re}\Bigg]
\left(\frac{40V_{\rm end}}{g_*\pi^2}\right)^{1/4}.
\label{2.5.8}
\end{eqnarray}
Using Eqns. (\ref{rhoend}) and (\ref{NreFormula}), we can also relate $\rho_{\rm re}$ to $V_{\rm end}$,
\begin{eqnarray}
  \rho_{\rm re}={4\over 3}V_{\rm end}\left({a_{\rm re}\over a_{\rm end}}\right)^{-3(1+\wrehbar)}=
{4\over 3}V_{\rm end}e^{-3 N_{\rm re} (1+\wrehbar)} \,.
\end{eqnarray}
Inserting this into Eq. (\ref{2.5.4}) yields
\begin{eqnarray}
\ln\left({a_{\rm re}\over a_0}\right)= {1\over 3}\ln\left(\frac{43}{11g_{s*}}\right)
+{1\over 4}\ln\left({\pi^2 g_{*}\over 30}\right) 
+{1\over 4}\ln\left({3T_0^4\over 4 V_{\rm end}}\right)+{3N_{\rm re}(1+\wrehbar)\over 4} \,. 
\label{2.5.XX}
\end{eqnarray}
Using Eqns. (\ref{CMBvsSlowRoll}) and (\ref{2.5.XX}) into Eq. (\ref{NNN2}) gives a useful expression for  $N_{\rm re}$,
\begin{eqnarray}
N_{\rm re}=\frac{4}{3\wrehbar-1}\Bigg[N_{k}+\ln\left(\frac{k}{a_{0}T_{0}}\right)+\frac{1}{4}\ln\left(\frac
{40}{\pi^2g_{*}}\right)+\frac{1}{3}\ln\left(\frac{11g_{s*}}{43}\right) 
-\frac{1}{2}\ln\left(\frac{\pi^2M_{\rm pl}^2\:r\:A_{s}}{2V_{\rm end}^{1/2}}\right)\Bigg] \, 
\label{2.5.7}
\end{eqnarray}
and, using the definition of $\Rrad$ given by Eq. \eqref{someRformula}, the above formula can be written as 
\begin{equation}
\ln\Rrad= N_{k}+\ln\left(\frac{k}{a_{0}T_{0}}\right)+\frac{1}{4}\ln\left(\frac
{40}{\pi^2g_{*}}\right)+\frac{1}{3}\ln\left(\frac{11g_{s*}}{43}\right) 
-\frac{1}{2}\ln\left(\frac{\pi^2M_{\rm pl}^2\:r\:A_{s}}{2V_{\rm end}^{1/2}}\right) \,. \label{2.5.7_Rrad}
\end{equation}
In Eqns. \eqref{2.5.7} and \eqref{2.5.7_Rrad}, $N_k$ can be expressed in terms of the potential parameters and CMB data in the following way. $N_k$ is given by Eq. (\ref{defNk}), 
which requires to specify $\phi_{\rm end}$ and $\phi_k$. 
$\phi_{\rm end}$ can be found in terms of the parameters in $V(\phi)$ by 
solving $\epsilon=\frac{1}{2}M_{pl}^2\big(\frac{\partial_{\phi} V}{V}\big)^2\Big|_{\phi_{\rm end}}=1$. 
$\phi_k$ can be expressed in terms of the CMB parameters $n_s$ and $r$ by solving first two equations of Eq. (\ref{CMBvsSlowRoll}) 
for $\phi_k$.  Using these $\phi_{\rm end}$ and $\phi_k$, 
Eq. (\ref{defNk}) gives $N_k$ in terms of the potential parameters and CMB data. 
Now inserting $N_k$ obtained this way into Eq. (\ref{2.5.7}) (or \eqref{2.5.7_Rrad}), which is to be plugged into Eq. (\ref{GammaConstraint}) (or \eqref{GammaConstraint-Rrad}), 
one can find an expression for $\Gamma$ 
entirely in terms of inflaton potential parameters and observable quantities, 
provided that the averaged equation of state during reheating $\wrehbar$ is approximately determined by inflaton potential. 
This allows us to ``measure" the effective dissipation rate $\Gamma$ at the end of reheating, i.e. in the moment when $\Gamma=H$, 
from CMB data for a given inflaton potential.
Since $\Gamma$ and the parameters in $V(\phi)$ (in the slow roll regime) are simultaneously obtained from the same CMB data,
this is only possible within a fixed model.
Vice versa, reliable constraints on $V(\phi)$ in a given model can only be derived from CMB data if the effect of the reheating period is properly taken into account. 
This supports the viewpoint that a detailed study of an inflationary model should be connected to the particle physics models in which it can be embedded \cite{Martin:2016iqo}.

The previous considerations are largely independent of the microphysics of reheating. 
We assume that the perfect fluid description of the energy momentum tensor, which is the basis for the Friedmann equations, holds during reheating, so that we can parametrise the macroscopic properties of the material that fills the universe by a single parameter $w$ and use the relation $\rho \propto a^{-3(1+w)}$. This is a very weak assumption that certainly holds if the inflaton directly dissipates its energy into relativistic particles, as in the examples considered below. Basically, the only nontrivial physical assumption one has to make is that	 the energy density of the universe was dominated by radiation, which simply cools down according to the $T\propto a^{-1}$ law, between reheating and the beginning of the matter dominated era in the standard cosmology. 
While it is obvious that any knowledge about $\Gamma$ in principle enables us to constrain the inflaton couplings, it is not clear that one can establish a simple relation between the CMB observables
and the microphysical coupling constants. The reason is that the reheating process may be driven by highly non-linear far from equilibrium processes, such as a parametric \cite{Traschen:1990sw,Kofman:1994rk,Kofman:1997yn} or tachyonic \cite{Felder:2000hj} resonance. 
In such cases, the dependence of $\Gamma$ on model parameters cannot be extracted in a simple way. 
However, in Ref. \cite{Drewes:2015coa} it was pointed out that such a relation may be derived analytically if reheating is primarily driven by perturbative processes. This may not seem obvious at first sight because, due to feedback effects, the time evolution of $\Gamma$ during the reheating era in general depends not only on the inflaton couplings, but also on the interactions amongst the produced particles. This is the case even if reheating can be treated by perturbative methods  \cite{Drewes:2013iaa,Drewes:2014pfa}.
The reason why the determination of $\Gamma$ from the CMB is not affected by feedback effects is that, in the perturbative regime, these primarily modify the \emph{thermal history} of the plasma.
The CMB, on the other hand, is primarily sensitive to the \emph{expansion history}.

\subsection{Application to the \texorpdfstring{$\alpha$}{Lg}-attractor E-model}

\begin{figure}[htp!]
\begin{center}
\includegraphics[scale=0.57]{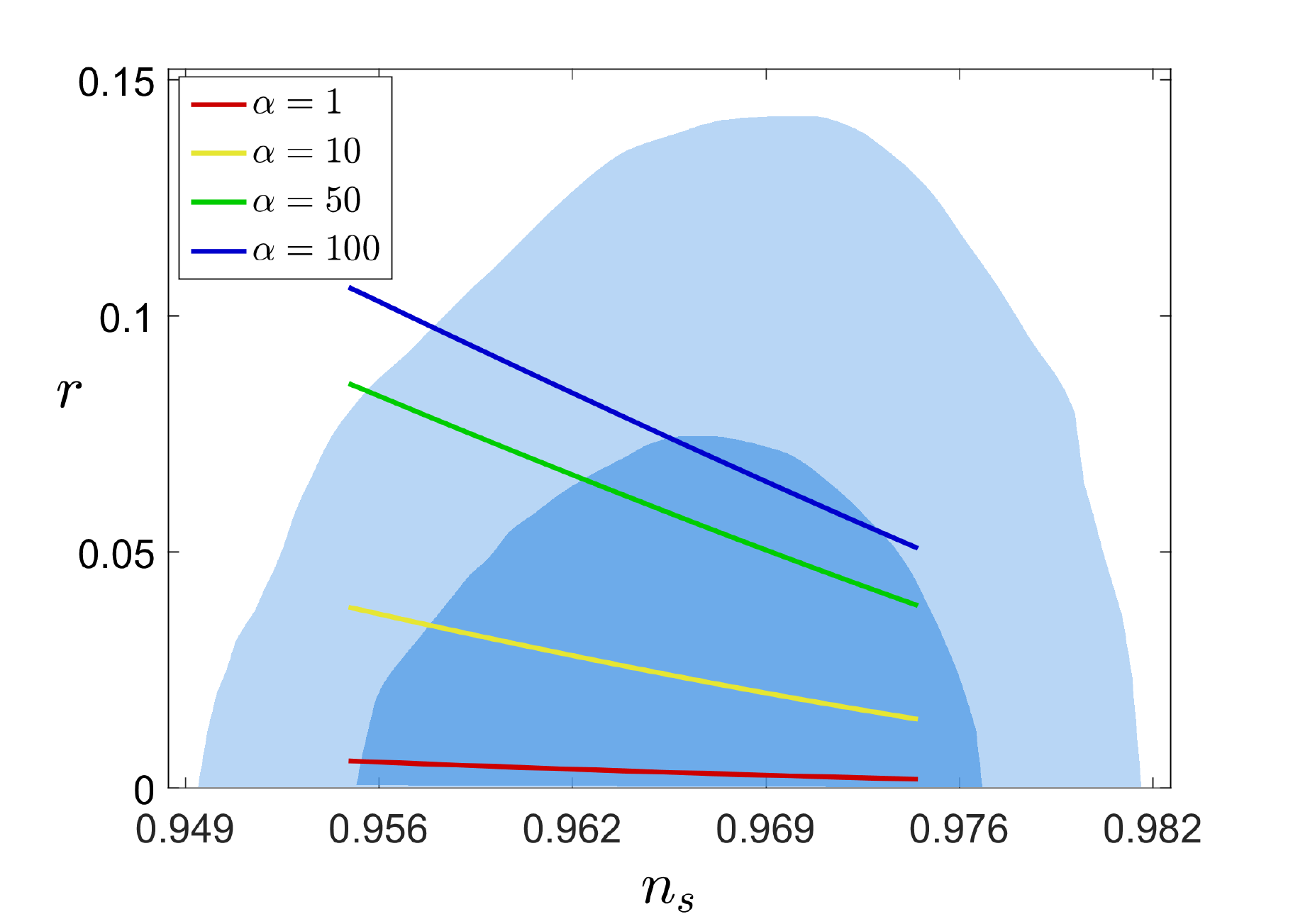}
\end{center}
\caption{Relationship between $r$ and $n_s$ for various values of $\alpha$ in the $\alpha$-attractor E-model with $n=1$.
The sky blue and light blue areas display marginalized joint confidence contours for $(n_s,r)$ at the $1\sigma$ ($68\%$) and 2$\sigma$ ($95\%$) CL \cite{Ade:2015lrj}, respectively.
The solid lines correspond to Eq. \eqref{scalartensor} for fixed  $\alpha$ and lie within 2$\sigma$ confidence interval of $n_s$ 
($n_s=0.9645 \pm 0.0098$ \cite{Ade:2015lrj}). The different values of $\alpha$ are indicated by different colours.}\label{planck_nsr}
\end{figure}

\begin{figure}[htp!]
\begin{subfigure}{0.5\textwidth}
	\includegraphics[width=1\linewidth, height=6.2cm]{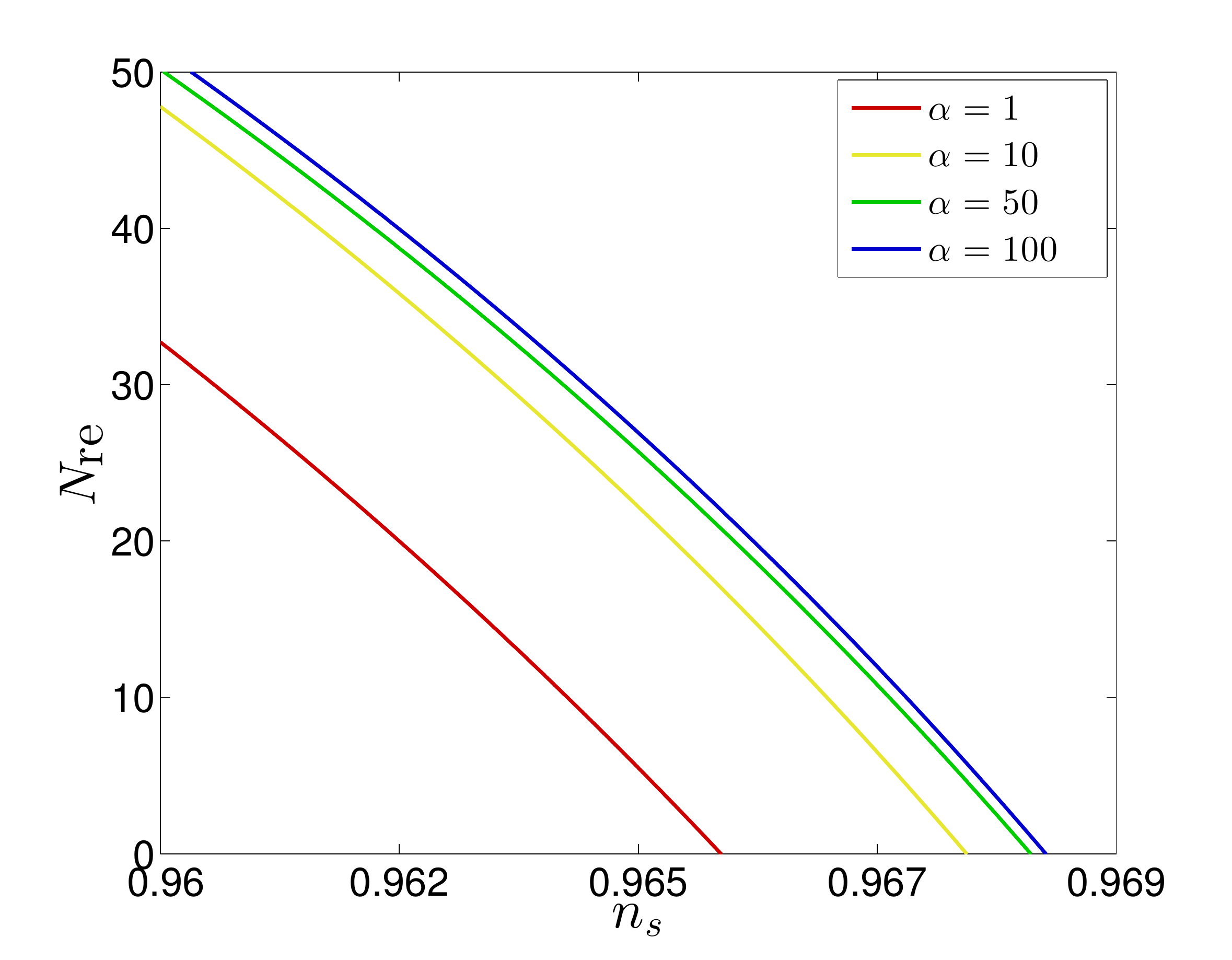}
	\caption{•}
\end{subfigure}
\begin{subfigure}{0.5\textwidth}
	\includegraphics[width=1\linewidth, height=6.2cm]{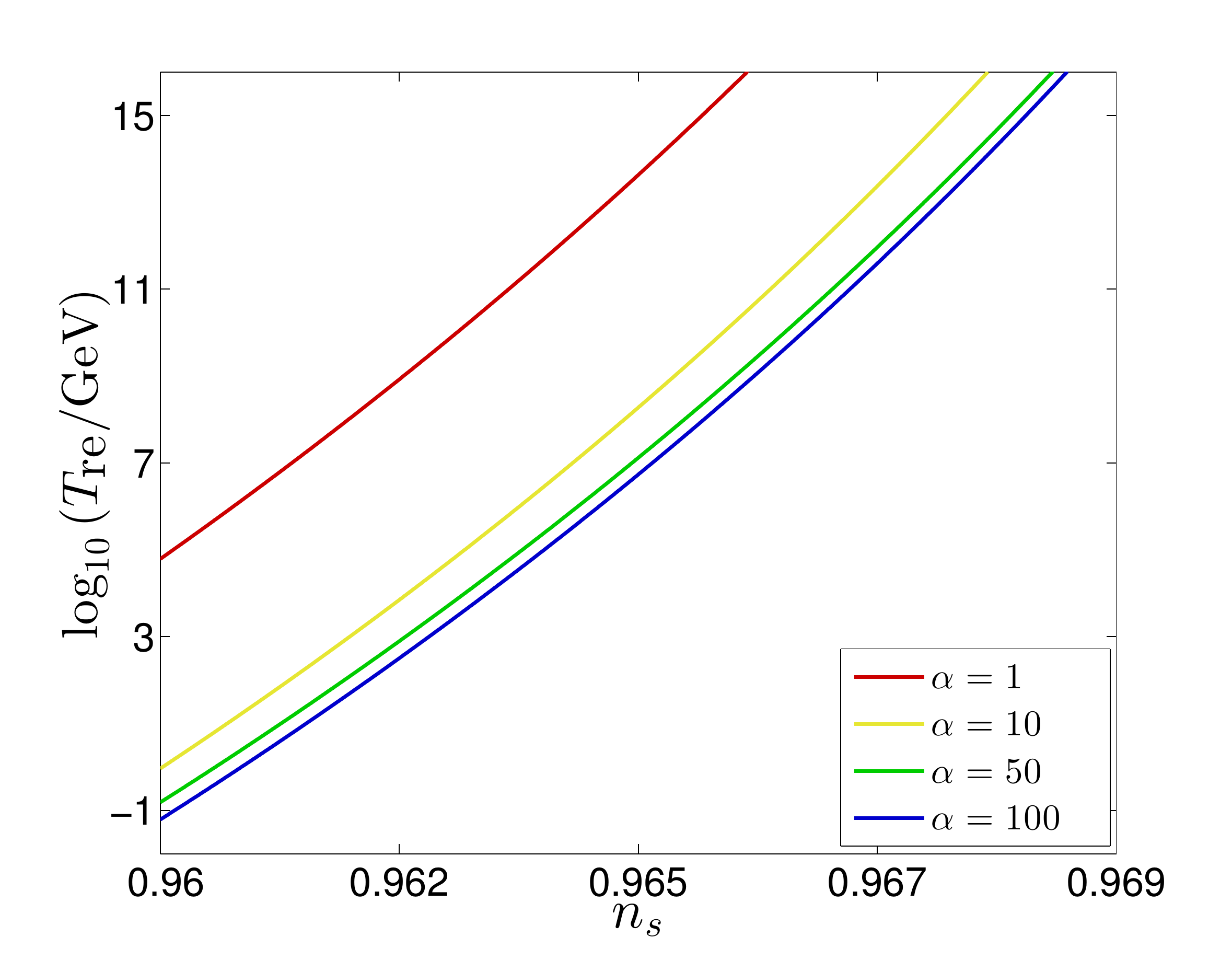}
	\caption{•}
\end{subfigure}
\caption{The spectral index dependence of $N_{\rm re}$ (left) and $T_{\rm re}$ (right) in $\alpha$-attractor E-model with $n=1$ for various values of $\alpha$.  
We have taken $\wrehbar=0$ and made use of Planck results \cite{Ade:2015lrj} $A_s=10^{-10}e^{3.094}$. For $T_{\rm re}$ we used Eq. \eqref{2.5.8} with $g_*=100$. The colour coding is used to denote different values of $\alpha$. 
}
\label{Nre-Tre}
\end{figure}

The $\alpha$-attractor E-model is specified by the potential  \begin{equation}\label{potential}
V=\Lambda^4\Big(1-e^{-\sqrt{\frac{2}{3\alpha}}\frac{\phi}{M_{pl}}}\Big)^{2n}.
\end{equation}
Let us first consider relationship between the potential parameters, the CMB data and reheating parameters ($N_{\rm re}$ or ${\rm R}_{\rm rad}$), applying the general recipe explained in the previous subsection \ref{RelaToCMB}.   
The e-folding number  $N_k$  between the horizon crossing of the perturbation with a comoving wave number $k$ and the end of inflation is obtained from Eq. (\ref{defNk}),
\begin{equation}\label{Nk}
N_k=\frac{3\alpha}{4n}\Bigg[e^{\sqrt{\frac{2}{3\alpha}}\frac{\phi_{k}}{M_{pl}}}-e^{\sqrt{\frac{2}{3\alpha}}\frac{\phi_{\rm end}}{M_{pl}}}-\sqrt{\frac{2}{3\alpha}}\frac{(\phi_k-\phi_{\rm end})}{M_{pl}}\Bigg]\;.
\end{equation}
Using $\epsilon=\frac{1}{2}M_{pl}^2\big(\frac{\partial_{\phi} V}{V}
\big)^2\Big|_{\phi_{\rm end}}=1$,  we find that inflation ends when the field value is
\begin{equation}\label{phi_end}
\phi_{\rm end}=\sqrt{\frac{3\alpha}{2}}M_{pl}\ln\Bigg(\frac{2n}{\sqrt{3\alpha}}+1\Bigg)\;,
\end{equation}
so that 
\begin{equation} \label{V_end}
V_{\rm end} = \Lambda^4\left(\frac{2n}{2n+\sqrt{3 \alpha}} \right)^{2n} \,.
\end{equation}
From Eqns. (\ref{SlowRollDef}) and \eqref{CMBvsSlowRoll} we find 
\begin{eqnarray}
n_s
=1-{8n\Bigl(e^{\sqrt{2\over 3\alpha}{\phi_k\over M_{\rm pl}}}+n\Bigr)
\over3\alpha \Bigl(e^{\sqrt{2\over 3\alpha}{\phi_k\over M_{\rm pl}}}-1\Bigr)^2 } \ ,
\end{eqnarray}
which can be inverted to give
\begin{equation}\label{phi_k} 
\phi_k=\sqrt{\frac{3\alpha}{2}}M_{pl}\ln{\Bigg(1+\frac{4n+\sqrt{16n^2+24\,\alpha\, n\,(1-n_s)(1+n)}}{3\alpha(1-n_s)}\Bigg)}\;,
\end{equation}
and
\begin{equation}\label{scalartensor}
r=\frac{64n^2}{3\alpha\Big(e^{\sqrt{\frac{2}{3\alpha}}\frac{\phi_k}{M_{pl}}}-1\Big)^2}
=\frac{192\, \alpha\, n^2\,(1-n_s)^2}{\left[4n+\sqrt{16n^2+24\,\alpha\, n\,(1-n_s)(1+n)}\right]^2}
\;,
\end{equation}
where we used Eq. \eqref{phi_k} in the second equality to express $r$ in terms of $n_s$, $n$ and $\alpha$.
If $r$ were measured, we could determine both, $\alpha$ and $\Gamma$ from data (for fixed $n$). 
Since observations currently only provide an upper bound for $r$, we at this stage can only determine $\Gamma$ if we fix both, $n$ and $\alpha$.\footnote{ Eq. \eqref{scalartensor} can of course be 
inverted to express $\alpha$ in terms of $r$, $n_s$ and $n$. 
Thus one can choose either $\alpha$ or $r$ as an input parameter.}
Fig. \ref{planck_nsr} illustrates this relationship for $n=1$. Plugging Eq. \eqref{phi_k} into Eq. \eqref{Feq} and using $H_k$ from Eq. \eqref{CMBvsSlowRoll}, we can obtain a relation between $\Lambda$ and the rest of the parameters,
\begin{align}\label{Lambda}
\Lambda=M_{pl}\Bigg(\frac{3\pi^2rA_s}{2}\Bigg)^{1/4}\Bigg[\frac{2n(1+2n)+\sqrt{4n^2+6\alpha(1+n)(1-n_s)}}{4n(1+n)}\Bigg]^{n/2}\;.
\end{align}
\\
 From this and Eq. \eqref{scalartensor}, we see that for given $\alpha$ and $n$ the normalisation $\Lambda$ of the potential is fixed 
from the Planck observations for $n_s$ and $A_s$. Hence, $V_{\rm end}$ given in 
Eq. \eqref{V_end} is determined by $A_s$, $n_s$, $\alpha$ and $n$.
One can also express
$N_k$ as a function of $n_s$, $\alpha$ and $n$, using Eqns. \eqref{phi_end} and \eqref{phi_k} into Eq. \eqref{Nk}. 
Given that $N_k$, $r$ and $V_{\rm end}$ are determined by $n_s$, $A_s$, $\alpha$ and $n$,  Eq. \eqref{2.5.7} tells us that $N_{\rm re}$ is also determined by the same parameter set 
once the averaged equation of state during reheating $\wrehbar$ is known.\footnote{The reheating parameter $\Rrad$, however, does not require knowledge of $\wrehbar$, see Eq. \eqref{2.5.7_Rrad}. In any case, it is necessary to specify $\wrehbar$ in order to constrain the dissipation rate of the reheating $\Gamma$, see Eqns. \eqref{GammaConstraint} and \eqref{GammaConstraint-Rrad}.} 
The same is true for $T_{\rm re}$, see Eq. \eqref{2.5.8}. 
Fig. \ref{Nre-Tre} (a) and (b) show $N_{\rm re}$ and $T_{\rm re}$, respectively, as a function of $n_s$ for $n=1$ and $\wrehbar=0$ for various values of $\alpha$. 

Let us now turn to the equation of state in this class of model. 
For 
$\sqrt{\frac{2}{3 \alpha}} \phi_{\rm end} < M_{pl}$, which amounts to $\alpha> {4n^2 \over 3(e-1)^2}\simeq 0.5 n^2$
from Eq. \eqref{phi_end},
we can approximate
\begin{eqnarray}\label{Vapprox}
V(\phi)&\simeq& \Lambda^4 \left(\frac{2}{3\alpha}\right)^n\left(\frac{\phi}{M_{pl}}\right)^{2n}
\end{eqnarray}
during the entire reheating era ($|\phi| < |\phi_{\rm end}|$). 
It is well known that if the  potential is of the form
$V\propto \phi^{2n}$,
then one can in good approximation use the averaged equation of state 
\begin{equation}\label{wrehbarforpolynomial}
\wrehbar\simeq \frac{n-1}{n+1} \, 
\end{equation}
during the oscillatory phase.
This has e.g. been confirmed numerically in the appendix of Ref. \cite{Martin:2010kz} (cf. also \cite{Lozanov:2016hid}).
Eq. \eqref{wrehbarforpolynomial} is valid when reheating is governed by the perturbative decay of the inflaton field rather than parametric resonance and the inflaton energy density is immediately transferred into radiation at the moment when the inflaton dissipation rate $\Gamma$ catches up the Hubble parameter. 

Another quantity relevant for reheating process is the inflaton mass. The inflaton mass $m_\phi$ is usually defined as the curvature of the effective potential at its minimum. For $n=1$ one can obtain the expression 
\begin{equation}\label{m_phi}
m_\phi=\frac{2\Lambda^2}{\sqrt{3\alpha}M_{pl}}
\end{equation}
from the expansion of Eq. (\ref{potential}) at the minimum. 
Note that $m_{\phi}$ in Eq. \eqref{m_phi} can be expressed entirely in terms of $\alpha$, $A_s$ and $n_s$ through Eqns. \eqref{Lambda} and \eqref{scalartensor} with $n=1$. 
For $n>1$ the effective potential $V(\phi)$ exhibits zero mass at its minimum. In order to make perturbative inflaton decays efficient enough to reheat the universe, which is the main interest of this work, the inflaton mass should not vanish. Therefore, we restrict ourselves to the case $n=1$ in the following considerations.

As shown in Eq. \eqref{Vapprox}, the $\alpha$-attractor E-model is approximated by a polynomial potential at the end of inflation when $\sqrt{\frac{2}{3\alpha}}\frac{\phi_{\rm end}}{M_{pl}}<1$, 
which boils down to 
$\alpha > 0.5$ for $n=1$. 
If this condition is fulfilled and the main reheating process is the perturbative decay of the inflaton as mentioned above, 
one can use Eq. \eqref{wrehbarforpolynomial} for $\wrehbar$ with $n=1$, i.e. $\wrehbar=0$. 
In addition,  $\alpha$ cannot be much larger than 100 for $n=1$ 
in order that the tensor-to-scalar ratio $r$ be consistent with the Planck result (1$\sigma$ result), see Eq. \eqref{scalartensor} and Fig. \ref{planck_nsr}. 
Thus, throughout the rest of this paper we stick to the potential Eq. \eqref{potential} with $1\leq \alpha\leq 100$ and $n=1$, so that we can use 
 $\wrehbar=0$ (during perturbative reheating) and Eq. \eqref{m_phi} for $m_{\phi}$. 
Then, using Eqns. \eqref{rhoend}, \eqref{V_end} and \eqref{m_phi} in Eq. \eqref{GammaConstraint} (or \eqref{GammaConstraint-Rrad}) the quantity
$\Gamma/m_{\phi}$ becomes 
\begin{equation}
\frac{\Gamma}{m_{\phi}}
= \left(\frac{2 \sqrt{\alpha}}{2 \sqrt{3} + 3 \sqrt{ \alpha}}\right) 
e^{-3 N_{\rm re}/2} 
= \left(\frac{2 \sqrt{\alpha}}{2 \sqrt{3} + 3 \sqrt{ \alpha}}\right) \, \Rrad^6 \,. \label{Gamma/mphiConstraint}
\end{equation}
A constraint on $\Gamma$ can be derived by inserting the expressions for $N_{\rm re}$ as a function of spectral index $n_s$ found in this subsection into Eq. \eqref{GammaConstraint} or \eqref{Gamma/mphiConstraint}. 
In the perturbative regime, the quantity $\Gamma/m_\phi$ has a clear physical interpretation because at leading order it can, up to some numerical factors, be identified with the squared inflaton (dimensionless) coupling, as illustrated in the next section, 
cf. 
Eqns. \eqref{gamma}, \eqref{gamma3} and \eqref{DP_yukawa}.
This means that one can directly ``measure" the inflaton coupling via the relation Eq. (\ref{Gamma/mphiConstraint}).

A practical problem lies in the complicated dependence of $\Gamma/m_\phi$ on the observational parameters. This in principle makes it very difficult to convert the observational error bar on $n_s$ and $A_s$ into an error bar on the coupling constant.
However, two facts can be used at our advantage. 
First, the dependence of our results on $A_s$ within the observational error bar is so weak that we can practically neglect the uncertainty in $A_s$. 
Second, the present error bar on $n_s$ is sufficiently small that the dependency of $\ln(\Gamma/ m_{\phi}) \sim  - N_{\rm re}$ on $n_s$ is  in good approximation linear in the observationally allowed regime, as can e.g. be seen in Fig. 
\ref{Nre-Tre} (a). This means that the conversion of the observational error bar for $n_s$ into an error bar for the coupling constant is trivial.
To further illustrate this, we expand $\ln{(\Gamma/m_{\phi})}$ around some base value $\bar{n}_s$.
To linear order, we obtain
\begin{equation}\label{LogGamma}
\ln{(\Gamma/m_{\phi})} \simeq C_0 +
(n_s-\bar{n}_s)\, C_1 
\end{equation}
with 
\begin{eqnarray}
C_0 = \ln{Q}+\bigg(3-\frac{9\alpha}{2}\bigg)\ln{\bigg(\frac{s+3}{s-1}\bigg)}+\frac{18\alpha}{s-1} \qquad {\rm and} \qquad
C_1 = \frac{18\alpha (s+6\alpha-1)}{s(3+s)(s-1)^2} \,.
\end{eqnarray}
Here $s=\sqrt{1+3 \alpha(1-\bar{n}_s)}$, and the $n_s$-independent quantity $Q$ is
\begin{eqnarray}
Q =  \frac{9}{8 \sqrt{2} \pi^3 (3 A_s)^{\frac{3}{2}}} 
\left(\frac{2+\sqrt{3\alpha}}{\sqrt{3\alpha}}\right)^{\frac{9\alpha-8}{2}} \,e^{6\kappa-3\sqrt{3\alpha}} \    \;,
\end{eqnarray}
where $\kappa=\ln\left(\frac{k}{a_{0}T_{0}}\right)+\frac{1}{4}\ln\left(\frac
{40}{\pi^2g_{*}}\right)+\frac{1}{3}\ln\left(\frac{11g_{s*}}{43}\right)$.
While the observation that Eq. (\ref{LogGamma}) is a good approximation is an important result, the explicit expressions for $C_0$ and $C_1$ are not particularly illuminating in the present form. We therefore provide a few numerical values of $C_0$ and  $C_1$  for  $\alpha=\{1,\;10,\;100\}$, setting $\bar{n}_s$ and $A_s$ to the Planck best fit values \cite{Ade:2015lrj} $\bar{n}_s=0.9645$, $A_s=10^{-10}e^{3.094}$ at the pivot scale $k/a_0=0.05\, {\rm Mpc}^{-1}$ :
\begin{eqnarray}
C_0 &=&\{-9.4,\; \;-34.0,\; \;-39.1\}\;,\label{C0value}\\
C_1 &=& \{9487.3,\; \;8933.9, \; \;8506.3\}\;,
\end{eqnarray}
which, in turn, can be plugged into Eq. \eqref{LogGamma} to determine $\log_{10}(\Gamma/m_\phi)$ at 1$\sigma$ CL:
\begin{eqnarray}\label{LogGammaOverM}
\log_{10}(\Gamma/m_\phi)=\{-4.1\pm 20.2,\, -14.8\pm 19.0,\,-17.0\pm 18.1 \},
\end{eqnarray}
provided $n_s=0.9645\pm 0.0049$ (68\% CL \cite{Ade:2015lrj}).
Here we set $g_*=g_{s*}=100$. We adopt this choice in all plots presented in this paper. In principle the fact that  $g_*$ and $g_{s*}$ appear in $\kappa$ implies that our results are sensitive to the entire particle mass spectrum of the underlying particle physics model up to the scale of inflation. This dependence is, however, very weak and can be neglected compared to other uncertainties.
The large error bars on the quantities in 
Eq.~(\ref{LogGammaOverM}) show that present CMB data does not allow one to impose a meaningful constraint on $\Gamma/m_\phi$. However, reducing the uncertainty of the spectral index $n_s$ by slightly more than an order of magnitude would allow one to pin down the order of magnitude of this ratio, which would provide very valuable insight into the mechanism of reheating.

\section{Constraining the inflaton coupling} 
\label{Section3}
In this section, we apply the general logic and procedures presented in the previous section and 
study constraints on inflaton couplings to other scalars $\chi$ via renormalisable interactions of the form $\phi\chi^m$ and to fermions $\psi$ via Yukawa interaction. 
We assume that reheating is dominantly driven by one of these interactions; if several terms contribute at comparable level, then one can obviously only constrain a combination of the involved coupling constants. 
Since a simple relation between $N_{\rm re}$ (or $\Rrad$) and microphysical parameters can only be established if reheating can be treated perturbatively, we do not consider terms with higher powers in $\phi$, which do not allow 
perturbative decays for kinematic reasons.\footnote{Of course, such interactions could in principle be relevant even in perturbative scenarios if the perturbative decay has heated the primordial plasma to a sufficient temperature that scatterings are relevant \cite{Drewes:2013iaa,Drewes:2014pfa} or if $\phi$ has a non-trivial quasiparticle spectrum \cite{Drewes:2013bfa}. However, the results found in Ref.~\cite{Drewes:2015coa} suggest that this will not affect the CMB because it will not affect the expansion history if reheating is perturbative.}
For the reason mentioned in the last part of the previous section, we will use the $\alpha$-attractor potential Eq. \eqref{potential} 
with $1\leq \alpha\leq 100$ and $n=1$ in this section. This allows us to use the averaged equation of state $\wrehbar=0$ for the perturbative reheating.  

\subsection{Scalar $\phi\chi^2$ interaction}
In the simplest case, which has previously been studied in Refs.~\cite{Drewes:2015coa,Ueno:2016dim}, the inflaton couples to another scalar via interaction
\begin{equation}\label{interaction}
\mathcal{L}_{\rm int} = -g\phi\chi^2\;,
\end{equation}
where $\chi$ is a light scalar with mass $m_\chi\ll m_\phi$,\footnote{
We use $m_\chi$ and  $m_\phi$ to denote the physical particle masses in vacuum in the following. The effective quasiparticle masses in the early universe in principle differ from these. On one hand there are thermal corrections  to the dispersion relations (``thermal masses") from forward scattering. On the other hand, $m_\chi^2$ receives a correction $\sim g\varphi$ from the coupling to the background field $\varphi$. 
The requirement $m_\chi\ll m_\phi$ is therefore in principle insufficient to guarantee that the $\chi$-particles produced in $\phi$-decays behave like radiation, i.e., are ultra-relativistic. 
However, using the estimates $\phi_{\rm end}\sim M_{pl}$,  $V_{\rm end}\sim \frac{1}{2}m_\phi^2\phi_{\rm end}^2$ and the relation (\ref{PerturbativeRegime}), it is straightforward to show that the correction $\sim g\varphi$ does not significantly modify the decay rate (\ref{Tdr}) in the perturbative  regime.
Following Eq.~ (\ref{Tdr}), we show that the same applies to the thermal correction.
} where the inflaton mass $m_\phi$ is given in Eq. \eqref{m_phi}, and $g$ is a coupling constant.
It is convenient to introduce a dimensionless coupling constant,
\begin{equation}
\tilde{g}=\frac{g}{m_{\phi}} \,.
\end{equation}

\subsubsection{Perturbative reheating} \label{Vacuum-chi2}
\begin{figure}[htp]
\begin{subfigure}{0.53\textwidth}

\includegraphics[width=0.9\linewidth, height=5cm]{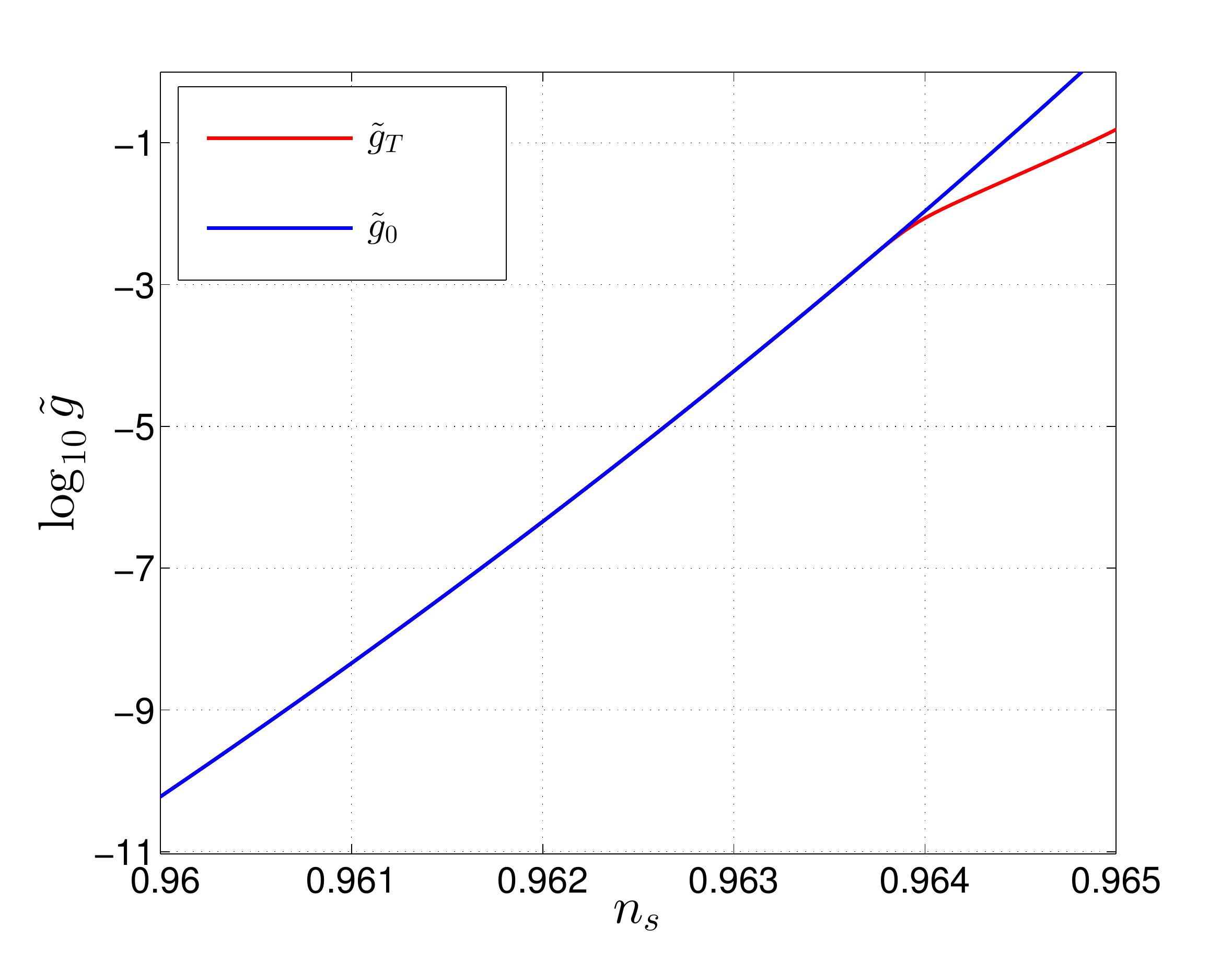}
	\caption{•}
\end{subfigure}
\begin{subfigure}{0.53\textwidth}
	\includegraphics[width=0.9\linewidth, height=5cm]{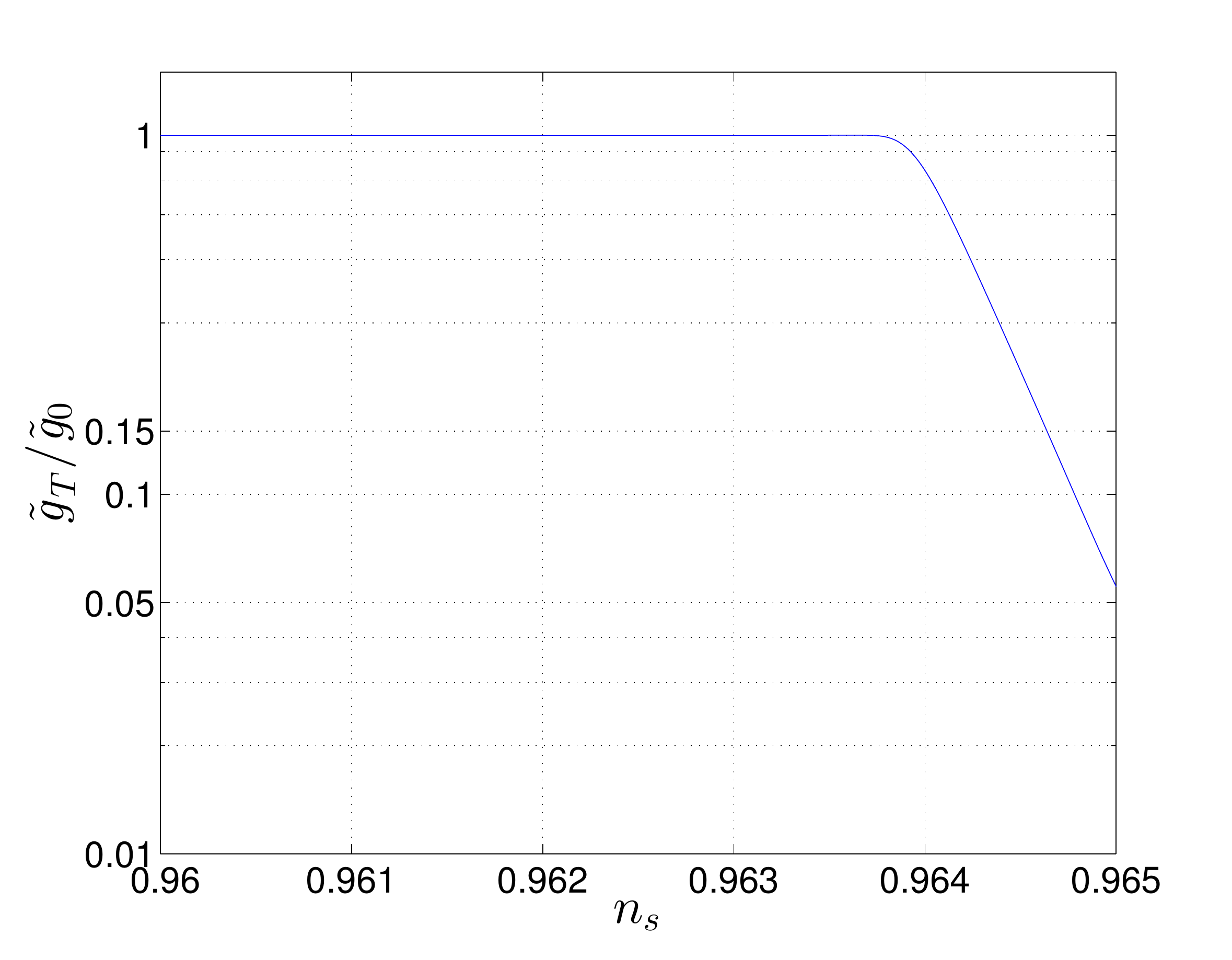}
	\caption{•}
\end{subfigure}
\caption{The spectral index dependence of the dimensionless coupling constant $\tilde{g}=g/m_{\phi}$ in $\alpha$-attractor E-model with $g \phi \chi^2$ interaction for $\alpha=1$ and $n=1$. In the left plot,  $\tildeg_0$ (blue line) and $\tildeg_T$ (red line) correspond to Eqns. \eqref{coupling} and \eqref{gt}, respectively.  
The right plot is to manifest the thermal effect relative to the vacuum decay.
We have chosen the self-coupling of $\chi$-particles as $\lambda=10^{-4}$.
}
\label{Fig.1}
\end{figure}

\paragraph{Vacuum decay} - 
The vacuum decay rate of inflaton field for the interaction Eq. (\ref{interaction}) is given by (cf. Ref. \cite{linder1990particle})
\begin{equation}\label{gamma}
\Gamma_{\phi\to\chi\chi}=\frac{g^2}{8\pi m_{\phi}}\sqrt{1-\left(\frac{2m_\chi}{m_\phi}\right)^2}\simeq \frac{g^2}{8\pi m_{\phi}} \,.
\end{equation}
At the beginning of the reheating process, $\Gamma$ is given by Eq. (\ref{gamma}). Once a certain number of $\chi$-particles have been produced, the interactions of these particles with $\phi$ and with each other can lead to different feedback effects. In Ref. \cite{Drewes:2015coa} it has been argued that this feedback has no effect on the CMB unless there is a parametric resonance. Let us for the moment assume that this statement is correct (we will check its consistency further below).
Using Eqns. \eqref{rhoend}, \eqref{V_end}, \eqref{gamma} and \eqref{wrehbarforpolynomial} 
into Eq. \eqref{Nre-GammaConstraint},
we get the expression for the $e$-folding number of reheating,
\begin{equation}\label{N_re}
N_{\rm re} =\frac{n+1}{3n}\,\ln{\Bigg[\frac{16\pi m_\phi}{3M_{pl}}\Bigg(\frac{\Lambda}{g}\Bigg)^2\Bigg(\frac{2n}{2n+\sqrt{3\alpha}}\Bigg)^n\Bigg]}\;,
\end{equation}
or from Eq. \eqref{GammaConstraint} (cf. Eq. \eqref{Gamma/mphiConstraint} for $n=1$) one obtains
\begin{equation}\label{coupling}
	g^2=\frac{16\pi m_\phi\Lambda^2}{3M_{pl}}\Bigg(\frac{2n}{2n+\sqrt{3\alpha}}\Bigg)^n\exp{\Bigg(\frac{-3nN_{\rm re}}{1+n}}\Bigg) \,.
\end{equation}
Here $\Lambda$ and $m_\phi$ are given in Eqns. \eqref{Lambda} and \eqref{m_phi}, respectively.
 Eq. \eqref{coupling} together with Eq. \eqref{2.5.7}  directly relates the coupling constant to the spectral index. 
Note that the lack of knowledge about $r$ does not introduce an uncertainty because $r$ is fixed 
for given $n_s$, $\alpha$ and $n$, cf. Eq. (\ref{scalartensor}). 
Thus,  Eq. \eqref{coupling} determines the coupling as a function of $n_s$. This is plotted (blue line) in Fig. \ref{Fig.1} (a). 
This plot shows that the coupling is a strictly increasing function of $n_s$, which is intuitive in view of Fig. \ref{Fig1} (see the caption thereof).

\paragraph{Thermal feedback} - 
Several authors have argued that different thermal effects, such as Bose enhancement, scatterings or the kinematic effect of 
``thermal masses" can modify the thermal history of the universe during reheating \cite{Kolb:2003ke,Yokoyama:2005dv,Drewes:2010pf,Bodeker:2006ij,Drewes:2013iaa,Drewes:2014pfa}.
For the interaction $g \phi \chi^2$, 
one has to take into account the effects of Bose enhancement and thermal mass. Scatterings and ``Landau damping" may come to dominate at very high $T\gg m_\phi$ (this e.g. happens for Yukawa interactions \cite{Drewes:2013iaa,Drewes:2015eoa}), but are of higher order in $g$ as long as the convergence of the loop expansion is not spoiled by infrared effects.
In order to include the effects of ``thermal masses" for $\chi$-particles,\footnote{We will ignore the thermal correction to $m_{\phi}$, since the potential Eq. \eqref{potential} 
for $n=1$ has negligible self-interaction at the minimum.}   
we have to specify the $\chi$ interactions. The in-medium decay rate has been studied for different types of $\chi$ interactions  \cite{Yokoyama:2004pf,Boyanovsky:2004dj,Yokoyama:2005dv,Anisimov:2008dz,Drewes:2010pf,Drewes:2013iaa,Drewes:2013bfa,Ho:2015jva}. 
For illustrative purposes we here assume that field $\chi$ has a quartic self-interaction, i.e., its dynamics is described by the Lagrangian 
\begin{equation}\label{L_chi}
\mathcal{L}_\chi=\frac{1}{2}\partial^{\mu}\chi\partial_\mu\chi-\frac{1}{2}m_\chi^2\chi^2-\frac{\lambda}{4!}\chi^4-g\phi\chi^2 \,,
\end{equation}
where $m_\chi\ll m_\phi$.
Let us assume that $\lambda\gg\tildeg$, so that the $\chi$-particles reach a smooth phase space distribution that can be described by an effective temperature $T$ on a time scale much shorter than $1/\Gamma$.  
The correction to the expression Eq. \eqref{gamma} is e.g. discussed in Ref. \cite{Drewes:2013iaa}, i.e. 
\begin{equation}\label{Tdr}
\Gamma_{\phi\to\chi\chi}=\frac{g^2}{8\pi m_\phi}\Bigg[1-\Big(\frac{2M_\chi}{m_\phi}\Big)^2\Bigg]^{1/2}\big[1+2f_B(m_\phi/2)\big]\;.
\end{equation}
Here $f_B(\omega)=(e^{\omega/T}-1)^{-1}$ is the equilibrium Bose-Einstein distribution function and $M_{\chi}$  the effective mass from the quartic self-interaction,
\begin{equation} \label{M_chi}
M_{\chi}^2=m_\chi^2+\frac{\lambda T^2}{24} \,.
\end{equation} 
The thermal mass correction from the $g\phi\chi^2$ is usually neglected; large logarithmic contributions that in principle may appear for $m_\chi\ll T$ \cite{Drewes:2013bfa} can be expected to be regulated by the contribution from the self-interaction.
As we see easily, we recover Eq. \eqref{gamma} from Eq. \eqref{Tdr} in the limit $T\rightarrow0$. If the temperature is much higher than the inflaton mass scale, Bose enhancement becomes very efficient, 
which may significantly modify the lower bound of the coupling constant.

We follow the same steps as we did in the case of vacuum decay above, but with the modified decay rate Eq. \eqref{Tdr}, to get the expression for the coupling constant
\begin{equation}\label{gt}
	g^2=\frac{16\pi m_\phi\Lambda^2}{3M_{pl}}\Bigg(\frac{2n}{2n+\sqrt{3\alpha}}\Bigg)^n
	\exp{\Bigg(\frac{-3nN_{\rm re}}{1+n}}\Bigg)
	\frac{\left(\exp{\left({m_\phi \over 2T_{\rm re}}\right)}-1\right)}{\left(1-\frac{\lambda T_{\rm re}^2}{6m_\phi^2}\right)^{1/2}\left(\exp{\left({m_\phi \over 2T_{\rm re}}\right)}+1\right)}\;.
\end{equation}
When the decay rate is independent of temperature, the maximum of the $e$-folding number $N_{\rm re}$ puts the lower bound of coupling constant, 
but this is not clear in the case of Eq. \eqref{gt} since $T_{\rm re}$ also depends on $N_{\rm re}$, see Eq. \eqref{2.5.8}.
Substituting Eq. \eqref{2.5.8} into Eq. \eqref{gt}, one obtains the coupling constant as a function of the e-folding number $N_{\rm re}$, which is, in turn, a function of spectral index $n_s$.

The coupling Eq. \eqref{gt} as a function of $n_s$ is plotted (red line) in Fig. \ref{Fig.1} (a). Fig. \ref{Fig.1} (b) shows the ratio between  Eqns. \eqref{gt} and \eqref{coupling}, 
which is the last factor depending on $T_{\rm re}$ in Eq. \eqref{gt}.  
The deviation due to the thermal effect becomes manifest when $n_s$ gets closer to 1. This is because the thermal effect becomes significant at high $T_{\rm re}$, which grows with $n_s$, 
see Fig. \ref{Nre-Tre} (b). Fig. \ref{Fig.1} shows that the thermal effect makes the reheating with smaller coupling as efficient as the one with larger coupling in the case of vacuum decay. 
Put differently, for a given value of coupling 
the thermal effect makes the reheating more efficient (smaller $N_{\rm re}$ and higher $T_{\rm re}$, see Fig. \ref{Nre-Tre}) 
in comparison with vacuum decay.  This is due to the Bose enhancement, which amplifies the decay rate $\Gamma$ in the presence of bosonic particles in the final state, 
see Eq. \eqref{Tdr}. It is also clear from Eq. \eqref{Tdr} that the growth of the thermal mass of $\chi$-particle with temperature in Eq. \eqref{M_chi}
decreases the decay rate due to the shrinking of the phase space allowed by the decay kinematics \cite{Drewes:2013iaa}.

\subsubsection{Resonances}
So far we have assumed that $\Gamma$ can be approximated by its vacuum value Eq. (\ref{gamma}) or thermally corrected one Eq. \eqref{Tdr}. 
These considerations were based on the assumption that the 
$\chi$-particles thermalise instantaneously ($\Gamma\ll\Gamma_\chi$). However, it is well-known that resonant particle production can outrun the thermalisation. 
The most prominent example is the \emph{parametric resonance}. In case a resonance occurs, the behaviour is highly non-linear, and it is in general not possible to establish a simple relation for the (time dependent) effective damping rate $\Gamma$ in terms of fundamental coupling constants. Since resonant particle production is very efficient, one could argue on a rather general basis that the reheating era should be 
very short ($N_{\rm re}\sim 1$), irrespective of the specific value of $g$ (as long as $g$ is large enough to trigger a resonant particle production). If so, the CMB would not contain much useful information about the reheating era. 
However, it has been shown in Ref.~\cite{GarciaBellido:2008ab} that interactions amongst the produced particles can delay the parametric resonance. 
While this means that $N_{\rm re}$ may be large enough to leave a detectable imprint in the CMB, it also means that it depends not only on the inflaton coupling, but also on the couplings of the produced particles amongst each other. To avoid all these complications, we in the present work restrict ourselves to perturbative reheating. 
It is clear that a parametric resonance can in principle be avoided by choosing sufficiently small $g$. It is, however, not obvious that this can be achieved in $\alpha$-attractor models in agreement with the observational constraints.
In the following, we briefly discuss the condition which favours perturbative decay over parametric resonance along the same lines as in Ref. \cite{Kofman:1997yn}.

The Heisenberg representation of the scalar field $\chi$ is 
\begin{equation}
\chi(t,\mathbf{x})=\frac{1}{(2\pi)^{3/2}}\int d^3k(\hat{a}_k\chi_k(t)e^{-i\mathbf{k}\mathbf{x}}+\hat{a}^{\dagger}_k\chi_k(t)e^{i\mathbf{k}\mathbf{x}})\;,
\end{equation}
where $\hat{a}_k$ and $\hat{a}^\dagger_k$ are annihilation and creation operators, respectively. 
Given that the inflaton field oscillates around the minimum of its potential, i.e. $\phi=\Phi\sin{m_\phi t}$ \footnote{This solution is obtained in $\alpha$-attractor E-model 
Eq. \eqref{potential} with $n=1$ which essentially exhibits the same behaviour as the chaotic inflation around the minimum of the potential.}, then the mode function $\chi_k$ obeys
\begin{equation}
\ddot{\chi}_k+[k^2+m_\chi^2+2\tildeg m_\phi\Phi\sin(m_\phi t)]\chi_k=0\;.
\end{equation}
This equation becomes Mathieu equation when we make a change of variables $m_{\phi}t=2z-\pi/2$, 
\begin{equation}
\chi^{\prime\prime}+(A_k-2q\cos2z)\chi_k=0\;,
\end{equation}
where $A_k=4(k^2+m_\chi^2)/m_\phi^2,\: q=4\tildeg\Phi/m_\phi$ and prime denotes  derivative with respect to $z$.
The important property of this equation is the existence of instability of its solution which appears in two different regimes according to the value of $q\;$.

\paragraph{Broad resonance} - 
In the regime with $q>1$, explosive particle production occurs for a broad range of momenta of $\chi$-particles whenever adiabaticity condition is violated.
The condition $q>1$  translates into 
\begin{equation}
\tildeg\Phi>m_\phi\;.
\end{equation}
Setting the amplitude of the oscillation around the minimum of its potential 
$\Phi$ to 
the value of inflaton field at the end of inflation 
$\phi_{\rm end}$, 
we obtain\footnote{This is a conservative estimate in the sense that in reality 
the broad resonance would kick off at a larger coupling  
than the one given in Eq. \eqref{g_Br} since $\Phi \ll \phi_{\rm end}$. In what follows, when considering the conditions for resonances, 
we make the replacement $\Phi \rightarrow \phi_{\rm end}$, 
which leads to the estimates of the lower bound of couplings for the resonance that is smaller than the actual one.   
}
\begin{equation}\label{g_Br}
\tildeg>\frac{m_\phi}{\phi_{\rm end}}\;.
\end{equation}
For a given inflaton potential, $\phi_{\rm end}$ is completely determined, thus Eq. \eqref{g_Br} directly gives a condition on the coupling constant in favour of broad resonance. 
In the $\alpha$-attractor E-model Eq. \eqref{potential},  $\phi_{\rm end}$ is given by Eq. \eqref{phi_end}. 
The inflaton mass $m_{\phi}$ is given in Eq. \eqref{m_phi}, which can be expressed entirely in terms of $\alpha$, $A_s$ and $n_s$ 
via Eqns. \eqref{scalartensor}-\eqref{Lambda} with $n=1$.  
The Planck data typically correspond to 
\begin{equation}
\frac{m_\phi}{\phi_{\rm end}} \sim 10^{-5} \quad {\rm for} \ \alpha=n=1 \,.
\end{equation}
Then Eq. \eqref{g_Br} implies that the broad resonance can occur when 
\begin{equation}
\tildeg>10^{-5} \quad {\rm for} \ \alpha=n=1  \,.
\end{equation}

\begin{figure}[htp]
\begin{center}
\includegraphics[scale=0.7]{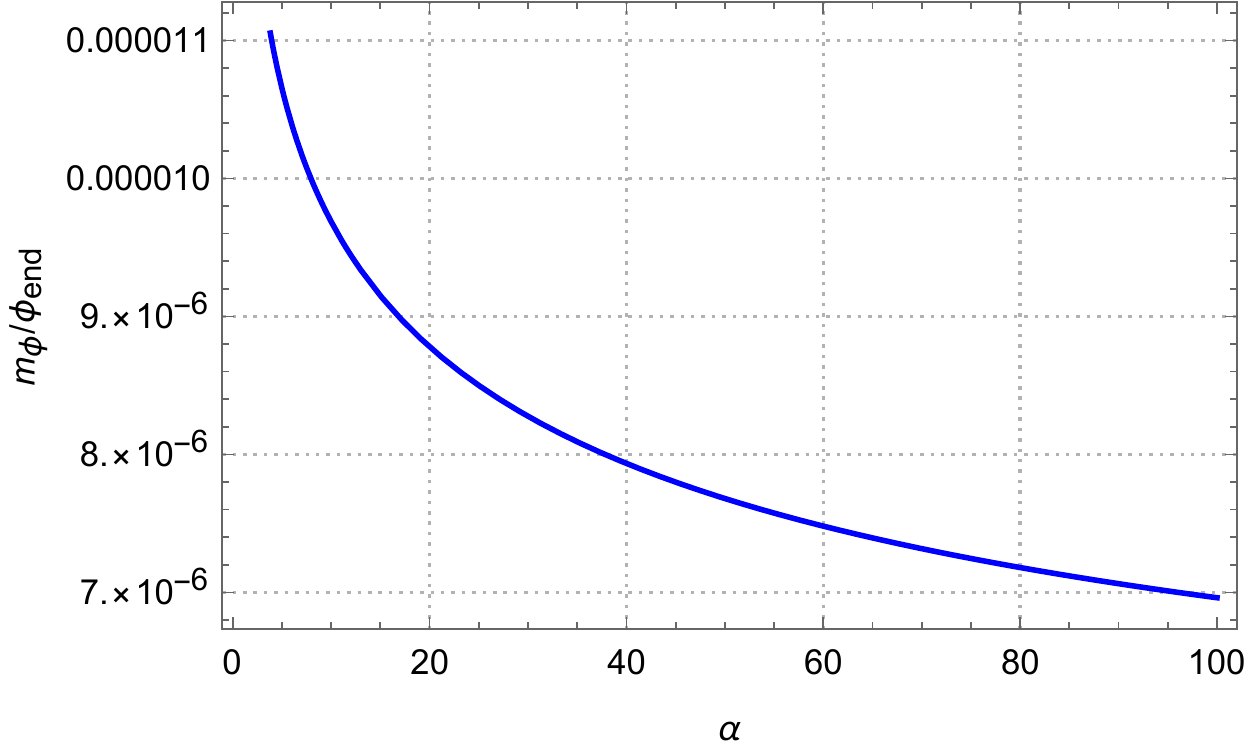}
\caption{$\alpha$ dependence of the ratio between $m_\phi$ and $\phi_{\rm end}$ obtained from Eq. \eqref{m_phi} and Eq. \eqref{phi_end}.  We have chosen $n=1$ and $n_s=0.9645$ as a pivot value.}
\label{Fig.3}
\end{center}
\end{figure}

\paragraph{Narrow resonance} - 
Violation of condition Eq. (\ref{g_Br}) guarantees the absence of non-perturbative particle production from the non-adiabatic evolution of the $\phi$-condensate.
This is necessary, but not sufficient for the applicability of our method. 
It is well-known that resonant particle production can also occur if the dissipation is driven by perturbative decays at an elementary level, but Bose enhancement amplifies the transition rate.
Such a \emph{narrow resonance} occurs when 
\begin{equation}\label{g_narrow1}
\frac{\tildeg m_\phi}{32\pi}<\Phi<\frac{m_\phi}{\tildeg} \,.
\end{equation}
Here the upper bound is given by the requirement that perturbation theory can be applied ($q<1$), while the lower bound reflects the condition   
that the $\chi$ occupation numbers in the mode with energy $m_\phi/2$ remain well above unity, so that Bose enhancement leads to exponential growth faster than the vacuum decay rate \cite{Kofman:1997yn}. 
Note that the condition Eq. \eqref{g_narrow1} does not take the redshifting by Hubble expansion into account, see below.
Using $\phi_{\rm end}$ for $\Phi$ in Eq. \eqref{g_narrow1}, we can obtain two conditions on the coupling constant,
\begin{eqnarray}
\tildeg&<&\frac{m_\phi}{\phi_{\rm end}} \,, 
\label{g_n1} \\
\tildeg&<&\frac{32\pi\phi_{\rm end}}{m_\phi} \,
. \label{g_n2}
\end{eqnarray}
Then the range of the coupling constant which favours narrow resonance over perturbative decay is determined by the ratio between $\phi_{\rm end}$ and $m_\phi$.
In $\alpha$-attractor E-model with $n=1$, $m_\phi$ never exceeds $\phi_{\rm end}$ for $\alpha<100$ as shown in Fig. \ref{Fig.3}, from which we easily 
see that Eq. \eqref{g_n2} is automatically satisfied once Eq. \eqref{g_n1} holds.

One might conclude that reheating always begins with narrow resonance or broad resonance depending on whether Eq. \eqref{g_n1} holds or not in the $\alpha$-attractor E-model under consideration.
However, notice that we have not considered so far the effect of the cosmic expansion. Even though the narrow resonance overwhelms the perturbative decay, it can only be efficient when the decay rate is greater than the Hubble parameter, i.e. $q^2m_\phi> H$ (cf. \cite{Kofman:1997yn}). This imposes another condition on the coupling constant 
for narrow resonance\footnote{In fact, Eq. \eqref{gnarrow_min} corresponds to $q^2m_\phi> H$ imposed at the beginning of the reheating (i.e. at the end of inflation), nevertheless, it gives the lower bound of the coupling constant for the narrow resonance in the most conservative case. Careful readers might be sceptical on this result, because both of the amplitude $\Phi$ of the inflaton oscillation and Hubble parameter $H$ evolve in time. 
However, it should be noted that $\Phi$ decreases faster than $\sqrt{H}$, which can be seen as follows.  
During the oscillator phase, the inflaton behaves like a dust (condensate of massive particles) in the absence of decay, 
so that one has solution $\Phi\propto\phi_{\rm end}t^{-1}$ and $\sqrt{H}\propto t^{-1/2}$ (cf. \cite{Mukhanov}). If the decay of the inflaton field is taken into account, $\Phi$ decreases more quickly. Hence, Eq. \eqref{gnarrow_min} can be taken as the estimator of the efficiency of the narrow resonance throughout the reheating. },
\begin{equation}\label{gnarrow_min}
\tildeg 
> \frac{V_{\rm end}^{1/4}}{\phi_{\rm end}} \left(\frac{m_{\phi}}{24 M_{pl}}\right)^{1/2} 
\;.
\end{equation}
Combined with Eq. \eqref{g_n1}, we see that narrow resonance becomes efficient under the condition
\begin{equation}\label{g_nar}
\frac{V_{\rm end}^{1/4}}{\phi_{\rm end}} \left(\frac{m_{\phi}}{24 M_{pl}}\right)^{1/2}<\tildeg<\frac{m_\phi}{\phi_{\rm end}}\;,
\end{equation}
 which, in turn, puts more stringent bound on the coupling constant for  successful perturbative reheating,
\begin{equation}\label{g_pert}
\tildeg< \frac{V_{\rm end}^{1/4}}{\phi_{\rm end}} \left(\frac{m_{\phi}}{24 M_{pl}}\right)^{1/2}
\;.
\end{equation} 
To sum up, we can distinguish the following regimes:
\begin{eqnarray}
&\tildeg< \frac{V_{\rm end}^{1/4}}{\phi_{\rm end}} \left(\frac{m_{\phi}}{24 M_{pl}}\right)^{1/2} & \quad \implies \quad {\rm perturbative \,\,regime} \label{PerturbativeRegime}\\  \nonumber \\
& \frac{V_{\rm end}^{1/4}}{\phi_{\rm end}} \left(\frac{m_{\phi}}{24 M_{pl}}\right)^{1/2} < \tildeg< \frac{m_\phi}{\phi_{\rm end}}& \quad \implies \quad {\rm narrow\,\, resonance}\\  \nonumber \\
&  \frac{m_\phi}{\phi_{\rm end}} < \tildeg & \quad \implies \quad {\rm broad\,\, resonance} 
\end{eqnarray} 
We again emphasise that the above bounds for resonances are conservative ones as they are estimated at the end of inflation as explained before,
and the perturbative regime can be extended in reality since scattering of produced particles, which were not taken into account in the above, 
may prevent narrow resonance from occurring.

\subsubsection{Results}

\begin{figure}[h!]
\begin{center}
\includegraphics[scale=0.6]{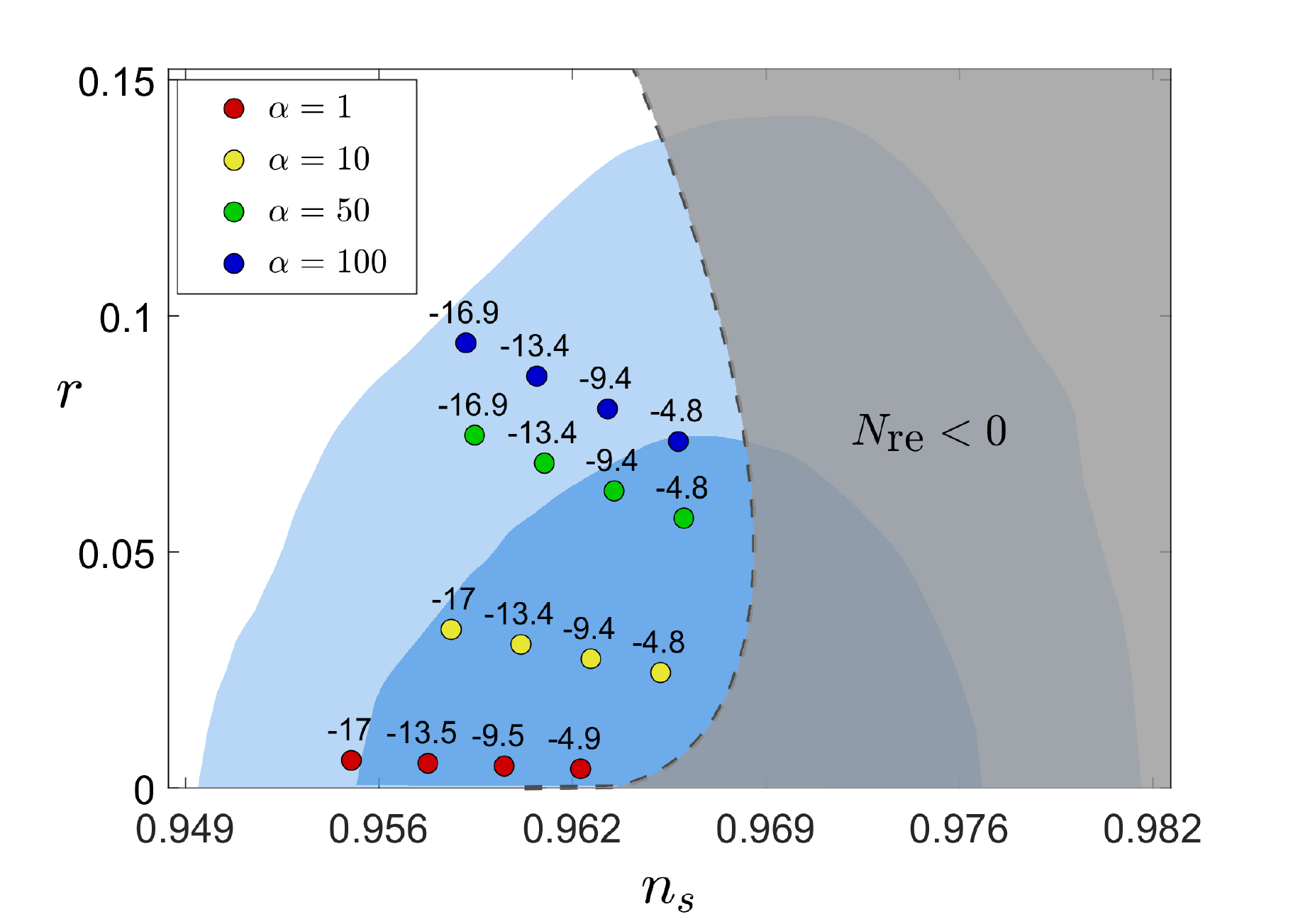}
\end{center}
\caption{
The effect that the reheating phase has on the predictions of $\alpha$-attractor E-models with $n=1$ for CMB is illustrated in the $n_s$-$r$ plane for a set of sample points.
We assume that reheating is primarily driven by an interaction of the form $g \phi \chi^2$.  
The predictions for $n_s$ and $r$ are indicated by the disks of a given colour for fixed $\alpha$, but with different values of $g$.
The numbers over disks are the logarithmic values of the coupling 
$\log_{10}(g/m_\phi)$ computed using Eq. \eqref{gt}.
The sky blue and light blue areas display 
marginalized joint confidence contours for $(n_s,r)$ at the $1\sigma$ ($68\%$) and 2$\sigma$ ($95\%$) CL \cite{Ade:2015lrj}, respectively. 
The dashed line corresponds to $N_{\rm re}=0$, so that $N_{\rm re}<0$ in the shaded region. 
The BBN constraint $T_{\rm re}>10$ MeV is respected on all chosen points in the plot. 
}\label{planck_phichi2}
\end{figure}

\begin{figure}[htp!]
\begin{subfigure}{0.5\textwidth}
	\includegraphics[width=1\linewidth, height=6.2cm]{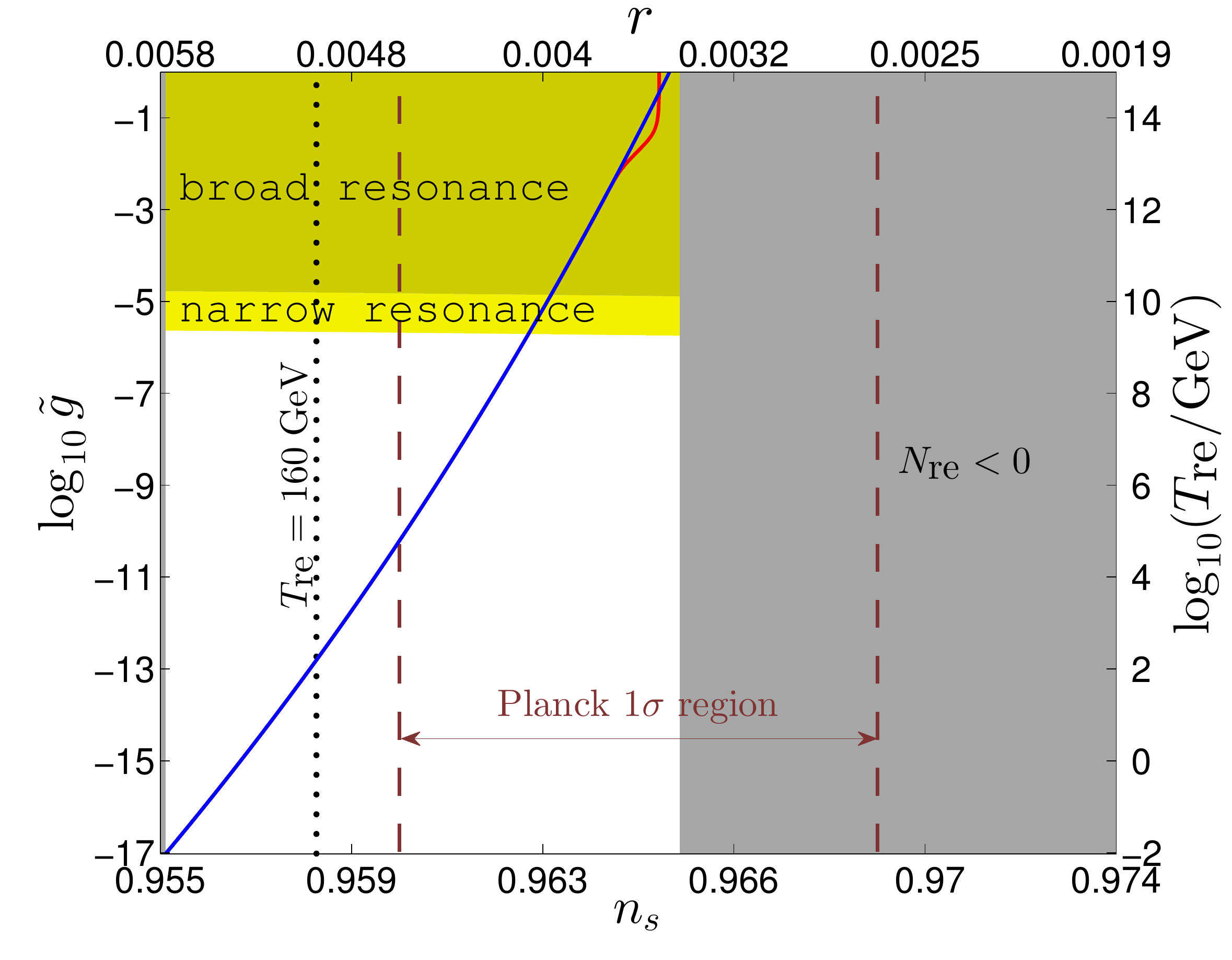}
	\caption{$\alpha=1$}
\end{subfigure}
\begin{subfigure}{0.5\textwidth}
	\includegraphics[width=1\linewidth, height=6.2cm]{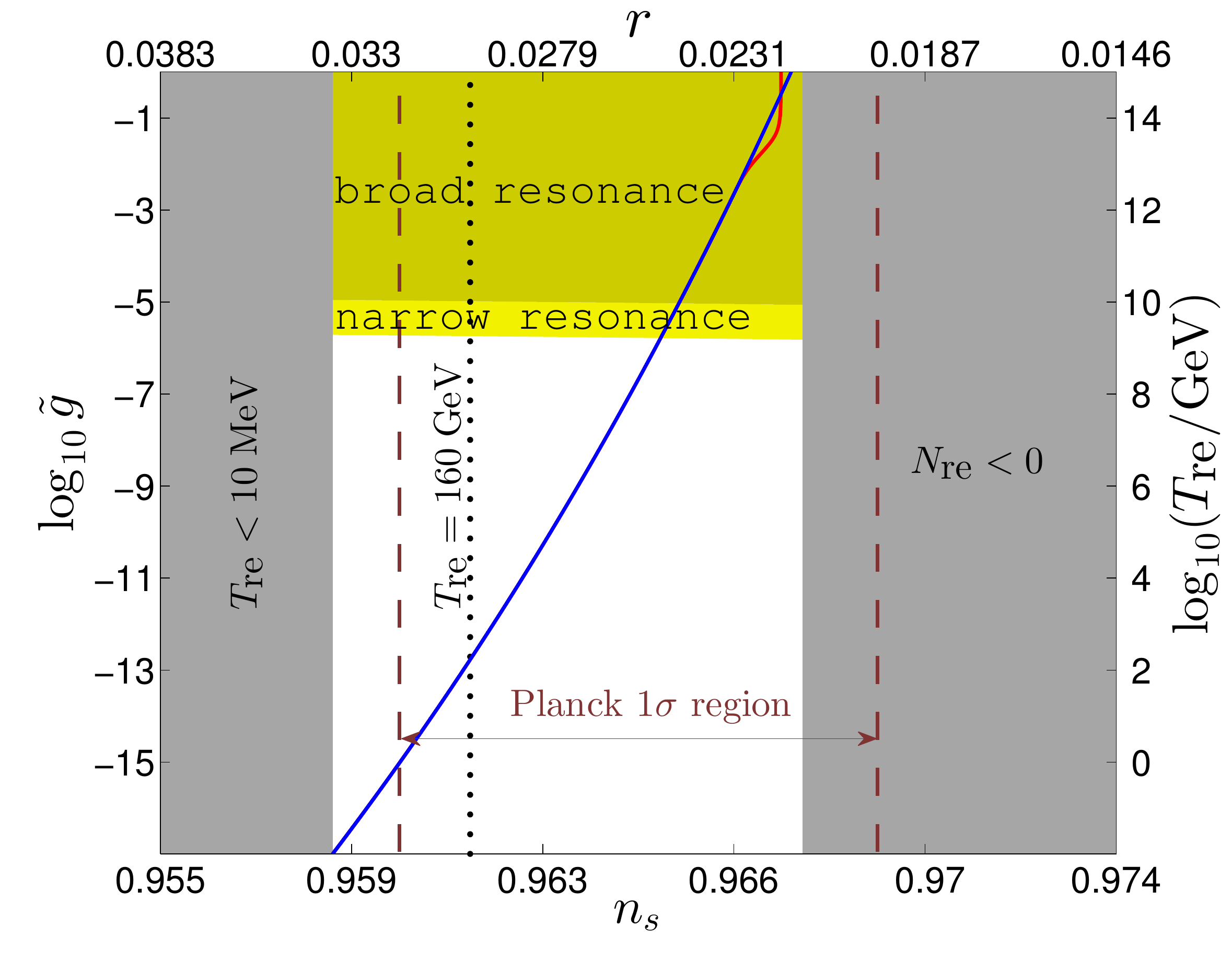}
	\caption{$\alpha=10$}
\end{subfigure}
\begin{subfigure}{0.5\textwidth}
	\includegraphics[width=1\linewidth, height=6.2cm]{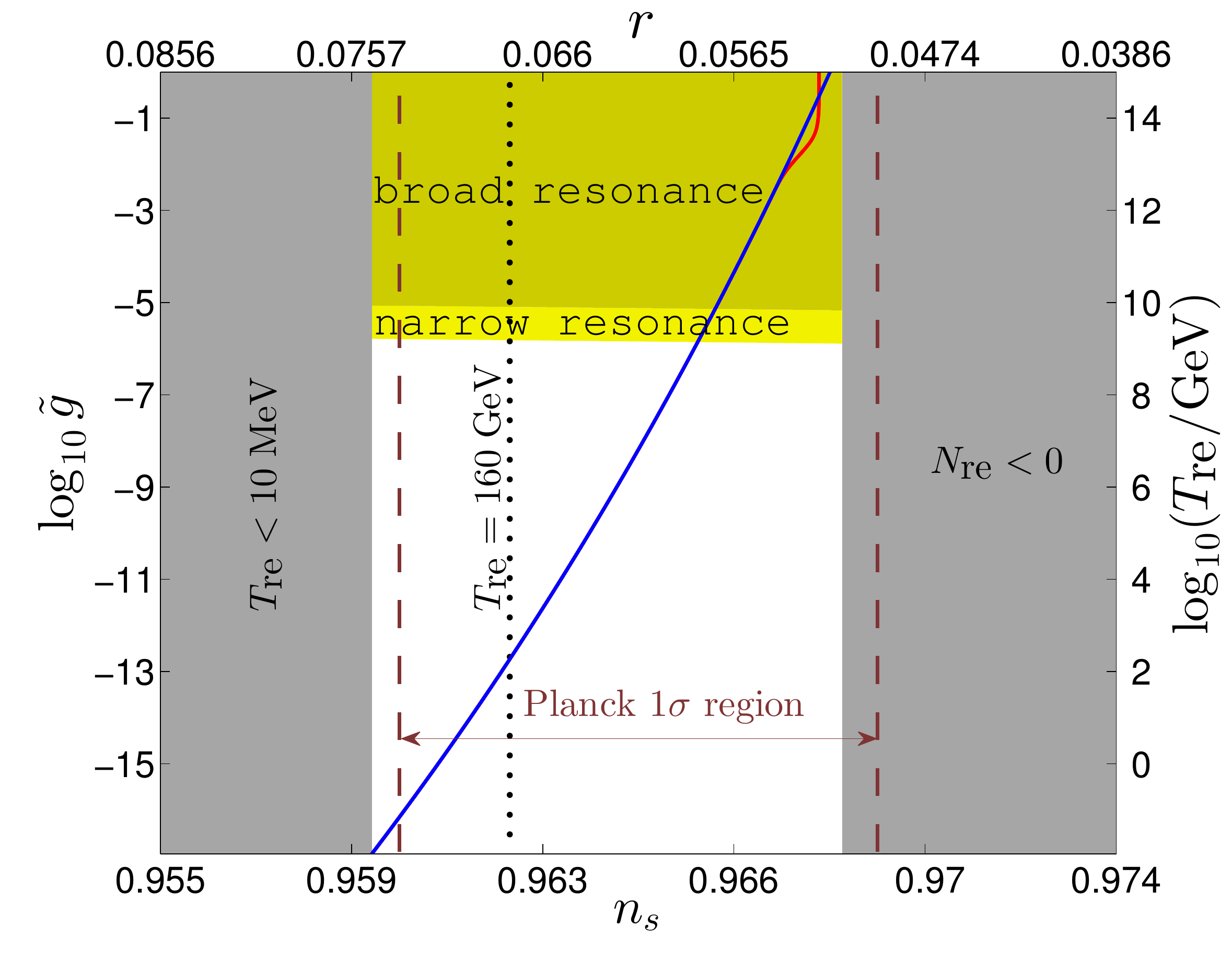}
	\caption{$\alpha=50$}
\end{subfigure}
\begin{subfigure}{0.5\textwidth}
	\includegraphics[width=1\linewidth, height=6.2cm]{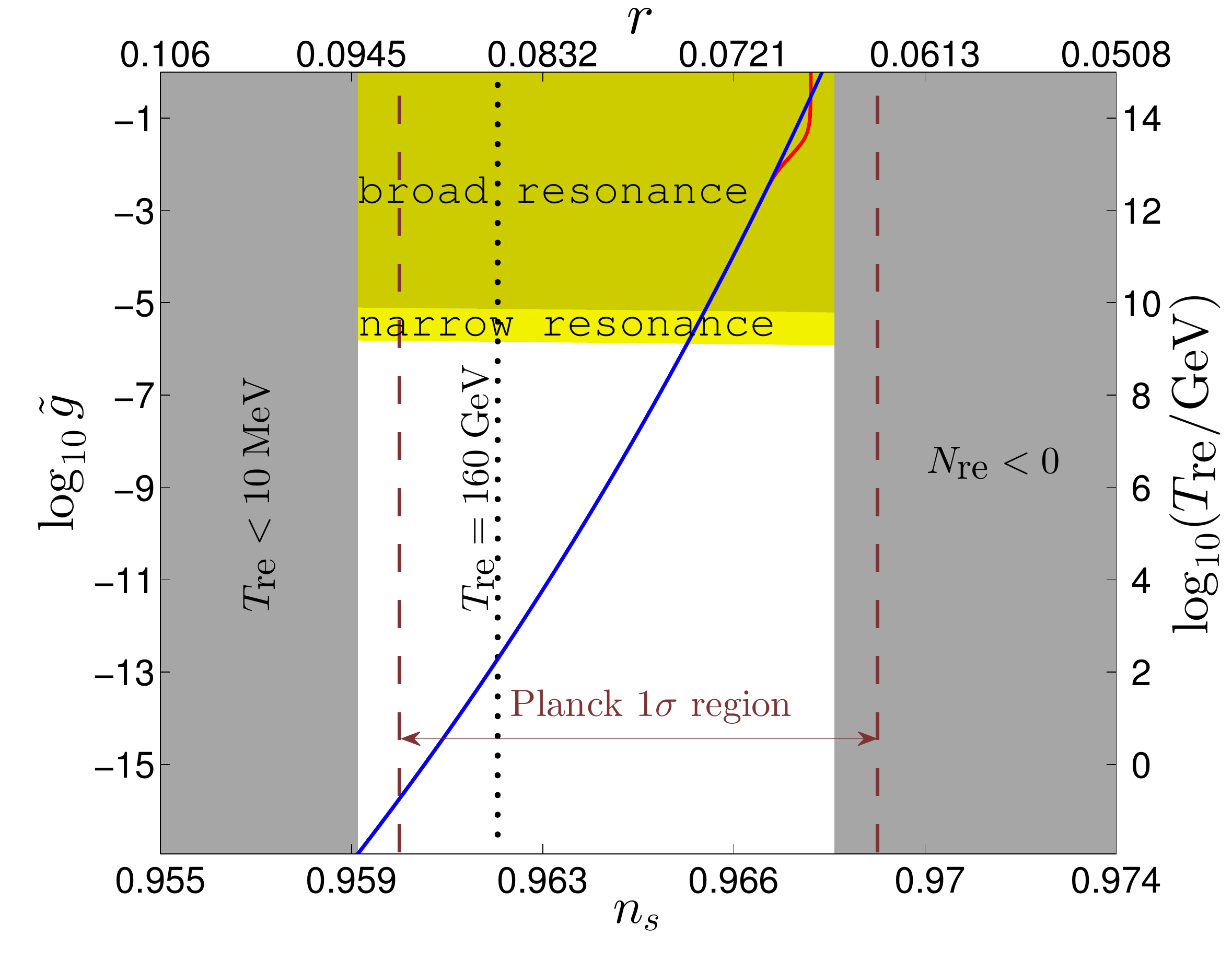}
	\caption{$\alpha=100$}
\end{subfigure}
\caption{For fixed $\alpha$ and $n$, the coupling constant $\tilde{g}=g/m_{\phi}$ in the interaction $g \phi \chi^2$ can be determined from $n_s$. 
As indicated in the plot, this also fixes reheating temperature $T_{re}$ and the scalar-to-tensor ratio $r$ (via Eq. \eqref{scalartensor}). 
The solid lines show $\tilde{g}$ as a function of $n_s$ for various values of $\alpha$. 
They are determined from Eqns. \eqref{coupling} (blue line) and \eqref{gt} (red line).
 We have chosen the self-coupling of $\chi$-particles as $\lambda=0.1$ for the red line. The uncertainty in the position of these lines due to the observational error bar on $A_s$ is so small that it would not be visible in this plot.
We take the wave number of perturbation $k/a_0=0.05 \, \text{Mpc}^{-1}$ as a pivot scale and use  $A_s=10^{-10}e^{3.094}$ \cite{Ade:2015lrj}. We approximate $\bar{w}_{\rm re}=0$ during reheating. 
The dashed lines indicate the 1$\sigma$ confidence region of the Planck measurement of $n_s$  ($n_s=0.9645\pm 0.0049$ \cite{Ade:2015lrj}).
The linear approximation (\ref{LogGamma}) holds very well within the perturbative regime. 
The vertically shaded regions on the right and left are physically ruled out because they give $N_{\rm re}<0$  or $T_{\rm re}<10\,\text{MeV}$ for these ranges of $n_s$ (see Fig. \ref{Nre-Tre}).
The dotted line corresponds to the reheating temperature at the electroweak scale ($T_{\rm re}=160$ GeV). 
In the yellow regions, a parametric resonance occurs and the relations \eqref{coupling} and \eqref{gt} cannot be applied. 
On the right of the each plot we also indicated $\log_{10}(T_{\rm re}/{\rm GeV})$ corresponding to the values of the coupling constant on the left of the plot.
Hereafter we keep on using the notations and shaded/coloured regions in subsequent figures with the same meaning as here.}
\label{phichi2_pert}
\end{figure}

In Figs. \ref{planck_phichi2} and \ref{phichi2_pert} 
we summarise all constraints found in this section. 
These plots show how the dimensionless coupling $\tildeg=g/m_\phi$ can be related to the CMB data (i.e. spectral index $n_s$ and tensor-to-scalar ratio $r$) 
for different values of $\alpha$. We use $\lambda=0.1$ for self-coupling of $\chi$-particles when taking into account thermal corrections.

Fig. \ref{planck_phichi2} uses the Planck results \cite{Ade:2015lrj} in the $n_s$-$r$ plane (1$\sigma$ and 2$\sigma$ CL), in which we indicate corresponding values of $\log_{10}\tildeg$ 
(numbers over disks) and $\alpha$ (distinguished by colour coding) at sample points (disks). 
 
The dashed line corresponds to the contour $N_{\rm re}=0$.   
We used Eq. \eqref{gt} to obtain $\log_{10}\tildeg$ in this figure.
Fig. \ref{phichi2_pert} shows the relation between $n_s$ and $\tilde{g}$ for different values of $\alpha$. 
The dark gray regions denote the regions excluded from $N_{\rm re}<0$ or inconsistency with BBN. 
The abundances of light elements in the intergalactic medium produced during BBN provide the strongest observational lower bound $T_{\rm re}>10$ MeV on the temperature in the early universe. However, the production of Dark Matter and baryogenesis in most particle physics models require much higher temperatures.
If the origin of the observed baryon asymmetry of the universe (see e.g. Ref. \cite{Canetti:2012zc}) relies on the  baryon number violation in the SM at high temperature \cite{Kuzmin:1985mm}, then the temperature should exceed $160$ GeV \cite{DOnofrio:2014rug}. We indicate the value $T_{\rm re}=160$ GeV by a vertical dotted line. In the yellow regions a resonance occurs and our formulae cannot be applied. 

The solid lines  in Fig. \ref{phichi2_pert} were plotted, assuming that reheating proceeded entirely in perturbative manner, i.e.,  by vacuum decay (blue) or thermally corrected perturbative decay (red), so that we used  $\wrehbar=0$. The blue lines and red lines are plotted from Eqns. \eqref{coupling} and \eqref{gt}, respectively.
The behaviour of the red lines in comparison to the blue lines in the plots is due to the Bose enhancement and the thermal blocking. 
As can be seen in Eq. \eqref{gt}, the effect of the Bose enhancement coming from $\exp{\left(\frac{m_\phi}{2 T_{\rm re}}\right)}$ becomes apparent when $T_{\rm re}\sim m_\phi$, while the thermal blocking occures when $T_{\rm re}\sim m_\phi/\sqrt{\lambda}$ at which the decay rate vanishes. Therefore, for $\lambda\ll 1$, as the reheating temperature $T_{\rm re}$ increases (i.e. as $n_s$ increases, see Fig. \ref{Nre-Tre} (b)), the red lines first deviate from the blue lines due to the Bose enhancement and later blow up due to the thermal blocking. 
The numerical evaluations shown in Fig. \ref{phichi2_pert} indicate  that in this model the thermal feedback (red lines) has no significant effect on the CMB observables unless $\tildeg$ 
is large enough to trigger a parametric 
resonance, in which case the relation Eq. (\ref{gt}) cannot be applied. 
In Ref. \cite{Drewes:2015coa}, it has been argued that thermal effects may leave no observable imprint in the CMB if the condition Eq. (\ref{g_pert}) is fulfilled, and our numerical results confirm this conclusion. 
  
\subsection{Scalar $\phi\chi^3$ interaction}
Now we consider an interaction of the form    
\begin{equation}\label{chi3}
\mathcal{L}_{\rm int}=-\frac{h}{3!}\phi\chi^3\;
\end{equation}
and discuss how the coupling $h$ can be constrained from CMB.  

\subsubsection{Perturbative reheating}

\paragraph{Vacuum decay} -
The rate of the three body decay $\phi\to\chi\chi\chi$ in vacuum can be found in the standard form for Dalitz plot  (see e.g. Sec. 47 of Ref. \cite{Olive:2016xmw}) 
as
\begin{equation}\label{gamma3}
\Gamma_{\phi\to\chi\chi\chi}=\frac{h^2}{3!} \int d(m_{12}^2)d(m_{23}^2) \frac{1}{32m_\phi^3} \frac{1}{(2\pi)^3}  
\;,
\end{equation}
where $m_{12}^2=(p_1+p_2)^2,\;m_{23}^2=(p_2+p_3)^2$, $p_i$ are the four-momenta of $\chi$-particles ($\phi$-particle is at rest). 
The integration limits (e.g. $(m_{23}^2)_{max}$ and $(m_{23}^2)_{min}$) are determined by Dalitz plot analysis and Eq. \eqref{gamma3} can be integrated 
when neglecting the mass of $\chi$-particle to yield the total decay rate
\begin{equation}\label{TG3}
\Gamma_{\phi\to\chi\chi\chi}=\frac{h^2m_\phi}{3!64(2\pi)^3}\;.
\end{equation}
Following the same steps as in the preceding subsection, one can relate the coupling constant to the reheating $e$-folding number $N_{\rm re}$ 
as
\begin{equation}\label{g3_0}
h^2=\frac{3!128(2\pi)^3\Lambda^2}{3M_{pl}m_\phi}\Bigg(\frac{2n}{2n+\sqrt{3\alpha}}\Bigg)^{n}\exp{\Bigg(\frac{-3n}{1+n}N_{re}}\Bigg)\;,
\end{equation}
which determines the coupling constant as a function of the spectral index $n_s$ via $n_s$-dependence of $N_{\rm re}$, $\Lambda$ and $m_\phi$. 
The relationship between $h$ and $n_s$ is illustrated by the blue lines plotted in Figs. \ref{phichi3_2}-\ref{phichi3_3}, 
in which the notations and parameter choices are the same as in Fig. \ref{phichi2_pert}. 
The red lines in those figures reflect the thermal effect, which we discuss below.  

\paragraph{Thermal feedback} - Now let us find the thermal correction to Eq. \eqref{g3_0}.
 In order to see the thermal effects, one should know the dissipation rate of the inflaton field 
 in the thermal background. 
One may immediately promote the vacuum rate of the 3-body decay Eq. \eqref{TG3} to the thermally corrected one 
 by including Bose enhancement factor for final products of decay and replacing the vacuum mass $m_\chi$ by the thermal effective mass $M_{\chi}$.
This thermally corrected rate for the 3-body decay $\phi\to\chi\chi\chi$ is given in Eq. \eqref{SD} of Appendix \ref{AppendixA}.
However, this does not capture the full dissipation rate 
in the presence of a finite density of $\chi$-particles in the thermal background because  
$\phi$ can also lose energy in scatterings $\phi\chi\to\chi\chi$ (similar to ``Landau damping").\footnote{For the $\phi \chi^2$ interaction in Eq. (\ref{L_chi})
we did not consider scatterings because these only occur at higher order in coupling constants.}
Scatterings can play an important role when the three body decay is kinematically forbidden by the  
growing thermal masses of the daughter particles as universe heats up \cite{Drewes:2013iaa}. 
The total rate that includes both 3-body decay $\phi\to\chi\chi\chi$ and scatterings $\phi\chi\to\chi\chi$ is presented in Appendix \ref{On-shell}.
It is shown in Fig. \ref{GammaOfT} how the contribution of the each channel to the total dissipation rate changes as the temperature increases.
\begin{figure}[htp!]
\begin{center}
\includegraphics[scale=1.1]{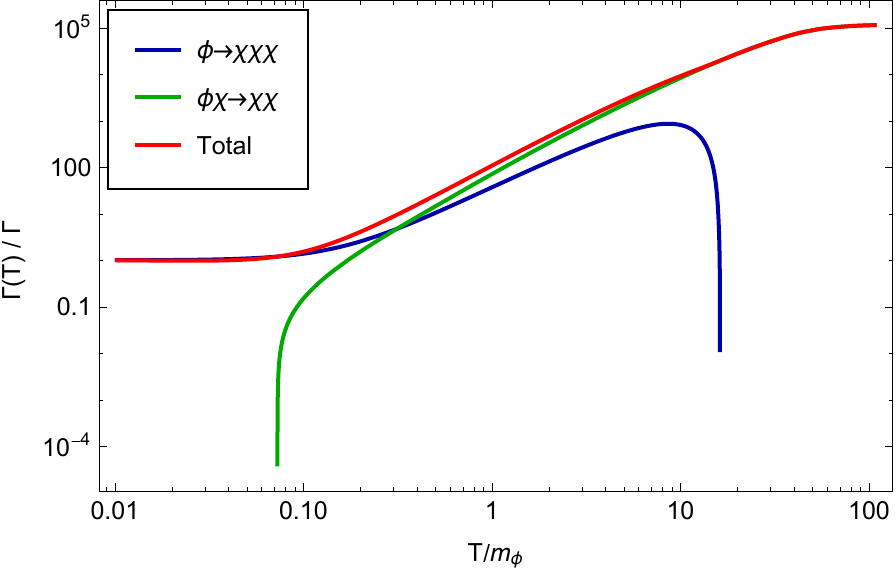}
\caption{The temperature dependence of the dissipation rate of the inflaton field for the interaction $h \phi \chi^3/3!$ and the self-interaction $\lambda \chi^4/4!$, see Eq. (\ref{L_chi}). 
The red line indicates the total dissipation rate, the blue and green line indicate the 3-body decay $\phi\to\chi\chi\chi$ and scatterings $\phi\chi\to\chi\chi$, 
respectively. 
Here we choose the self-coupling constant of $\chi$-particles as $\lambda=0.01$.
The temperature is normalised by inflaton mass and the decay rate by its vacuum value (i.e. at zero temperature).}
\label{GammaOfT}
\end{center}
\end{figure}\\
Fig. \ref{phichi3_2} establishes the relation between $n_s$ and $h$, using the total rate as calculated in Appendix \ref{On-shell}. 
When one increases the self-coupling constant $\lambda$,  
the effect of the growing thermal mass, which suppresses the dissipation rate, becomes more significant. This can be seen by comparing the behaviours of the red lines in Fig. \ref{phichi3_2} ($\lambda=10^{-4}$) and Fig. \ref{phichi3_3} ($\lambda=0.1$). 
Fig. \ref{planck_phichi3} is analogous to Fig. \ref{planck_phichi2}, but the numbers over the disks denote $\log_{10}h$, which is computed using Eq. \eqref{gamma3_t1}.   
In Figs.  \ref{phichi3_2} and \ref{phichi3_3} the yellow regions denoted by ``preheating'' correspond to the resonance regimes where the perturbative treatment given above 
does not seem applicable, 
see the next subsection for the discussion of the resonances. 
In Figs. \ref{phichi3_2} and \ref{phichi3_3} one can clearly see that there 
exists a range of the inflaton couplings in which our perturbative method can be used to determine $h$ from $n_s$. As for the $g\phi\chi^2$ interaction, thermal effects are visible (see the red lines) only outside of the perturbative regimes.

\begin{figure}[h!]
\begin{subfigure}{0.5\textwidth}
	\includegraphics[width=1\linewidth, height=6.2cm]{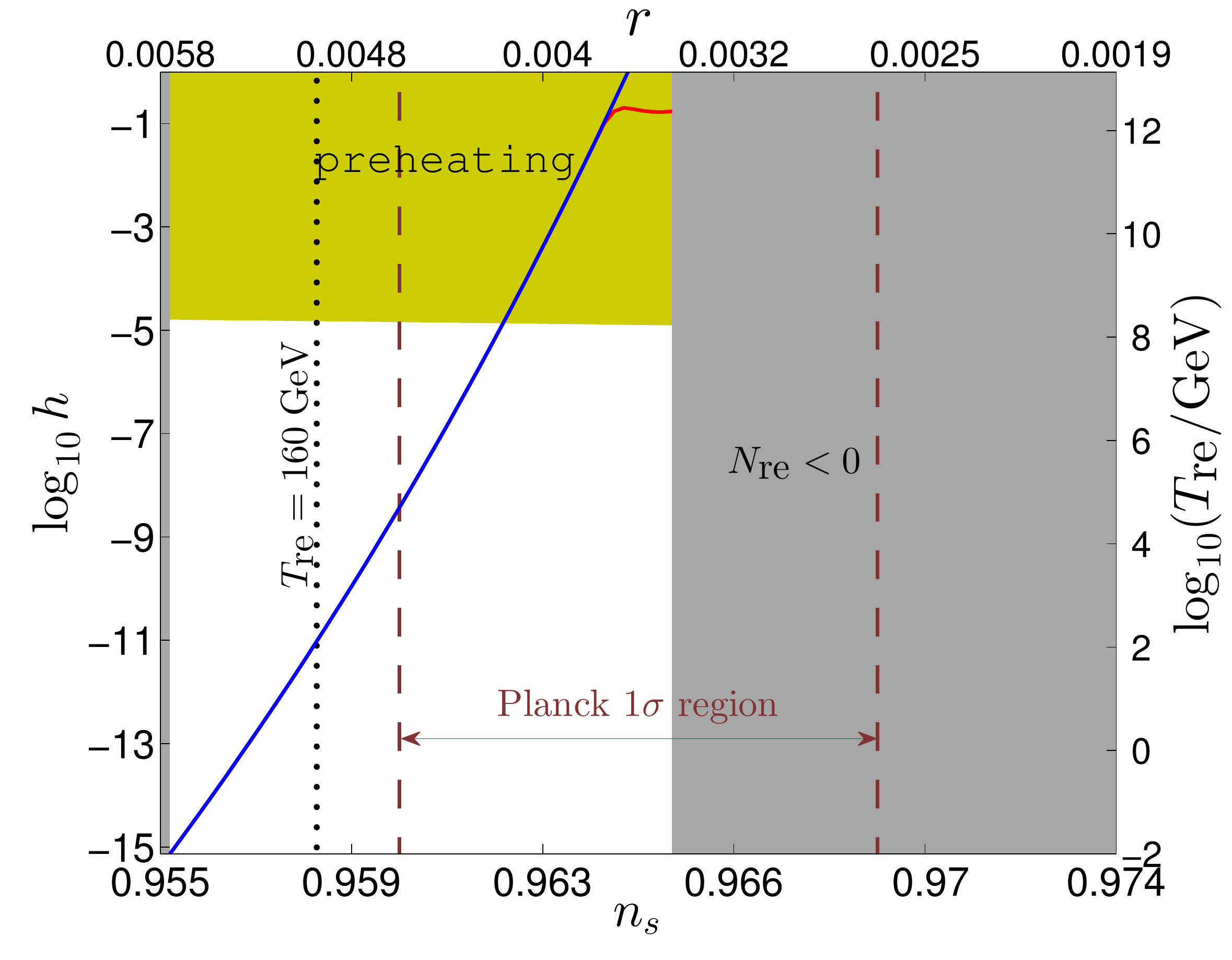}
	\caption{$\alpha=1$}
\end{subfigure}
\begin{subfigure}{0.5\textwidth}
	\includegraphics[width=1\linewidth, height=6.2cm]{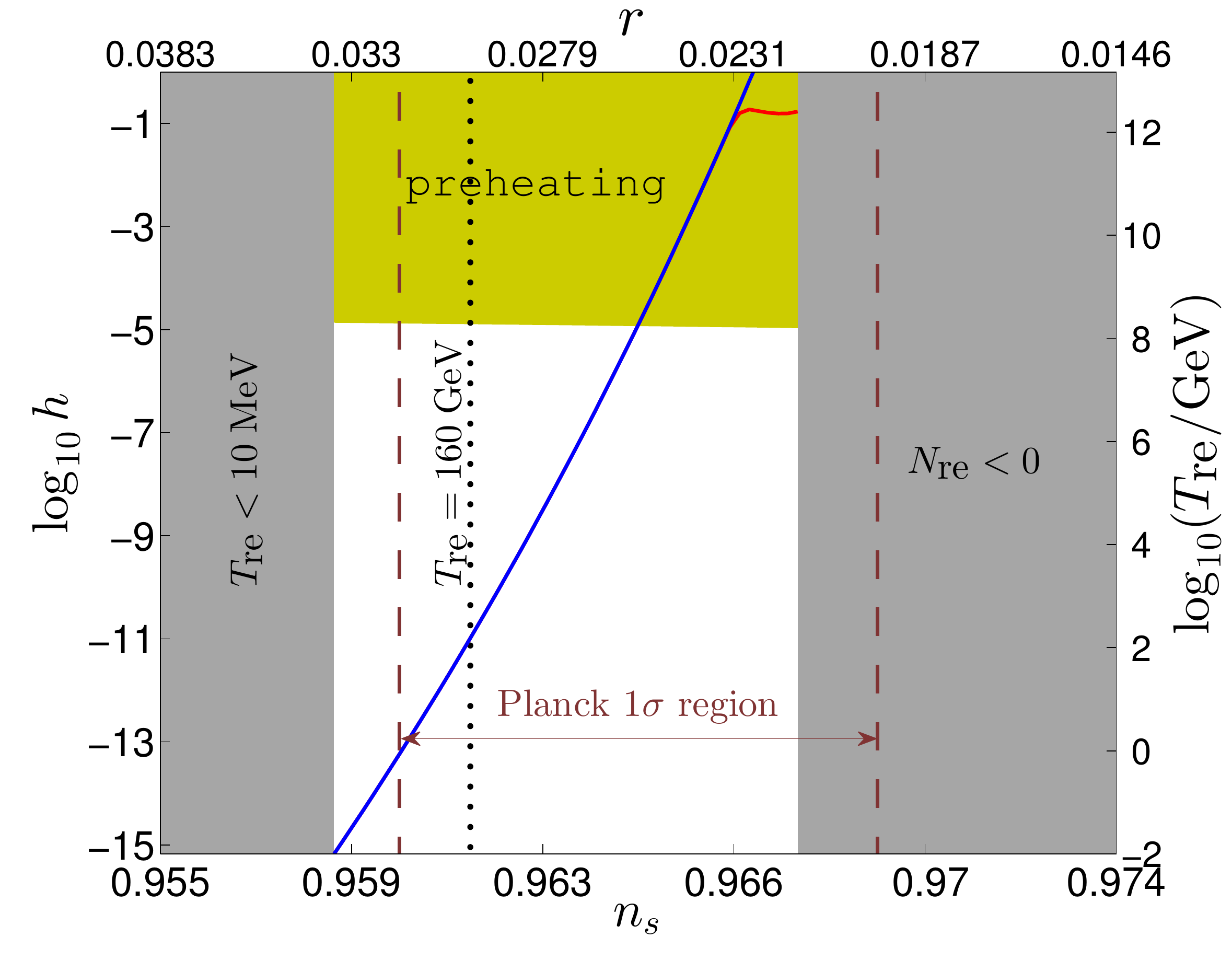}
	\caption{$\alpha=10$}
\end{subfigure}
\begin{subfigure}{0.5\textwidth}
	\includegraphics[width=1\linewidth, height=6.2cm]{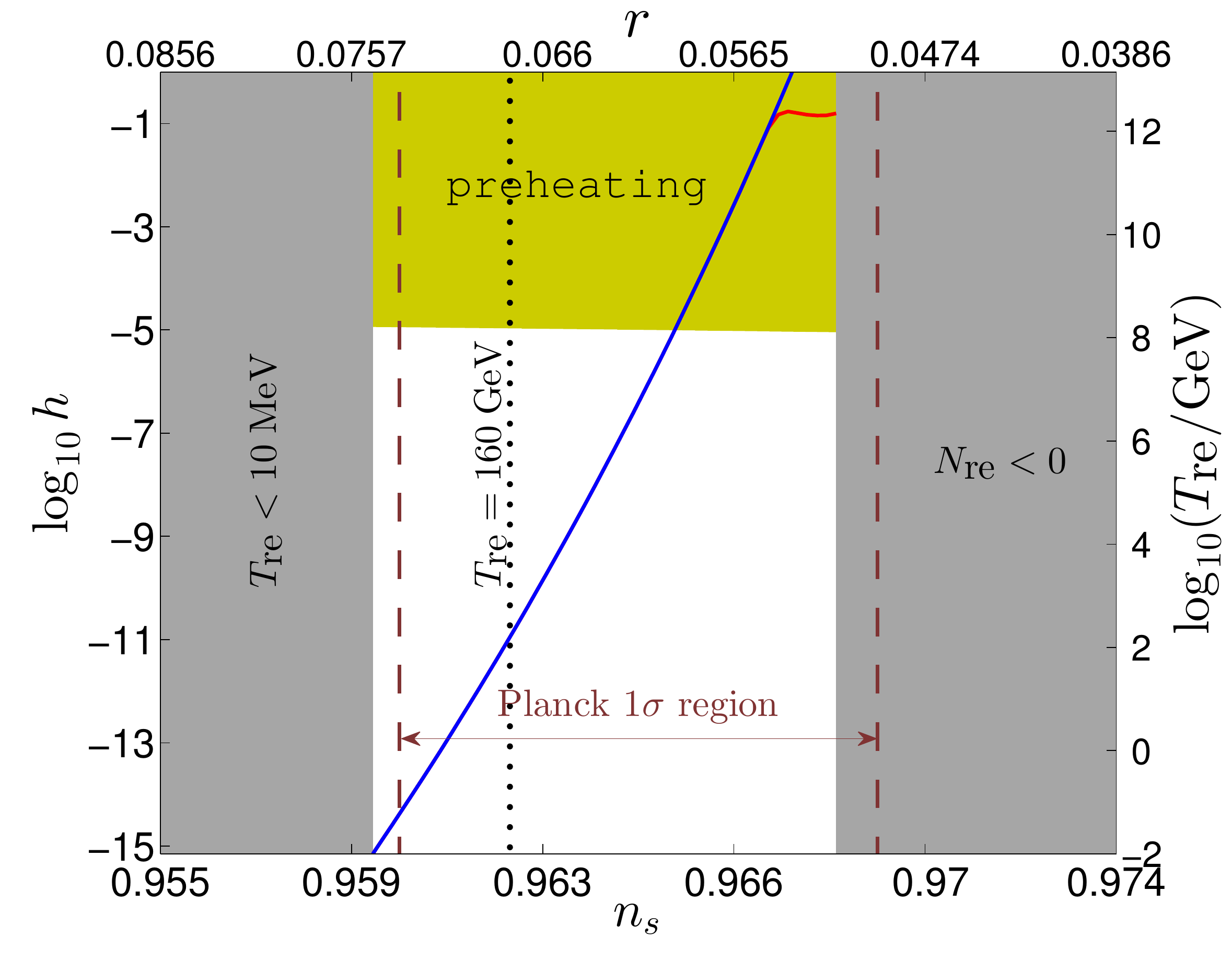}
	\caption{$\alpha=50$}
\end{subfigure}
\begin{subfigure}{0.5\textwidth}
	\includegraphics[width=1\linewidth, height=6.2cm]{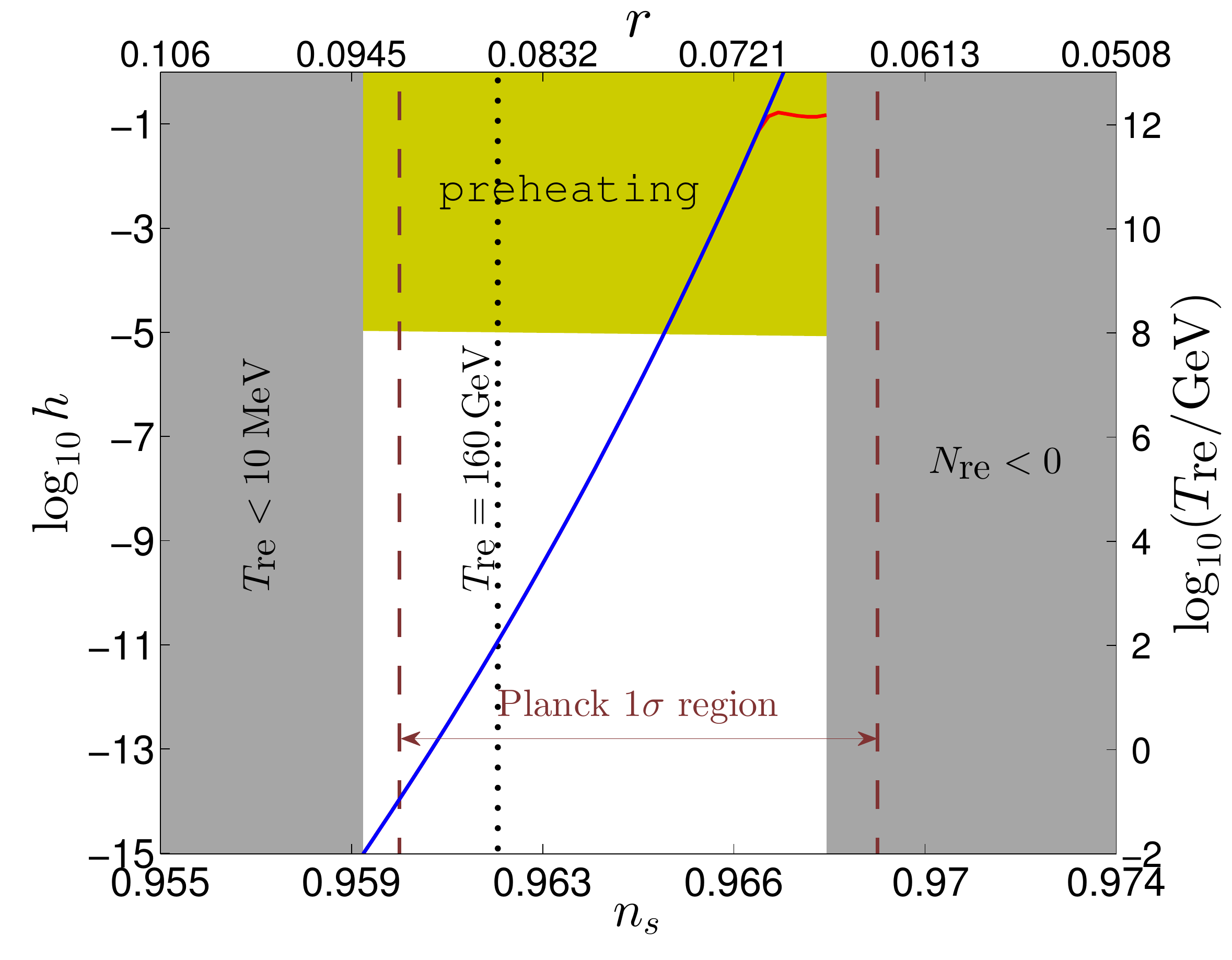}
	\caption{$\alpha=100$}
\end{subfigure}
\caption{For fixed $\alpha$ and $n$, the coupling constant $h$ in the interaction $h \phi \chi^3/3!$ can be determined from $n_s$. 
The solid lines are plotted from Eqns. \eqref{g3_0} (blue line) and \eqref{gamma3_t1} (red line) for various values of $\alpha$. 
The same notations and set of parameters as in Fig. \ref{phichi2_pert} are used. 
The self-coupling of $\chi$-particles is chosen as $\lambda=10^{-4}$.}
\label{phichi3_2}
\end{figure}

\begin{figure}[h!]
\begin{subfigure}{0.5\textwidth}
	\includegraphics[width=1\linewidth, height=6.2cm]{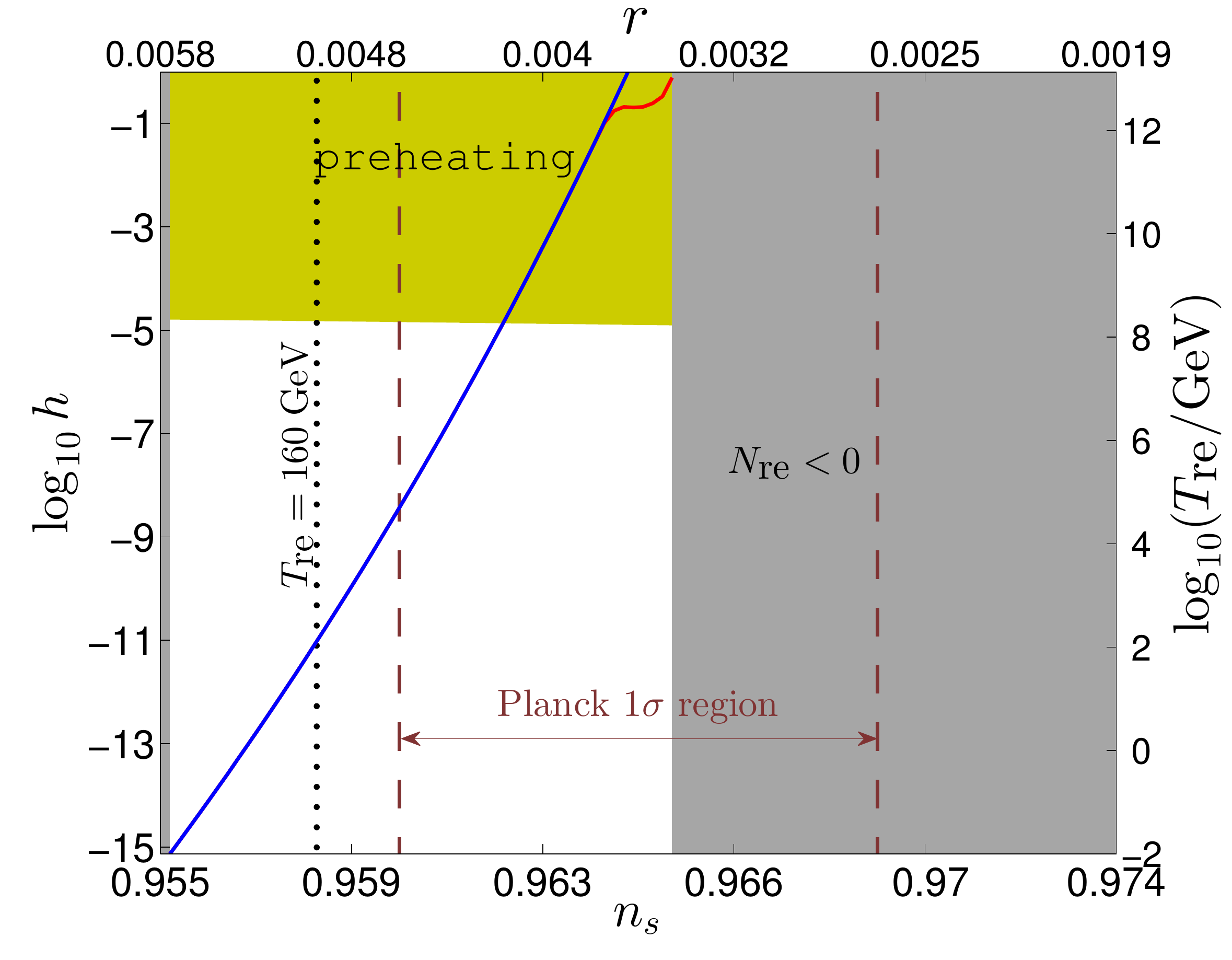}
	\caption{$\alpha=1$}
\end{subfigure}
\begin{subfigure}{0.5\textwidth}
	\includegraphics[width=1\linewidth, height=6.2cm]{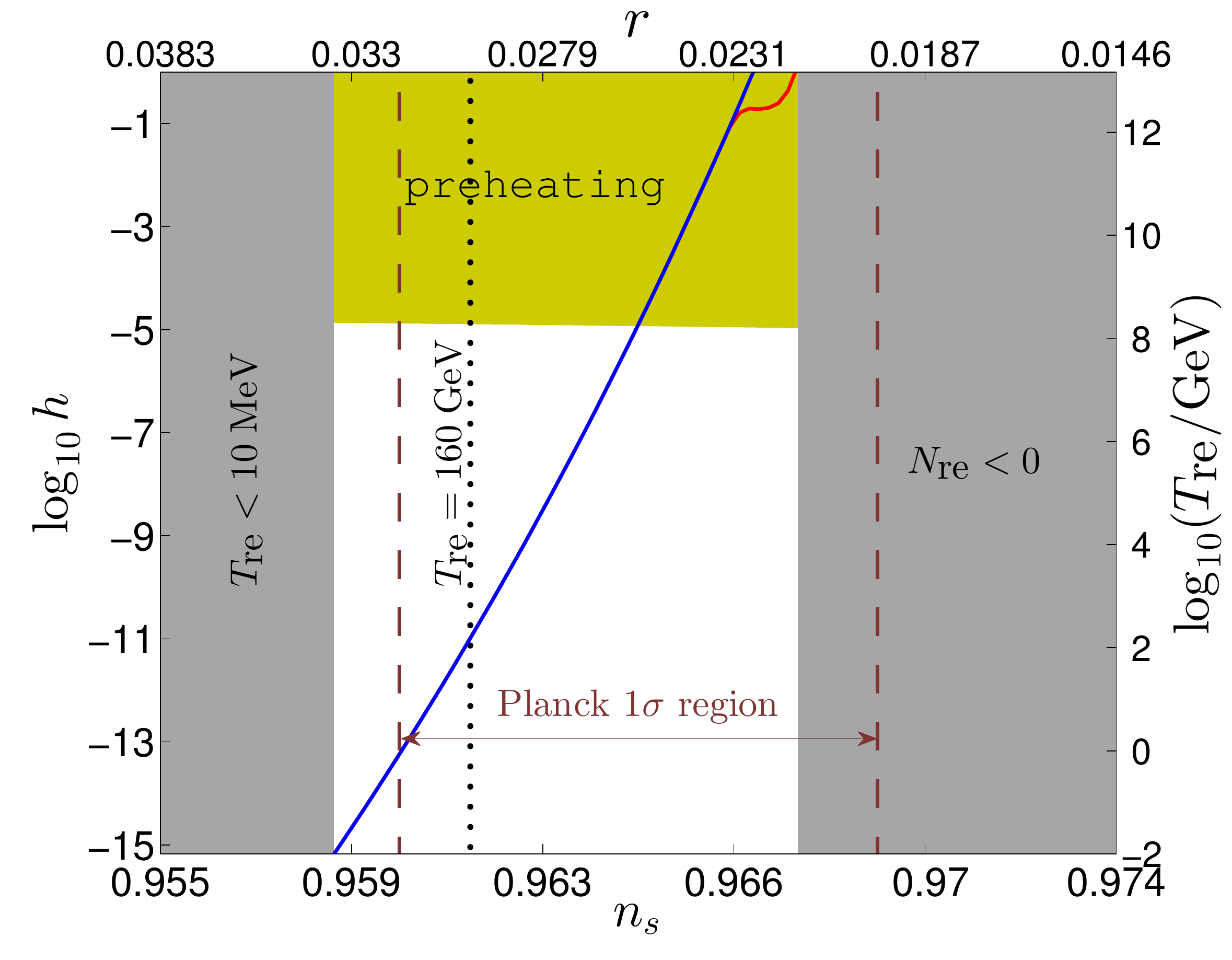}
	\caption{$\alpha=10$}
\end{subfigure}
\begin{subfigure}{0.5\textwidth}
	\includegraphics[width=1\linewidth, height=6.2cm]{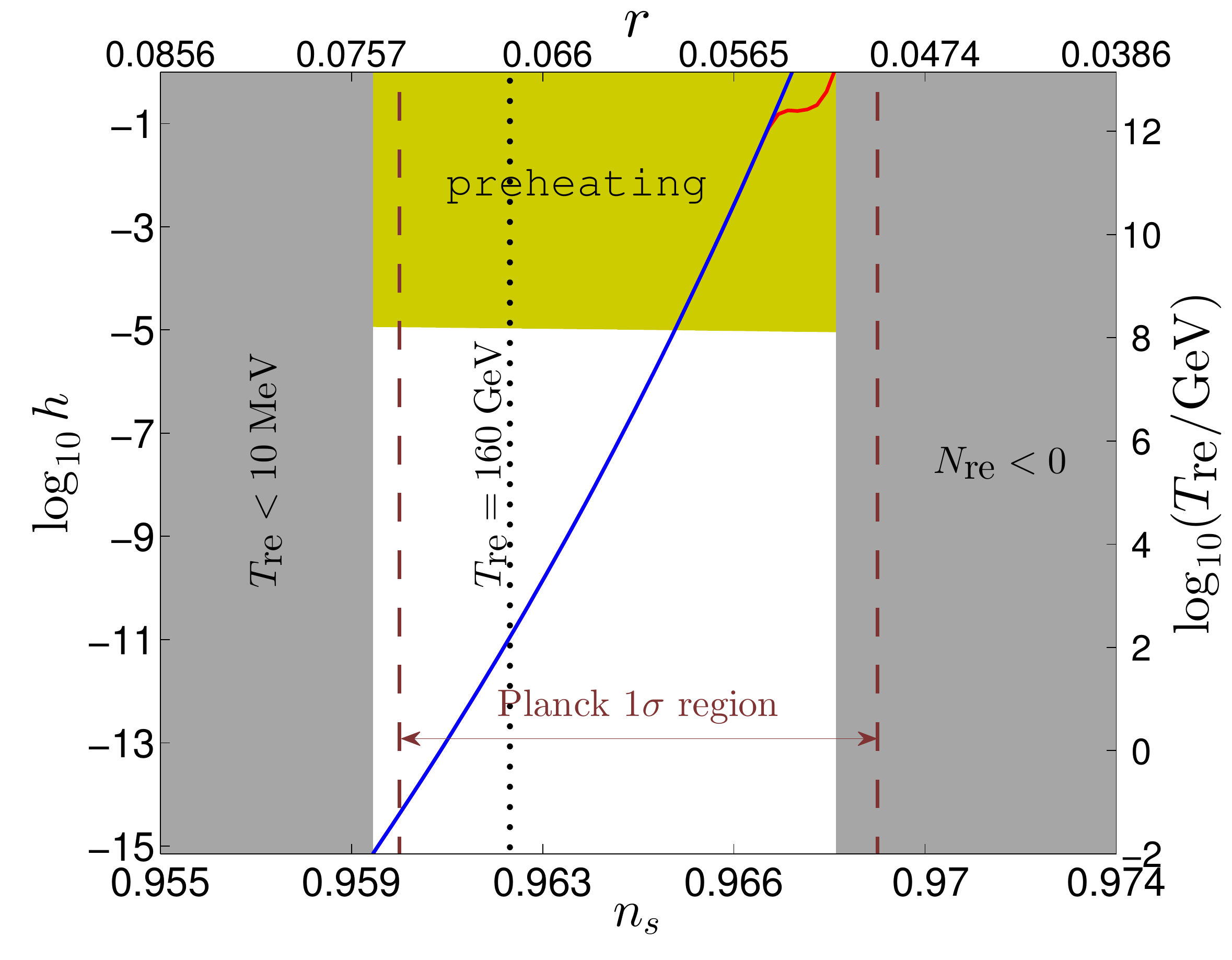}
	\caption{$\alpha=50$}
\end{subfigure}
\begin{subfigure}{0.5\textwidth}
	\includegraphics[width=1\linewidth, height=6.2cm]{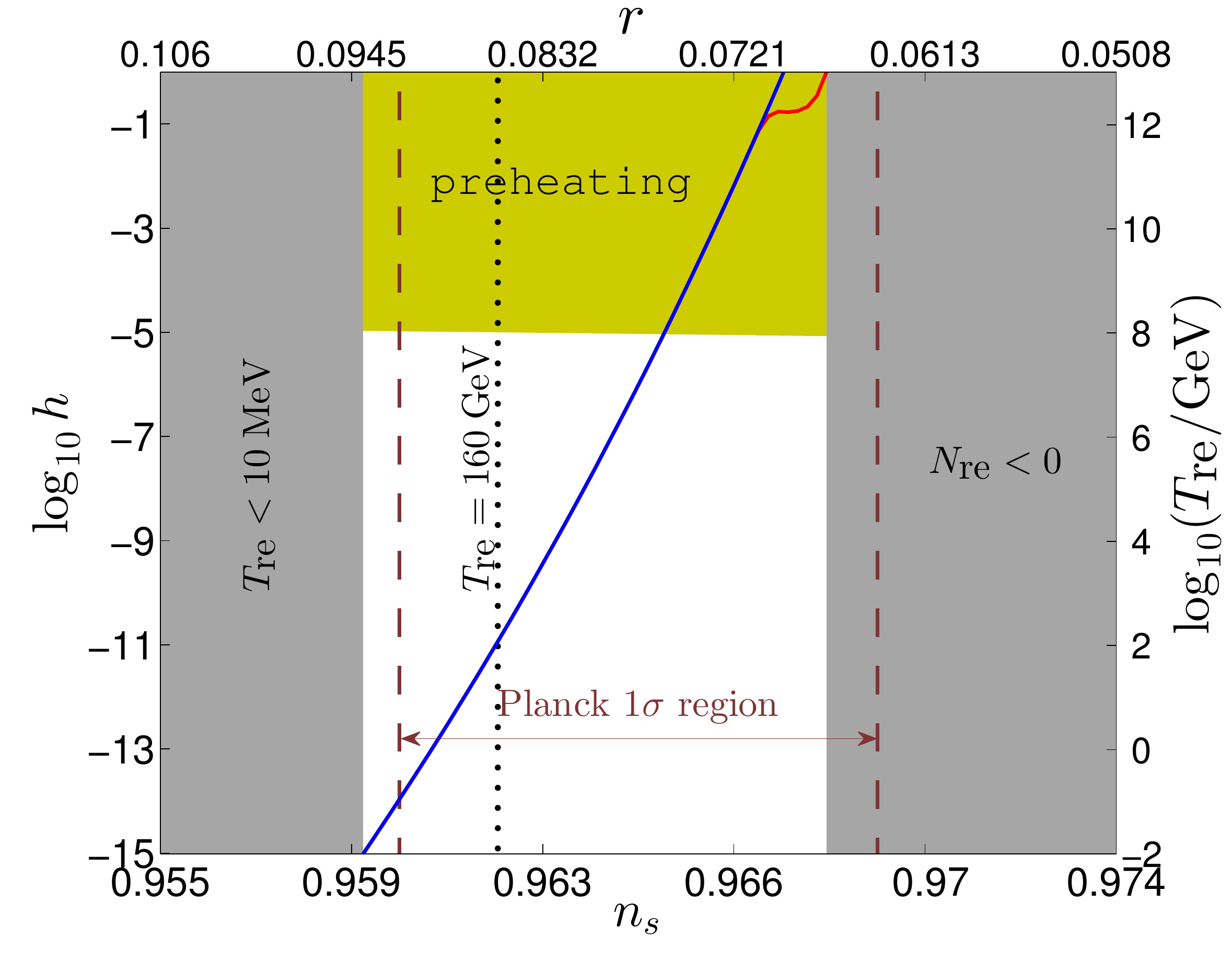}
	\caption{$\alpha=100$}
\end{subfigure}
\caption{The same as Fig. \ref{phichi3_2}, but with $\lambda=0.1$.}
\label{phichi3_3}
\end{figure}

\begin{figure}[htp]
\begin{center}
\includegraphics[scale=0.6]{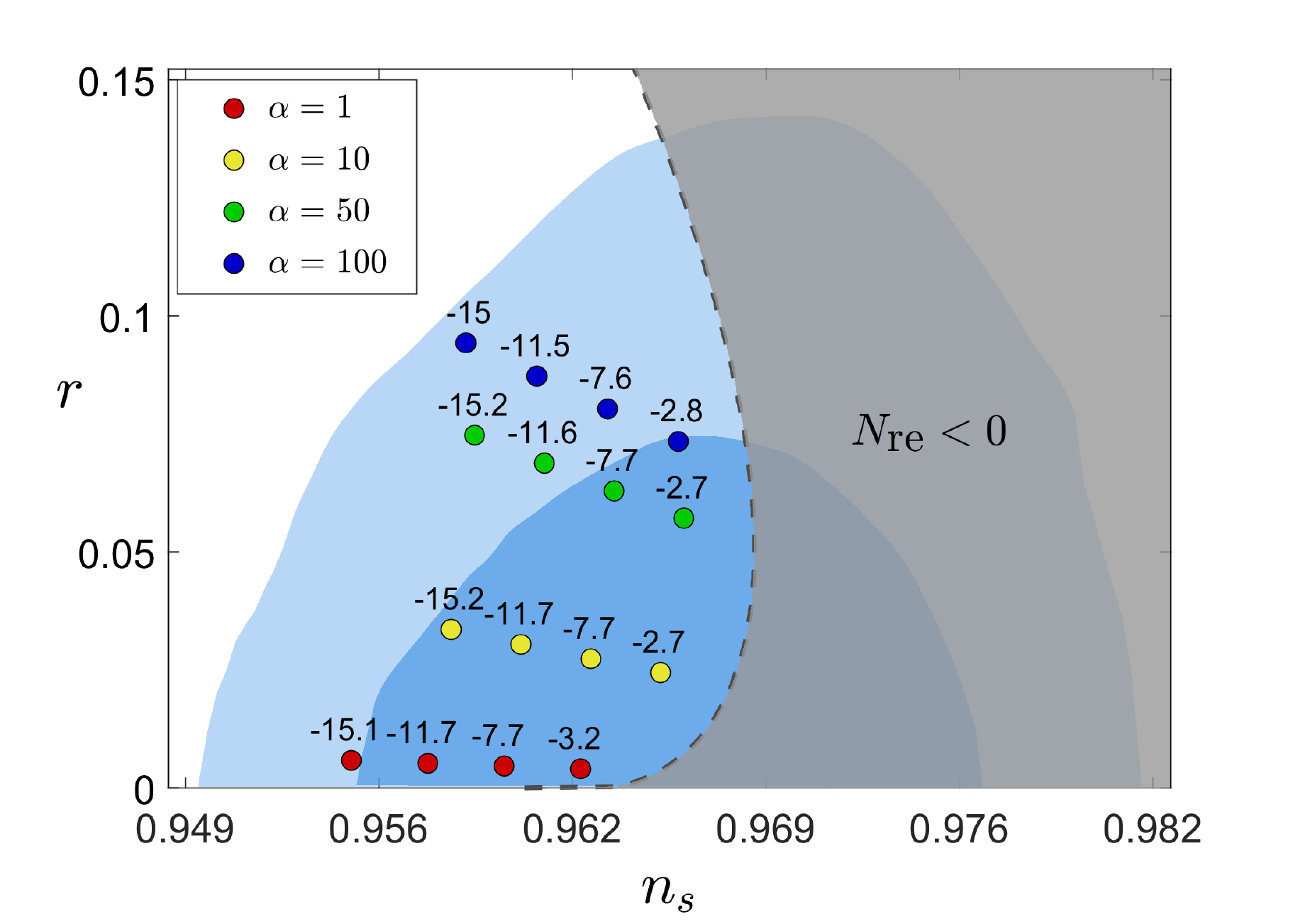}
\end{center}
\caption{The same as Fig. \ref{planck_phichi2}, but for the interaction $h \phi \chi^3/3!$. 
The numbers over the disks denote $\log_{10}h$ computed using Eq. \eqref{gamma3_t1}.
}\label{planck_phichi3}
\end{figure}

\subsubsection{Resonances}
We shall again investigate to which degree 
the applicability of the previous perturbative result is affected or limited 
by resonances.

\paragraph{Broad resonance} - 
At tree level, the equation of motion for the mode function $\chi_k$ is obtained as
\begin{equation}
\ddot{\chi}_k+(k^2+m_\chi^2)\chi_k=0\;.
\end{equation}
Notice that the mass term is completely time independent, as a result, parametric resonance due to the time dependent mass term does not happen by the interaction Eq. \eqref{chi3}. A time dependent correction to the mass term is generated radiatively at order $h^2$. Leaving aside some numerical prefactors, we can estimate that the considerations from the previous subsection can be applied if one replaces $\tildeg m_\phi\Phi\rightarrow h^2\Phi^2$, i.e. $q\sim h^2\Phi^2/m_\phi^2$. Hence, the condition for broad resonance (assuming $\Phi \sim M_{pl}$) is roughly
\begin{equation}\label{BroadResonance2}
h \Phi> m_\phi \qquad \implies \qquad h > m_\phi / M_{pl} \,.
\end{equation}
In reality, one may need somewhat larger lower bound for coupling than the one given in the inequality \eqref{BroadResonance2} because we have neglected the numerical factors in $q$.

\paragraph{\emph{Thermal Resonance} due to the Bose enhancement} - 
The condition for the occurrence of a narrow resonance for the $\phi\chi^2$-interaction can not be directly applied here 
because, in contrast to the two body decay, the particles produced in a three body decay do not all have the same magnitude of momentum. 
In the limit $m_\chi\rightarrow0$ 
it is straightforward to show from Eq. (\ref{gamma3}) that the spectrum of produced $\chi$-particles is proportional to $|\textbf{p}|$ and ends at $|\textbf{p}|=m_\phi/2$.
This corresponds to a phase space distribution function 
$f_\chi(|\textbf{p}|) = \frac{c}{|\textbf{p}|}\theta(m_\phi/2 - |\textbf{p}|)$,
where $c$ is a normalisation constant.
A simple criterion for the occurrence of a resonance in some mode $|\textbf{p}|$ is $f_\chi(|\textbf{p}|)\sim1$.
Let us consider a time interval $\tau$ at the onset of reheating. If $\tau$ is sufficiently short, then we can neglect Hubble expansion and rescatterings. 
The total energy dissipated by the inflaton during the time $\tau$ is then $\varepsilon=\frac{4}{3}V_{\rm end}(1-e^{-\Gamma_{\phi\rightarrow\chi\chi\chi}\tau})
\simeq \frac{4}{3}V_{\rm end}\Gamma_{\phi\rightarrow\chi\chi\chi}\tau$.
Since each decay produces three $\chi$-particles, the total number of produced particles is $n_\chi=3\varepsilon/m_\phi$, and we can determine 
the coefficient $c$ from the requirement that the phase space integral over $f_{\chi}$ is $n_\chi$. We find
\begin{equation}\label{fchi}
f_\chi(|\textbf{p}|) = \frac{12\varepsilon(2\pi)^2}{m_\phi^3|\textbf{p}|}\theta(m_\phi/2 - |\textbf{p}|)
=\frac{16V_{\rm end}\Gamma_{\phi\rightarrow\chi\chi\chi}\tau(2\pi)^2}{m_\phi^3|\textbf{p}|}\theta(m_\phi/2 - |\textbf{p}|) \,.
\end{equation}
$f_\chi(|\textbf{p}|)$ exceeds unity for arbitrarily short $\tau$ if $|\textbf{p}|$ is sufficiently small, but the phase space for small momenta is also small. To asses the question whether induced transitions cause an exponential growth of $n_\chi$ we therefore first focus on small momenta.

For $|\textbf{p}|\ll T$, the Bose-Einstein distribution scales as $\sim T/|\textbf{p}|$, i.e., has the same momentum-dependence 
as Eq. (\ref{fchi}).
We can therefore address the question whether a resonance occurs that is strong enough to affect the CMB by comparing to the idealised case that the $\chi$-particles thermalise instantaneously,\footnote{See Refs. \cite{Mazumdar:2013gya,Harigaya:2013vwa,Mukaida:2015ria} and references therein for a recent discussion on the issue of thermalisation.} 
which has been studied in Ref. \cite{Drewes:2015coa}. 
There it was found that, if $\Gamma$ has the form $\Gamma=\Gamma_0+\Gamma_2(T/m_\phi)^2$, then feedback from thermally distributed particles only affects the CMB if
\begin{equation}\label{Gamma2}
\Gamma_2>\frac{m_\phi^2}{M_{pl}}\sqrt{\frac{8\pi^3 g_{*}}{90}} \, .
\end{equation}
To identify $\Gamma_2$ for the case under consideration here, one can use the thermal damping rate given in Appendix \ref{AppendixA}. 
For a rough estimate, we may use the approximation \cite{Drewes:2013iaa}
\begin{equation}\label{gammaTapprox}
\Gamma_2 \simeq \frac{h^2 m_\phi}{768\pi} \,.
\end{equation}
Inserting Eq. (\ref{gammaTapprox}) into (\ref{Gamma2}), one finds that induced transitions caused by thermally distributed $\chi$-particles only affect the CMB if
\begin{equation} \label{therm-cond}
h>200 \sqrt{m_\phi/M_{pl}} \,.
\end{equation}  
Since this condition  for $m_\phi \ll M_{pl}$ is stronger than (\ref{BroadResonance2}), 
it is clear that thermal feedback would only affect the CMB if the coupling $h$ is so large that a perturbative treatment of reheating cannot be applied in any case.
In order to ensure that the induced decays of $\phi$ into infrared $\chi$ modes do not leave an imprint in the CMB in the perturbative regime, it is therefore sufficient to show that the occupation numbers remain below their thermal would-be values. 
Let us now find a condition for $f_\chi(|\textbf{p}|)/f_B(|\textbf{p}|)<1$, using that 
the temperature associated with an energy density $\varepsilon$ is $T=(30\varepsilon/\pi^2 g_*)^{1/4}$ with the assumption of instantaneous thermalisation for the $\chi$-particles. 
For $|\textbf{p}|\ll T$ we can approximate
\begin{eqnarray}
\frac{f_\chi(|\textbf{p}|)}{f_B(|\textbf{p}|)} 
\simeq 360 g_*^{1/4} \left(\frac{\varepsilon^{1/4}}{m_\phi}\right)^{3} \,.
\end{eqnarray}
With $\phi_{\rm end}\simeq M_{pl}\sqrt{2}$
and 
$\varepsilon\simeq\frac{4}{3}V_{\rm end} \Gamma_{\phi\rightarrow\chi\chi\chi}\tau
\simeq \frac{4}{3}\frac{1}{2}m_\phi^2\phi_{\rm end}^2\Gamma_{\phi\rightarrow\chi\chi\chi}\tau
\simeq \frac{4}{3}m_\phi^2M_{pl}^2\Gamma_{\phi\rightarrow\chi\chi\chi}\tau$, we find
\begin{eqnarray}\label{Route66}
\frac{f_\chi(|\textbf{p}|)}{f_B(|\textbf{p}|)} 
\simeq 445 g_*^{1/4} \left(\frac{M_{pl}}{m_\phi}\right)^{3/2}(\Gamma_{\phi\rightarrow\chi\chi\chi}\tau)^{3/4}\,.
\end{eqnarray}
We can now investigate for what values of coupling the ratio  (\ref{Route66}) remains small for some relevant time interval $\tau$ so that no resonance occurs. If we assume that the $\chi$-particles have additional interactions that drive them to equilibrium on a typical scattering time scale $1/\Gamma_\chi$, we may set $\tau=1/\Gamma_\chi$. This leads to the condition 
$\sqrt{\Gamma_{\phi\rightarrow\chi\chi\chi}/\Gamma_\chi} < 0.02 g_*^{-1/6} m_\phi/M_{pl}$.
For $g_*\sim 100$ this roughly implies $h < m_\phi/M_{pl}$ (assuming the coupling constant that governs $\Gamma_\chi$ is of order one). 
If this condition is fulfilled, then the occupation numbers never exceed their equilibrium values because scatterings redistribute them efficiently before they do so.
Hubble expansion also helps to reduce the occupation numbers, so we alternatively may set $\tau=H^{-1}\simeq \sqrt{3}M_{pl}/\sqrt{V_{\rm end}}$, which leads to condition $h< 1.6 g_*^{-1/6} m_\phi/M_{pl}$. For $g_*\sim 100$ this again roughly implies $h < m_\phi/M_{pl}$. 

For sufficiently large momenta ($|\textbf{p}|\gtrsim T$) one finds $f_\chi(|\textbf{p}|)/f_B(|\textbf{p}|) > 1$, but in this regime 
$f_\chi(|\textbf{p}|)<1$, so that there is no resonance. 
This can be seen by evaluating $f_\chi(|\textbf{p}|)$ at $|\textbf{p}|=T$
and at the maximal momentum $|\textbf{p}|=m_\phi/2$. From Eq. \eqref{fchi} we obtain $f_\chi(T)= 445 g_*^{1/4}(
M_{pl}/m_\phi 
)^{3/2}(\Gamma_{\phi\rightarrow\chi\chi\chi}\tau)^{3/4}$, which
obviously remains below unity under the same conditions as the ratio (\ref{Route66}).
For $|\textbf{p}|=m_\phi/2$, $f_\chi(m_\phi/2)=128\pi^2(M_{pl}/m_\phi)^2 \Gamma\tau$ can again be evaluated at $\tau=1/\Gamma_\chi$ or $\tau=1/H$. 
Both roughly reproduce the same condition $h < m_\phi/M_{pl}$.

To sum up, we can use $h < m_\phi/M_{pl}$ as a condition to treat reheating perturbatively.
This condition is indicated in Figs.  \ref{phichi3_2}-\ref{phichi3_3}, in which the yellow regions denoted by ``preheating'' correspond to the preheating (resonance) regimes 
where $h > m_\phi/M_{pl}$.
 
 \subsection{Yukawa interaction}

As a final example, 
we discuss a Yukawa interaction 
with massless fermions\footnote{Here by massless fermions we mean that their vacuum masses are zero. Thermal effects and the coupling to $\varphi=\langle \phi \rangle$ can generate a fermion mass in the early universe, see below.} $\psi$ given by
\begin{equation}\label{yukawa}
\mathcal{L}_{\rm int}=-y\phi\bar{\psi}\psi\;,
\end{equation}
where $y$ stands for the coupling constant.
The decay rate of the inflaton in vacuum 
is
\begin{equation}\label{DP_yukawa}
\Gamma_{\phi\to\bar{\psi}\psi}=\frac{y^2m_\phi}{8\pi}\;.
\end{equation}
We employ the same method as previously used to express the coupling constant in terms of the spectral index as
\begin{equation}\label{h_0}
y^2=\frac{16\pi\Lambda^2}{3m_\phi M_{pl}}\Bigg(\frac{2n}{2n+\sqrt{3\alpha}}\Bigg)^{n}\exp{\Bigg(\frac{-3n N_{re}}{1+n}}\Bigg)\;.
\end{equation}

At finite temperature, there are thermal corrections to Eq. \eqref{DP_yukawa} \cite{Drewes:2013iaa, Drewes:2015eoa}, and the decay rate is given by
\begin{equation} \label{DPT_yukawa}
\Gamma_{\phi\to\bar{\psi}\psi}=\frac{y^2}{8\pi m_\phi}\Bigg[1-\Big(\frac{2M_\psi}{m_\phi}\Big)^2\Bigg]^{1/2}(1-2f_F(m_\phi/2))\;,
\end{equation}
where $f_F(m_\phi/2)=(1+e^{\frac{m_{\phi}}{2T}})^{-1}$ is the Fermi-Dirac distribution function. 
The last factor with $f_F$ in Eq. \eqref{DPT_yukawa} suppresses the rate and manifests the Pauli blocking effect. 
$M_\psi=\frac{\alpha_G T}{2}$ is  the asymptotic effective mass of the fermion induced from the  thermal effect through a gauge interaction 
characterised by constant $\alpha_G$ \cite{Weldon:1982bn}. We set $\alpha_G=0.1$ in the plots.    
Using  Eq. \eqref{DPT_yukawa} we get, instead of Eq. \eqref{h_0},
\begin{align}\label{h_T}
y^2=\frac{16\pi\Lambda^2}{m_\phi M_{pl}}\Bigg(\frac{2n}{2n+\sqrt{3\alpha}}\Bigg)^{n}\exp{\Bigg(\frac{-3nN_{\rm re}}{1+n}\Bigg)}
\cdot\Bigg[1-\Big(\frac{2M_\psi}{m_\phi}\Big)^2\Bigg]^{-1/2}(1-2f_F(m_\phi/2))^{-1}\;.
\end{align}
\begin{figure}[htp!]
\begin{center}
\includegraphics[scale=0.6]{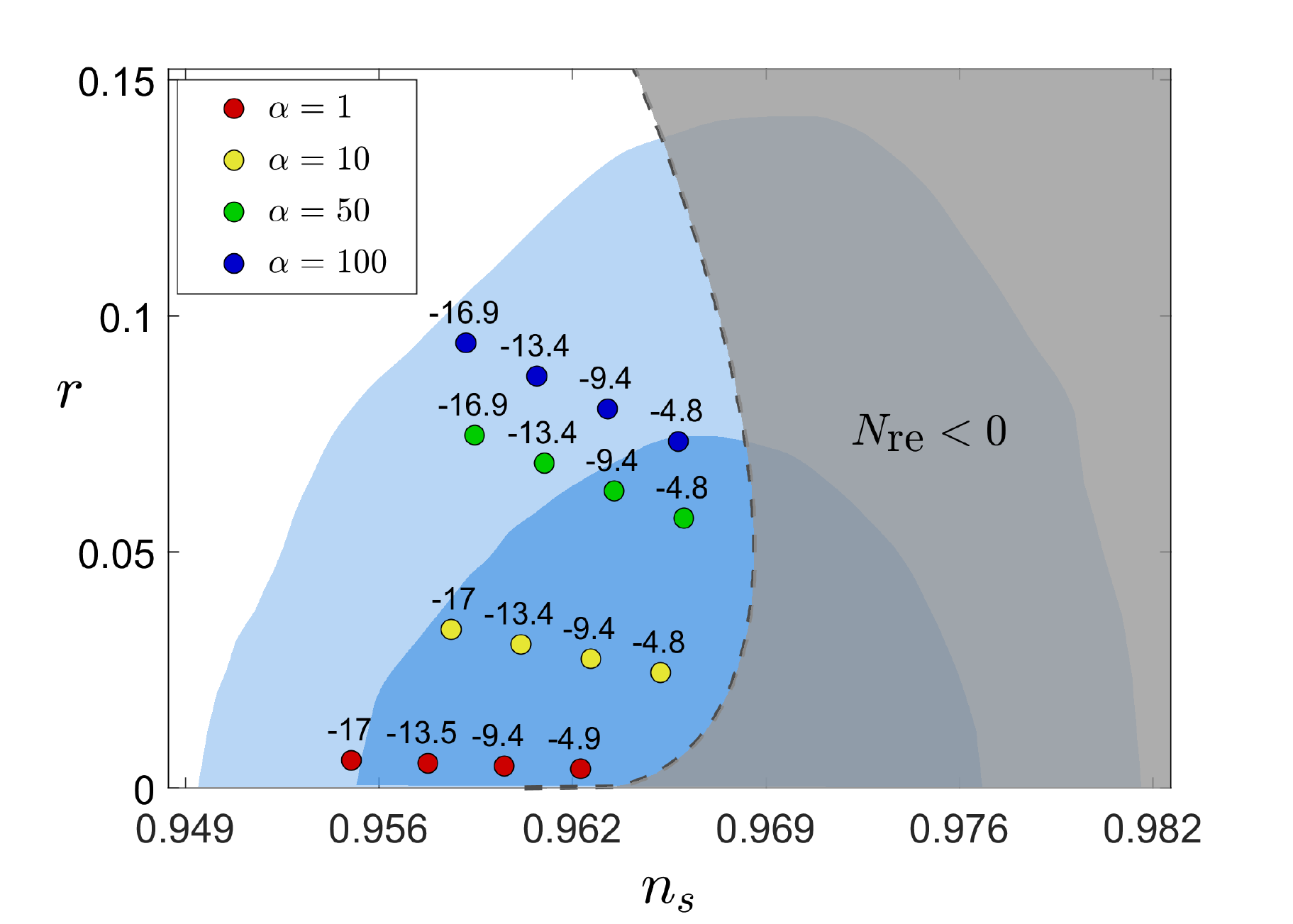}
\end{center}
\caption{The same as  Fig. \ref{planck_phichi2}, but for Yukawa interaction $y \phi \bar{\psi} \psi$.
The numbers over disks denote $\log_{10} y$ computed using Eq. \eqref{h_T}. }\label{planck_Yukawa}
\end{figure}
Using Eqns. \eqref{h_0} and \eqref{h_T}, we can plot the Yukawa coupling constant $y$ as a function of the CMB data, in particular spectral index $n_s$.  
Fig. \ref{planck_Yukawa} shows the effect of the reheating phase on the predictions of CMB parameters. 
The numbers over the disks indicate $\log_{10}y$, using Eq. \eqref{h_T}.
Fig. \ref{Fig.9} shows the spectral index dependence of the Yukawa coupling given by Eqns. \eqref{h_0} (blue line) and \eqref{h_T} (red line) for various values of $\alpha$.
The red line is steeper than the blue line when thermal effects are relevant, which is in sharp contrast to the bosonic cases (e.g. compare Fig. \ref{Fig.9} with Fig. \ref{phichi2_pert}), 
is the consequence of the fact that the rate of the inflaton decay into fermions is Pauli-suppressed at finite temperature. 
\begin{figure}[h!]
\begin{subfigure}{0.5\textwidth}
	\includegraphics[width=1\linewidth, height=6.2cm]{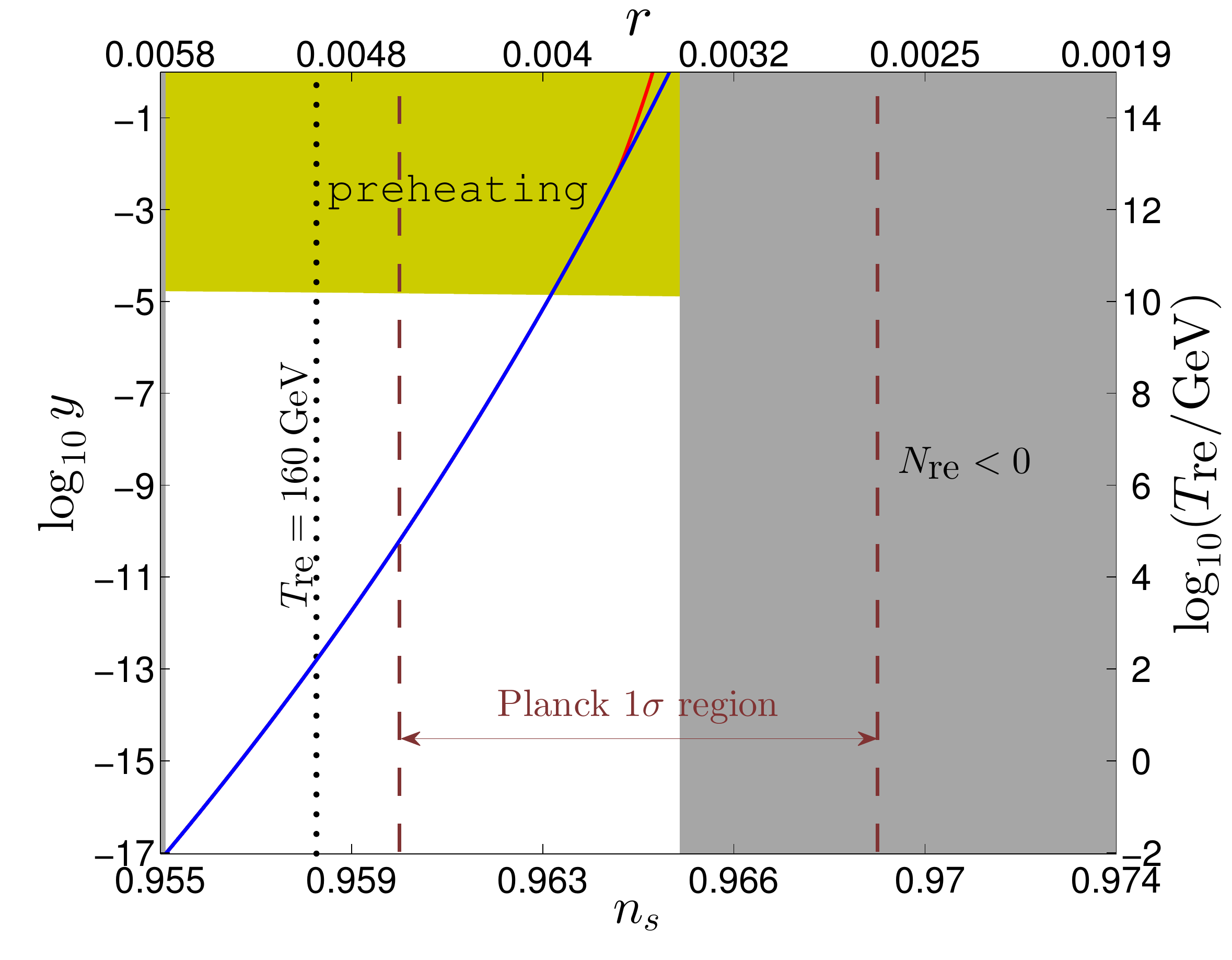}
	\caption{$\alpha=1$}
\end{subfigure}
\begin{subfigure}{0.5\textwidth}
	\includegraphics[width=1\linewidth, height=6.2cm]{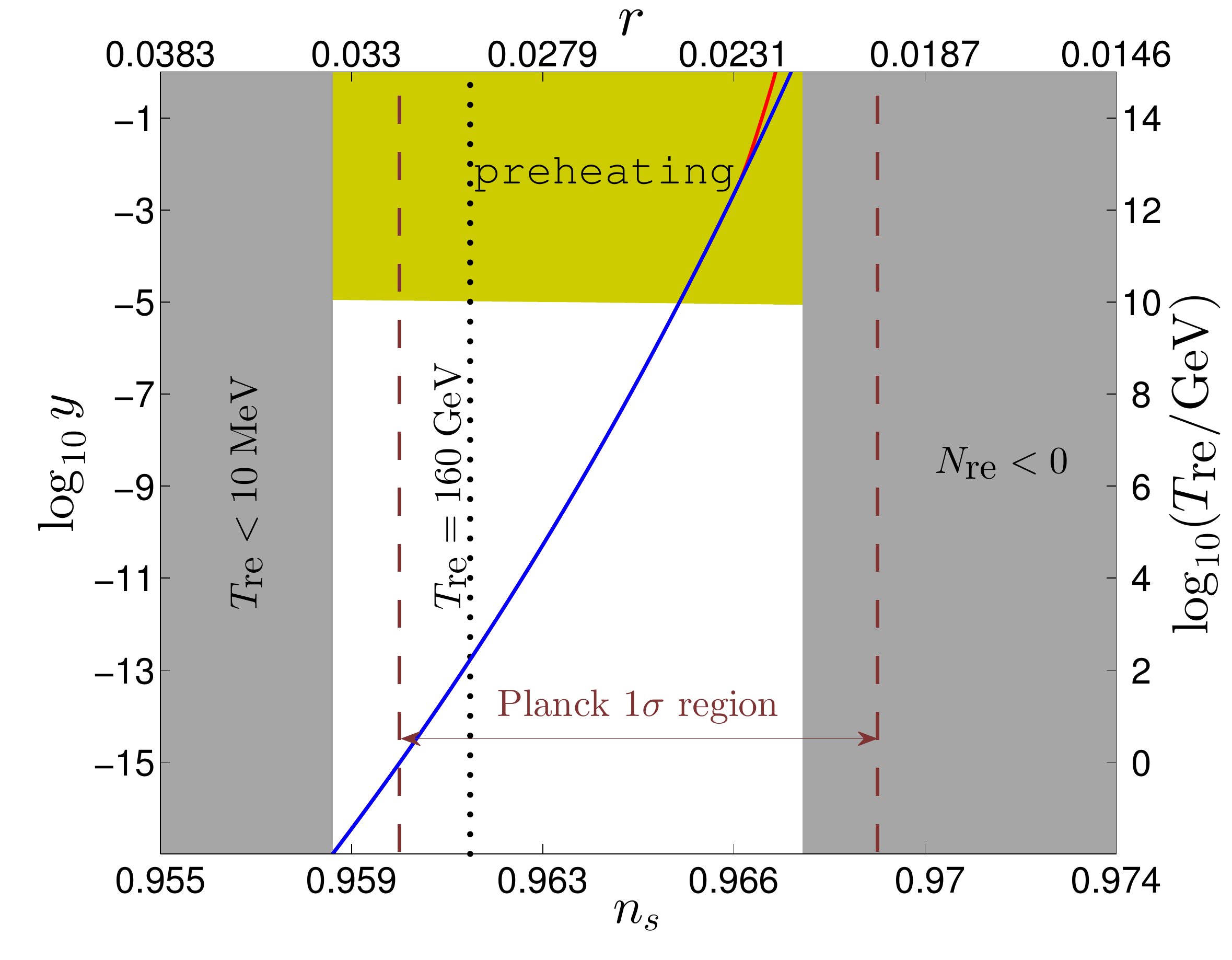}
	\caption{$\alpha=10$}
\end{subfigure}
\begin{subfigure}{0.5\textwidth}
	\includegraphics[width=1\linewidth, height=6.2cm]{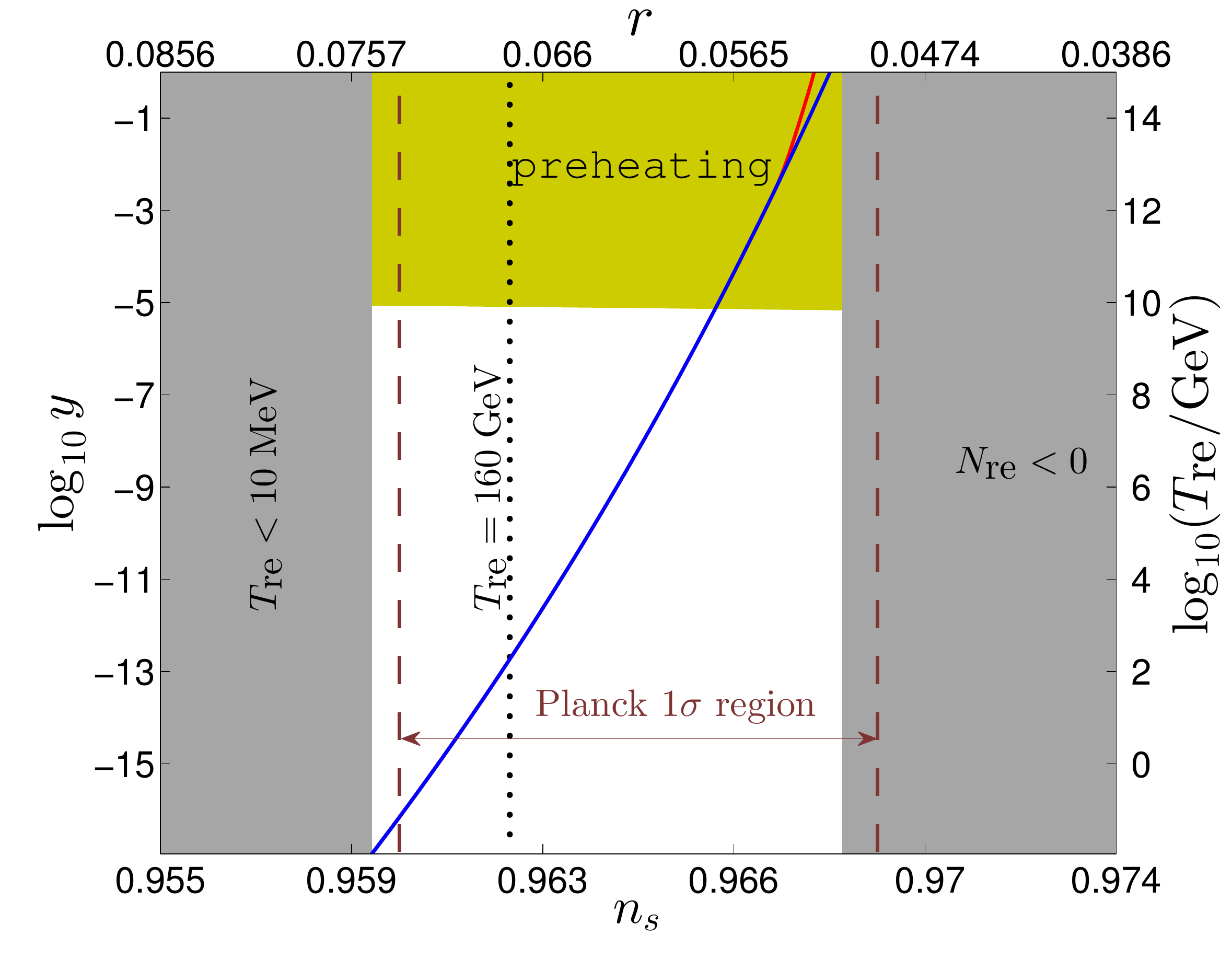}
	\caption{$\alpha=50$}
\end{subfigure}
\begin{subfigure}{0.5\textwidth}
	\includegraphics[width=1\linewidth, height=6.2cm]{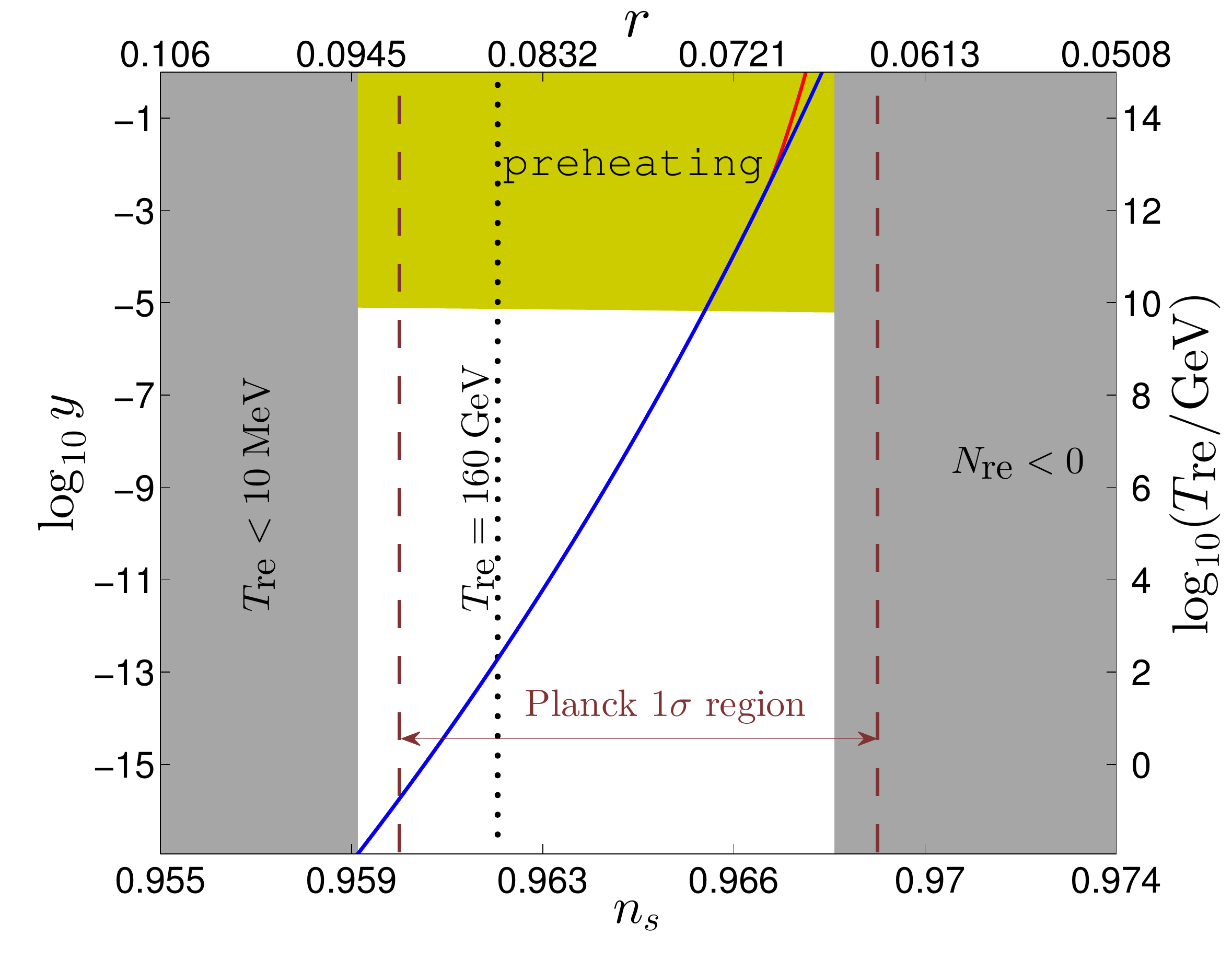}
	\caption{$\alpha=100$}
\end{subfigure}
\caption{Spectral index dependence of the Yukawa coupling. The solid lines are plotted from Eqns. \eqref{h_0} (blue line) and \eqref{h_T} (red line) for various values of $\alpha$. The parameter choices and notations are 
the same as in Fig. \ref{phichi2_pert}. 
The yellow regions denoted by ``preheating'' correspond to resonance regimes where the perturbative treatment of reheating is not applicable. 
}
\label{Fig.9}
\end{figure}
However, in addition to the thermal mass $\sim\frac{\alpha_G T}{2}$ from forward scattering, $\psi$ also receives a time dependent contribution  $\sim y\varphi$ to its effective mass from its coupling to $\phi$ via Eq. (\ref{yukawa}). It has been found in Ref. \cite{Greene:1998nh} that the reheating can not be treated perturbatively when $q=\frac{y^2\Phi^2}{m_\phi^2} >1$. In this regime the time dependent fermion mass leads to non-perturbative \emph{fermionic preheating}.
This condition\footnote{As in the bosonic case, here we replace $\Phi \rightarrow \phi_{\rm end}$, which leads to the lower bound of the coupling for the preheating that is smaller than the actual one 
since $\Phi \ll \phi_{\rm end}$ in reality.} 
for preheating of fermions is indicated in Fig. \ref{Fig.9}.
In principle, we should impose the stronger condition $y\varphi \ll \frac{\alpha_G T}{2}$ to guarantee that the fermion masses are dominated by thermal contributions, as assumed in Fig. \ref{Fig.9}. However, since the effective fermion masses are only relevant if their magnitude is comparable to $m_\phi$, these two conditions are more or less equivalent.

Each plot in Fig. \ref{Fig.9} shows that there exists a range of the coupling constant in which Eq. \eqref{h_0} is valid. 
In this regime, $y$ can be determined from $n_s$ by using the relation  \eqref{h_0}.
From Fig. \ref{Fig.9} one also observes that as in the cases of the other interactions considered above the thermal effects can be neglected as long as one deals with perturbative reheating.

\newpage

\section{Conclusions}\label{Conclusions}
In inflationary cosmology, the reheating era is probably the least understood epoch of the cosmic history. It is, however, of enormous importance for both, cosmology and particle physics. The expansion history during reheating affects the relation between observed CMB modes and physical scales during inflation. This effect can considerably modify the predictions of inflationary models for CMB observables. This modification is sensitive to the inflaton's couplings to other fields, as illustrated in Figs. \ref{planck_phichi2}, \ref{planck_phichi3} and \ref{planck_Yukawa}. In the present work, we have studied the perspectives to make use of this dependency to  measure (or at least constrain) the inflaton coupling(s) from CMB observations. Any information on these couplings is of tremendous importance for particle physics because this can help to understand how the mechanism that drove inflation may be embedded into a more fundamental theory of nature. Since the scale of inflation may be many orders of magnitude larger than the electroweak scale, the CMB may be the only probe of these fundamental parameters, which were crucial to set the stage for the ``hot big bang'', i.e., the radiation dominated era.  

The effect of reheating on the CMB can easily be understood at a qualitative level: For larger coupling constants, the duration of the reheating era decreases 
because larger couplings generally imply a more efficient transfer of energy from the inflaton to other degrees of freedom. Accordingly, the $e$-folding number $N_k$ between the horizon crossing of the perturbation $\phi_k$ of wave number $k$ and the end of inflation increases (see Fig. \ref{Fig1}). 
In other words, horizon crossing occurs at larger field values $\phi_k$, 
where the slow roll parameters are smaller, and therefore the spectral index $n_s$ is closer to 1. 
This argument is rather universal, in the sense that no matter what kind of interactions drives the reheating, the coupling constant should be a monotonically increasing function of $n_s$. 
In practice it is, however, in general very difficult to establish a simple relation between the coupling constants and CMB observables because reheating is a highly non-linear far-from-equilibrium process that may be driven by non-perturbative feedback effects, such as parametric resonances. A detailed understanding of this process requires expensive numerical simulations, which have to be repeated for many choices of values for a potentially large number of unknown parameters in each model of inflation. 

Most of the complications occur due to effects that can broadly be described as feedback of the produced particles on the ongoing reheating process. Since these particles are diluted by Hubble expansion, which competes with the particle production during reheating, one may expect that the feedback effects become negligible if the particle production proceeds sufficiently slow. This suggests that there may be a regime of small inflaton coupling constants in which one can find simple relations between these fundamental parameters and CMB observables.
In the present work, we have shown the existence of this regime in a particular class of inflationary models, $\alpha$-attractor models of the E type, and for three types of inflaton couplings to other scalars and fermions. Roughly speaking, this requires coupling constants that are smaller than $10^{-5}$, where the exact numerical value depends on the type of interaction and the values of other parameters in the inflaton potential.
Hence, if the coupling of other fields to the inflaton is not stronger than the coupling of the electron to the Higgs field, then there exist simple relations between the coupling constant and the spectral index of CMB perturbations in the class of models we consider.
These are given in equations (\ref{coupling}), (\ref{g3_0}) and (\ref{h_0}) and illustrated in Figs. \ref{phichi2_pert}, \ref{phichi3_2}-\ref{phichi3_3} and \ref{Fig.9}.
For the interactions we consider, the linear approximation (\ref{LogGamma}) holds in good approximation across the observationally allowed range of values for the spectral index.

The figures \ref{phichi2_pert}, \ref{phichi3_2}-\ref{phichi3_3} and \ref{Fig.9} also illustrate that,
under the present assumptions,
a measurement of the spectral index $n_s$ uniquely determines the scalar-to-tensor ratio $r$ and the reheating temperature $T_{\rm re}$.
If the parameter $r$ is measured in the future, this measurement can therefore be used as an independent cross-check. Alternatively it can be used to obtain a best fit value for the parameter $\alpha$ in the inflaton potential, which we had to fix in order to uniquely relate the inflaton coupling to $n_s$.
Our results can directly be applied if the reheating temperature is smaller than $10^8$ - $10^9$ GeV. Larger reheating temperatures require inflaton coupling constants larger than $10^{-5}$, which typically lead to parametric resonances. In this case the CMB may still contain valuable information about the reheating era, but the simple relations 
(\ref{coupling}), (\ref{g3_0}) and (\ref{h_0}) do not hold.

If taken at face value, our results indicate that the inflaton coupling can be 
determined from CMB data. Such a measurement would be a major breakthrough for both, particle physics and cosmology. However, several comments are in place.
In order to obtain a unique relation between the coupling constant and observable quantities, we had to assume that reheating is primarily driven by one specific interaction term in the inflaton Lagrangian. If several types of interactions play a role, then one can only constrain a combination of the respective coupling constants. We also had to assume that no significant amounts of entropy were injected into the primordial plasma after reheating.
If this assumption is dropped, one can only impose an upper bound on the coupling constant. 
Even if these assumptions are fulfilled, the range of values for the inflaton coupling that is consistent with the Planck observation of $n_s$ at the $1 \sigma$ level still spans several orders of magnitudes. 
In order to ``measure'' at least the order of magnitude of the inflaton coupling reliably, 
the present error bar for $n_s$ would have to be reduced by an order of magnitude.  
Finally, the biggest caveat in our results lies in their model dependence. The relations between the inflaton couplings and CMB observables can only be established within a fixed model of inflation, and for given values of the parameters in the inflaton potential. 
If one leaves these parameters (or even the choice of model) free and aims to obtain best fit values from CMB data, then there are considerable parameter degeneracies that make it difficult to extract meaningful constraints on the inflaton coupling.

In spite of these practical difficulties, we consider the results of the present work very encouraging. They show that it is in principle possible to constrain  the inflaton couplings (and thereby the reheating temperature) from cosmological data. These fundamental microphysical parameters have an immense impact on the evolution of the cosmos by setting the stage for the hot big bang, and they contain information about fundamental physics at an energy scale that may never be accessible to any laboratory experiments. 
We expect that at least some of the practical limitations can be overcome in the future when the error bars on the existing cosmological observables further shrink, and when other cosmological probes become available, such as B-mode polarisation, non-Gaussianities or gravitational waves.

\section*{Acknowledgements}
We are grateful to Christophe Ringeval and Jan Hamann for inspiring discussions and comments on our work.
We also thank Sebastien Clesse, Paolo Creminelli, Eichiro Komatsu, Fernando Quevedo and Antonio Racioppi for very helpful discussions and comments on the manuscript. This  research  was  supported  by  the  DFG  cluster  of  excellence  Origin and Structure of the Universe (www.universe-cluster.de).

\begin{appendix}

\section{Thermal damping rate for the $h\phi\chi^3$ interaction}
\subsection{Thermally corrected rate of 3-body decay $\phi \rightarrow \chi\chi\chi$} \label{AppendixA}

In this appendix we compute the thermal correction to Eq. \eqref{gamma3} for the zero momentum mode of $\phi$.
This can be done by including in the vacuum rate (before the 4-momenta integrations) the Bose enhancement factor for the final states
and replacing the vacuum mass $m_\chi$ by thermal effective mass $M_\chi$. After integration over energies of three $\chi$-particles and one spatial momentum, one obtains 
\begin{align} \label{SSE}
\Gamma_{\phi \rightarrow \chi\chi\chi}=&\frac{h^2}{12 m_\phi}\int{\frac{d^3\p_1}{(2\pi)^5 2\Omega_1}\frac{d^3\p_3}{2\Omega_2 2\Omega_3}}
\,\delta(m_\phi-\Omega_1-\Omega_2-\Omega_3) \nonumber\\
& \quad \times \Big[\left(1+f_B(\Omega_1)\right)\left(1+f_B(\Omega_2)\right)\left(1+f_B(\Omega_3)\right)
-f_B(\Omega_1) f_B(\Omega_2) f_B(\Omega_3)\Big]\Big\vert_{-\p_2=\p_1+\p_3}\;.
\end{align}   
Here $\Omega_{i}^2=M_\chi^2+\p_i^2$ with $\p_i$ being 3-momentum of $i$th $\chi$-particle, and $f_B(\omega)=(e^{\omega/T}-1)^{-1}$ is the equilibrium Bose-Einstein distribution function.
Let $\theta$ be an angle between $\p_1$ and $\p_3$ and introduce new variable 
\begin{equation}
u^2\equiv \abs{\p_1+\p_3}^2=\abs{\p_1}^2+\abs{\p_3}^2+2\abs{\p_1}\abs{\p_3}\cos\theta\;,
\end{equation}\label{u_pm}
which takes the value between $u_-$ and $u_+$ defined as
\begin{align}
u_-&=\abs{\abs{\p_1}-\abs{\p_3}}\;,\\
u_+&=\abs{\p_1}+\abs{\p_3}\;.
\end{align}
We rewrite Eq. \eqref{SSE} in the spherical coordinates of $\p_3$ using $u$ instead of $\theta$ as
\begin{align}\label{D0}
\Gamma_{\phi \rightarrow \chi\chi\chi}&=\frac{h^2}{12 m_\phi}\int{\frac{d^3\p_1}{(2\pi)^5 2\Omega_1} 
\frac{p_3^2 dp_3}{2\Omega_3}\frac{1}{p_1p_3}}\Bigg\{\int_0^{2\pi}{d\varphi}\int_{u_-}^{u_+}{\frac{udu}{2\Omega_2}}\, 
\delta(m_\phi-\Omega_1-\Omega_2-\Omega_3)\nonumber\\ 
& \times \Big[\left(1+f_B(\Omega_1)\right)\left(1+f_B(\Omega_2)\right)\left(1+f_B(\Omega_3)\right)
-f_B(\Omega_1) f_B(\Omega_2) f_B(\Omega_3)\Big]\Bigg\}\;,
\end{align}
where $\Omega_2$ is given by a function of $u$
\begin{equation}
\Omega_2^2=M_\chi^2+u^2\;.
\end{equation}
This integral has a non-zero value only if there exists some $u\in [u_-,u_+]$ satisfying
\begin{equation}
m_\phi-\Omega_1-\Omega_3-\sqrt{M_\chi^2+u^2}=0\;,
\end{equation}
which confines the phase spaces such that
\begin{equation}\label{om2}
m_\phi-\Omega_1-\sqrt{u_+^2+M_\chi^2}\geq\Omega_3\geq m_\phi-\Omega_1-\sqrt{u_-^2+M_\chi^2}\;.
\end{equation}
We introduce Lorentz invariant variables $m_{ij}^2=(p_i+p_j)^2$, then we have
\begin{equation} \label{Omega_k-m_ij}
\Omega_k=\frac{m_\phi^2+M_\chi^2-m_{ij}^2}{2m_\phi}\;,\;i\neq j\neq\ k\;.
\end{equation}
Furthermore, it turns out that Eq. \eqref{om2} is translated to the kinematic limits of the Lorentz invariant quantities
\begin{align}\label{kinemactics}
(m_{23}^2)_{max}&= (m_\phi-M_\chi)^2\;,\\
(m_{23}^2)_{min}&= 4M_\chi^2\;,\\
(m_{12}^2)_{max}&= 2M_\chi^2+\frac{1}{2m_{23}^2}\Big[(m_\phi^2-m_{23}^2-M_\chi^2)m_{23}^2\nonumber\\&\;\;\;\;\;\;\;\;+ 
\sqrt{f(m_{23}^2,m_\phi^2,M_\chi^2)\cdot f(m_{23}^2,M_\chi^2,M_\chi^2)}\,\Big]\;,\\
(m_{12}^2)_{min}&= 2M_\chi^2+\frac{1}{2m_{23}^2}\Big[(m_\phi^2-m_{23}^2-M_\chi^2)m_{23}^2\nonumber\\&\;\;\;\;\;\;\;\;
- \sqrt{f(m_{23}^2,m_\phi^2,M_\chi^2)\cdot f(m_{23}^2,M_\chi^2,M_\chi^2)}\,\Big]\;,
\end{align}
where $f(x,y,z)=x^2+y^2+z^2-2xy-2yz-2zx$.
Then Eq. \eqref{D0} becomes  
\begin{eqnarray}\label{SD}
\Gamma_{\phi \rightarrow \chi\chi\chi}&=& \frac{h^2}{6}\int d(m_{12}^2) d(m_{23}^2) \frac{1}{32m_\phi^3(2\pi)^3} \nonumber \\
&& \times\Big[\left(1+f_B(\Omega_1)\right)\left(1+f_B(\Omega_2)\right)\left(1+f_B(\Omega_3)\right)
-f_B(\Omega_1) f_B(\Omega_2) f_B(\Omega_3)\Big]\;, \nonumber \\
\end{eqnarray}
with $\Omega_k$ given by Eq. \eqref{Omega_k-m_ij}. One can easily check that at zero temperature limit the above boils down to Eq. \eqref{gamma3}. 
Eq. \eqref{SD} can also be derived by taking only the decay part of the spectral self-energy given in Appendix \ref{On-shell}.

\subsection{The self-energy \texorpdfstring{$\Pi_\p^-$}{Lg} from setting-sun diagram} \label{On-shell}

\begin{figure}
	\center
	\includegraphics[scale=0.3]{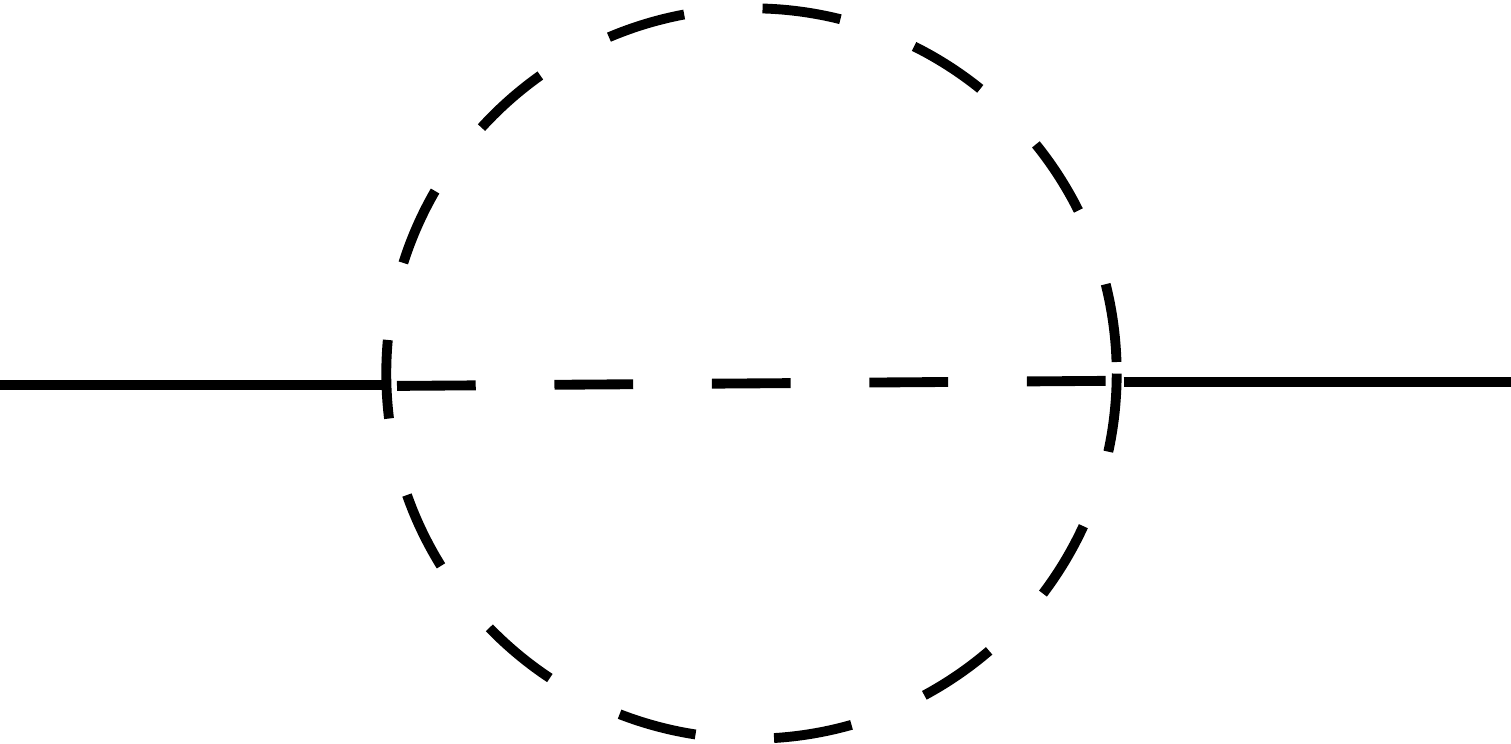}
	\caption{Setting sun diagram for $h \phi \chi^3/3!$ interaction. Solid and dashed lines correspond to $\phi$ and $\chi$, respectively.}
	\label{fig:diag_phi3sig}
\end{figure}

In this appendix we present the expression for the spectral self-energy $\Pi_\p^-$ for the inflaton coming from the interaction with the scalar field $\chi$ via  Eq. \eqref{chi3}. This self-energy is represented by the setting-sun diagram shown in Fig. \ref{fig:diag_phi3sig} and includes both, contributions from scatterings as well as decays and inverse decays. 

The dissipation rate can be obtained from the spectral self energy as
\begin{equation}\label{gamma3_t1}
\Gamma=\frac{\textmd{Im}{\Pi_0^-(m_\phi)}}{2m_\phi}\;.
\end{equation}
The spectral self-energy $\Pi_\p^-$ from the setting-sun diagram in Fig. \ref{fig:diag_phi3sig} has a general expression \cite{Cheung:2015iqa, Buchmuller:1997yw} 
\begin{eqnarray}
\Pi^-_{ \p}(p_0)&=& -\frac{i h^2}{6} \int \frac{d^4q}{(2 \pi)^4} \frac{d^4k}{(2 \pi)^4} \frac{d^4 l}{(2 \pi)^4} (2 \pi)^4 \delta^{(4)}(p-q-k-l) 
\rho_{\chi}(q) \rho_{\chi}(k) \rho_{\chi}(l) \nonumber \\
&& \times \Big[\left(1+f_B(q_0)\right) \left(1+f_B(k_0)\right) \left(1+f_B(l_0)\right)
-f_B(q_0)f_B(k_0)f_B(l_0)\Big]\,, \nonumber \\ \label{SettingSunB} 
\end{eqnarray}
where $\rho_{\chi}$ is the spectral density for $\chi$-particles. 
Using the zero-width limit approximation for $\rho_{\chi}$ and after integrations over $q$ and some angles of momenta $\k$ and $\l$, we get\footnote{An overall factor 2 is missing in Eqns- (95) and (124) of Ref. \cite{Cheung:2015iqa}.}
\begin{eqnarray}
\Pi^-_{ \p}(p_0)=- 2i \big(\mathcal{D}_{\p}(p_0)+\mathcal{S}_\p(p_0)-\mathcal{D}_\p(-p_0)-\mathcal{S}_{\p}(-p_0)\big) \label{SettingSunB1}
\end{eqnarray}
with
\begin{eqnarray}
\mathcal{D}_{\p}(p_0)&=&\pi\frac{h^2}{24(2\pi)^5}\theta(p_0)
\int_{M_\chi}^{p_0-2M_\chi} d\Omega_{\l} \int_{M_\chi}^{\Omega^-} d\Omega_{\k} \, F_d(\Omega_{\k},\Omega_{\l},-A)\,\mathcal{I}(a,b,c) \nonumber \\
\label{D}\\
\mathcal{S}_{\p}(p_0)&=&3\pi\frac{h^2}{24(2\pi)^5}\theta(p_0)
\int_{M_\chi}^\infty d\Omega_{\l} \int_{\Omega^+}^\infty d\Omega_{\k}\, F_s(\Omega_{\k},\Omega_{\l},A)\,\mathcal{I}(a,b,c). \label{S}
\end{eqnarray}
Here
\begin{eqnarray}
F_d(\Omega_{\k},\Omega_{\l},A)&=&
\big(1+f_B(\Omega_{\k})\big) \big(1+f_B(\Omega_{\l})\big)
\big(1+f_B(A)\big)-
f_B(\Omega_{\k}) f_B(\Omega_{\l})  f_B(A), \\
F_s(\Omega_{\k},\Omega_{\l},A)&=&
\big(1+f_B(\Omega_{\k})\big) \big(1+f_B(\Omega_{\l})\big)
f_B(A)-
f_B(\Omega_{\k}) f_B(\Omega_{\l})  \big(1+f_B(A)\big), \label{F_s} \\
\Omega^\pm&=&{\rm max}\big[p_0-\Omega_{\l}\pm M_\chi,M_\chi\big], \\
A&=&\Omega_{\k}+\Omega_{\l}-p_0, \label{A} \\
\mathcal{I}(a,b,c)&=&\int_{-1}^{1}dx \int_{-1}^1 dy\, \frac{\theta\left(I(x,y,z)\right)}{\sqrt{I(x,y,z)}}, \label{I_B}\\
z&=& a + b\, y + c\, x, \\
I(x,y,z)&=&(1-x^2)(1-y^2)-(z-xy)^2, \\
a&=& \frac{ A^2 - \p^2-\k^2-\l^2-M_\chi^2}{2 |\k| |\l|}, \,\, b= \frac{|\p|}{|\k|}, \,\, c= \frac{|\p|}{|\l|}. 
\end{eqnarray}
The $\mathcal{D}$ terms represent decay and inverse decay $\phi\leftrightarrow \chi\chi\chi$ already discussed in Appendix \ref{AppendixA}. The integration limits imply that $\mathcal{D}_{\p}(p_0)$ vanishes for $p_0<3M_\chi$ due to the energy-momentum conservation and the decay processes $\phi\rightarrow \chi\chi\chi$ is forbidden. $\mathcal{S}$ terms correspond to the damping by scattering processes, $\phi \chi \leftrightarrow \chi \chi$. The variables $x$ and $y$ are the cosines of the two nontrivial angles, i.e. $x=\frac{\p \cdot \k}{|\p||\k|}$, 
$y=\frac{\p \cdot \l}{|\p||\l|}$. $\mathcal{I}(a,b,c)$ is the integral over these angles and was computed in appendix C of Ref. \cite{Cheung:2015iqa}. 

There are two limiting cases in which analytic approximations exist,
\begin{eqnarray}
\Gamma&\simeq& \frac{h^2T^2}{768\pi m_\phi} 
\end{eqnarray}
for  $M_\chi\simeq m_\phi$ \cite{Parwani:1991gq} and 
\begin{eqnarray}
 \Gamma&\simeq& \frac{h^2m_\phi }{6(2\pi)^4}\frac{T^2}{M_\chi^2}\left[
1 + \log\left(\frac{81}{8}\frac{M_\chi}{m_\phi}\right)\right]
\end{eqnarray}
for $M_\chi\gg m_\phi$ \cite{Drewes:2013iaa}.  

\end{appendix}

\newpage

\bibliographystyle{JHEP}
\bibliography{ref} 
\end{document}